\documentclass[10pt,prd,twocolumn,preprintnumbers,superscriptaddress,amsmath,amssymb,aps,longbibliography,nofootinbib]{revtex4-2}

\usepackage{dsfont}
\usepackage[utf8]{inputenc}
\usepackage{graphicx}   
\usepackage[table,dvipsnames]{xcolor}
\usepackage{colortbl}
\usepackage{ragged2e}
\usepackage{url}
\usepackage[normalem]{ulem}
\usepackage[english]{babel}
\usepackage{slashed,centernot}
\usepackage{bbold}
\usepackage[noindentafter]{titlesec} 
\usepackage{epigraph}
\usepackage{booktabs} 
\usepackage{upgreek}
\usepackage{physics}
\usepackage{pifont}
\usepackage{booktabs}
\usepackage{braket}
\usepackage{bm}
\usepackage{soul}
\usepackage{cancel}
\usepackage{changes}

\usepackage{array}
\usepackage{tabularx}
\newcolumntype{Y}{>{\raggedright\arraybackslash}X}

\definecolor{red}{rgb}{0.6,.0706,.1373}
\definecolor{blue}{rgb}{0,0.396,0.741}
\definecolor{realblue}{rgb}{0,0,1}
\definecolor{green}{rgb}{0.25,0.6,0.2}
\definecolor{rossoc}{cmyk}{0,1,1,0.2}

\definecolor{teal}{rgb}{0, 0.5,0.5}
\colorlet{mylinkcolor}{teal}
\colorlet{mycitecolor}{teal}
\colorlet{myurlcolor}{teal}

\usepackage{hyperref}
\usepackage[capitalise, english]{cleveref}
\hypersetup{
  linkcolor  = mylinkcolor,
  citecolor  = mycitecolor,
  urlcolor   = myurlcolor,
  colorlinks = true
}

\def\Sigmatwo{\ensuremath{\Sigma}}

\newcommand{\QCD}{\mathrm{SU}(3)_{\textrm{QCD}}}
\newcommand{\SU}[1]{\mathrm{SU}(#1)}
\newcommand{\Uone}{\mathrm{U}(1)}
\newcommand{\UPQ}{\mathrm{U}(1)_{\textrm{PQ}}}
\newcommand{\alphas}{\alpha_s}
\newcommand{\alphaem}{\alpha_{\textrm{em}}}
\newcommand{\LambdaQCD}{\Lambda_{\mathrm{QCD}}}
\newcommand{\barLambdaQCD}{\bar{\Lambda}_{\mathrm{QCD}}}
\newcommand{\LambdaQCDbar}{\bar{\Lambda}_{\mathrm{QCD}}}

\begin{document}
\preprint{ IFT-UAM/CSIC-26-58 }

\title{The structure of multi-axion solutions to the strong CP problem}

\author{Mario Fern\'andez Navarro}
\email{mario.fernandeznavarro@physik.uzh.ch}
\affiliation{Physik-Institut, Universitat Z\" urich, 8057 Z\" urich, Switzerland}

\author{Marta F. Zamoro}
\email{marta.zamoro@uam.es}
\affiliation{Departamento de F\'isica Te\'orica and Instituto de F\'isica Te\'orica UAM/CSIC, \\
Universidad Aut\'onoma de Madrid, Cantoblanco, 28049, Madrid, Spain}

\author{Marko Pesut}
\email{marko.pesut@psi.ch}
\affiliation{Physik-Institut, Universitat Z\" urich, 8057 Z\" urich, Switzerland}
\affiliation{PSI Center for Neutron and Muon Sciences,
5232 Villigen PSI, Switzerland}

\author{Xavier Ponce D\'iaz}
\email{xavier.poncediaz@unibas.ch}
\affiliation{Department of Physics, University of Basel, Klingelbergstrasse 82,  CH-4056 Basel, 
Switzerland}

%=============================================================================
\begin{abstract}

A broad experimental program is targeting the QCD axion band predicted by single-axion solutions to the strong CP problem. Multi-axion theories provide a well-motivated departure from this canonical picture, since additional states generically modify the mass-photon-coupling relation. We investigate the general structure of multi-axion solutions to the strong CP problem and study the different qualitative mass-coupling patterns that arise, including axions to the right of the QCD band, axions in the experimentally accessible region to its left, and scenarios in which the QCD axion band itself is displaced. This general treatment reveals a broad set of phenomenological possibilities that are not captured by more restrictive assumptions. In particular, we identify the structure of Peccei-Quinn symmetry breaking and the relative alignment between the QCD and electromagnetic anomalies as key ingredients determining the location of the axions in parameter space. Combining these ingredients, we derive a general sum rule for $N$-axion systems that incorporates both general PQ breaking and non-universal anomaly coefficients. We apply the framework to the two-axion system and to general multi-axion setups, identifying UV-complete theories in which the different phenomenological regimes arise naturally. Our results motivate an extended axion search program and have implications for our understanding of fundamental physics and the ultraviolet completion of the Standard Model.

\end{abstract}

\maketitle

\tableofcontents

\section{Introduction}

The absence of CP violation in quantum chromodynamics (QCD) is one of the most robust and remarkable fine-tuning puzzles of the Standard Model (SM). Although CP violation in the Standard Model arises from an $\mathcal{O}(1)$ complex phase in the Yukawa sector, the non-observation of a neutron electric dipole moment implies the stringent bound $|\bar{\theta}| = |\theta_\textrm{QCD} + \arg \det (Y_u Y_d)| \lesssim 10^{-10}$ \cite{Abel:2020pzs}. Since no symmetry enforces such a small value or parametric cancellation, this discrepancy gives rise to the so-called \emph{Strong CP problem}.

Despite recent discussions regarding its interpretation \cite{Ai:2020ptm,Ai:2024vfa, Ai:2024cnp,Kaplan:2025bgy,Benabou:2025viy,Ai:2025quf,Gamboa:2025hxa,Khoze:2025auv,Bhattacharya:2025qsk,Ringwald:2026apz,Aghaie:2026pkf}, the QCD axion \cite{Peccei:1977hh, Peccei:1977ur,Weinberg:1977ma,Wilczek:1977pj, DiLuzio:2020wdo} remains a compelling dynamical solution to the Strong CP problem. In this approach, a global $\mathrm{U}(1)_{\rm PQ}$ Peccei-Quinn (PQ) symmetry, anomalous under QCD and spontaneously broken at a scale $f_a$, provides a pseudo–Nambu–Goldstone boson, the axion, which acquires a mass through instanton effects and whose dynamics relax $\bar\theta$ to zero, thereby explaining the smallness of the QCD vacuum angle. The simplicity of this mechanism, along with its possible role as a dark matter candidate~\cite{Abbott:1982af,Dine:1982ah,Preskill:1982cy} and its generic realization in string theory~\cite{Witten:1984dg,Arvanitaki:2009fg,Svrcek:2006yi,Dine:1986bg,Halverson:2019cmy,Gendler:2023kjt}, makes it a compelling candidate for BSM physics.

In the canonical QCD axion solution, the mass of the axion $m_a$ is set by nonperturbative QCD effects and related to the PQ breaking scale $f_a$ via 
\begin{equation}
m_a f_a \simeq m_\pi f_\pi \frac{\sqrt{m_u m_d}}{m_u+m_d}\equiv \LambdaQCD^2 \,.
\label{eq:mass}
\end{equation}
Additionally, the axion mass is closely tied to the axion-photon coupling. In axion models, this coupling receives two distinct contributions. The first is model independent and arises from pion-axion mixing, a generic feature of the axion solution to the strong CP problem. The second depends on the implementation of the QCD and electromagnetic anomalies in the Peccei–Quinn model and is determined by the ratio of the corresponding anomaly coefficients, $E/N$ \cite{Cheng:1995fd}. Taking both contributions into account, the axion-photon coupling can be written as
\begin{equation}
\label{eq:axion_window}
    |g_{a\gamma}| = \frac{\alphaem}{2\pi}\left|\frac{E}{N}-C_\chi\right|\frac{m_a}{\LambdaQCD^2} \equiv \left| C_{a\gamma} \right| \frac{m_a}{\LambdaQCD^2}\, ,
\end{equation}
where $C_{\chi}=1.92(4)$~\cite{GrillidiCortona:2015jxo} (see also Refs.~\cite{Lu:2020rhp,Gao:2024vkw}). This relation defines the so-called QCD line in the $m_a-g_{a\gamma}$ plane, or rather QCD band, since it is defined by the class of UV-complete axion models for which the photon coupling is fully calculable. This band provides a well-motivated experimental target and includes most of the simplest axion constructions. Imposing, in addition, a viable cosmological history together with theoretical consistency conditions further restricts the parameter space to an axion window~\cite{DiLuzio:2016sbl,DiLuzio:2017pfr,Plakkot:2021xyx,Cheek:2023fht,DiLuzio:2024xnt}, which typically corresponds to $|C_{a\gamma}|\in \left[0.03,\, 1.5\right]\times10^{-2}$.

\begin{figure}
    \centering
    \includegraphics[width=1\linewidth]{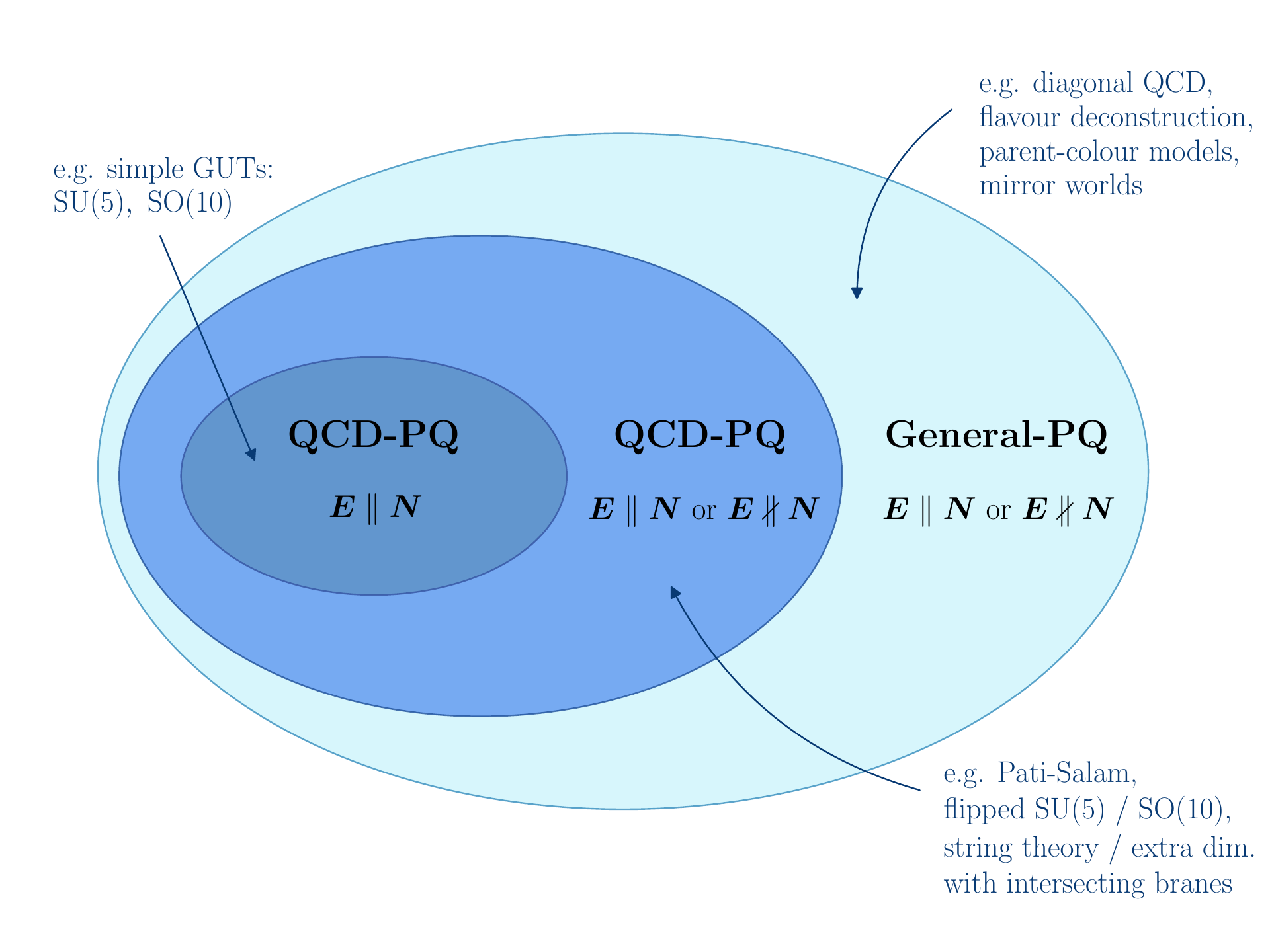}
    \caption{Theory space of (multi-)axion solutions to the strong CP problem. We show that (multi-)axion systems are governed by two structural conditions: whether there exists a PQ symmetry only broken by QCD (QCD-PQ) or more general PQ breaking is present (General-PQ), and whether anomaly coefficient ratios $E_i/N_i$ are universal for all axions (then the QED and QCD anomaly vectors $\boldsymbol{E}$ and $\boldsymbol{N}$ are \mbox{parallel}) or not. These two conditions connect the experimental phenomenology with the low-energy axion system and with possible ultraviolet completions, allowing a systematic study and a general classification. Representative UV realisations of each scenario are included as illustrative examples, not as an exhaustive list.}
    \label{fig:QCD-general-PQ}
\end{figure}

The possibility of identifying axion solutions to the Strong CP problem that deviate from~\cref{eq:mass,eq:axion_window} has motivated extended theoretical efforts \cite{Dine:1986bg,Choi:1988sy,Choi:1998ep,Hook:2018jle,Rubakov:1997vp,Berezhiani:2000gh,Albertus:2026fbe}. Specifically, the mass of the axion can be modified by enhancing instanton effects. This can be achieved, for instance, via modified coupling running \cite{Holdom:1982ex,Holdom:1985vx,Flynn:1987rs} or by embedding QCD into a higher dimensional theory~\cite{Poppitz:2002ac,Gherghetta:2020keg} in order to trigger the confinement at a higher scale. It has also been pointed out that instantons originating from particular classes of UV extensions, such as models with extended confining sectors~\cite{Rubakov:1997vp,Croon:2019iuh,Reig:2019vqh,Gavela:2018paw} or non-trivial embeddings of QCD into a UV gauge group~\cite{Gherghetta:2016fhp,Agrawal:2017ksf,Gaillard:2018xgk,Fuentes-Martin:2019bue,Csaki:2019vte,Dimopoulos:1979pp,Valenti:2022tsc}, could provide the dominant contribution to the axion mass, even if the UV theory remains weakly coupled. Therefore, they would modify the prediction of \cref{eq:mass} and open new regions of axion parameter space. Other mechanisms relying on the misalignment between the axion-mass and interaction bases have been explored e.g.~in~\cite{Gavela:2023tzu,Agrawal:2017cmd}.

A common feature of some of these different approaches is the presence of multiple axions, whose masses and couplings are typically shifted relative to the standard QCD axion relation, \cref{eq:axion_window}, often lying to the right of the canonical QCD line \cite{Rubakov:1997vp,Berezhiani:2000gh,Gianfagna:2004je,Hsu:2004mf,Hook:2014cda,Fukuda:2015ana,Hook:2014cda,Blinov:2016kte,Chiang:2016eav,Dimopoulos:2016lvn,Gherghetta:2016fhp,Kobakhidze:2016rwh,Agrawal:2017ksf,Agrawal:2017evu,Gaillard:2018xgk,Hook:2019qoh,Gherghetta:2020ofz,Csaki:2019vte,Kivel:2022emq} in parameter space, and, perhaps less generically, to its left \cite{Farina:2016tgd,Agrawal:2017cmd, Hook:2018jle,DiLuzio:2021gos,DiLuzio:2021pxd}. Moreover, multi-axion systems naturally appear in well-motivated extensions of the SM, such as in theories with additional confining sectors commuting with QCD~\cite{Rubakov:1997vp,Berezhiani:2000gh,Strassler:2006im,Hook:2014cda,DiLuzio:2021pxd,Chen:2021jcb,Kondo:2025hdc,Lee:2026umy}, non-trivial embeddings of QCD~\cite{Gherghetta:2016fhp,Agrawal:2017ksf,Gaillard:2018xgk,Fuentes-Martin:2019bue,Csaki:2019vte}, extra dimensions~\cite{Dienes:1999gw,Flacke:2006ad,deGiorgi:2024elx} or string theory~\cite{Witten:1984dg,Arvanitaki:2009fg,Svrcek:2006yi,Dine:1986bg,Halverson:2019cmy,Gendler:2023kjt}, among others. On the experimental side, the presence of multiple axion solutions motivates the search for these particles outside of the QCD axion band. In fact, axions lying to the left of the QCD line are already within the reach of current and future experiments, whereas those to the right might require entirely new experimental searches. In both regimes, multi-axion scenarios can qualitatively modify experimental signatures and discovery prospects, including laboratory experiments~\cite{deGiorgi:2025ldc,Dunsky:2025sgz}.

Although theories with multiple axions have been considered in several works, see Refs.~\cite{Farina:2016tgd,Agrawal:2017cmd,Agrawal:2022wyu,Gavela:2023tzu}, most existing analyses focus on specific classes of axion solutions or impose additional assumptions that do not capture the full structure of multi-axion theories. In this paper, we develop a general framework for multi-axion solutions to the Strong CP problem and identify the conditions under which axion states populate experimentally and theoretically relevant regions of parameter space, both to the left and to the right of the QCD band. In particular, we go beyond the assumptions that the axion potential preserves a PQ symmetry broken only by QCD and that all axions share a universal $E/N$ anomaly coefficient ratio. Relaxing these assumptions reveals a substantially broader and phenomenologically richer axion landscape. We show that this landscape can be organised into qualitatively distinct scenarios, whose mass-coupling patterns encode the structure of Peccei-Quinn symmetry breaking and point to specific classes of ultraviolet realisations. Understanding the underlying structure of multi-axion systems then allows a systematic study of their qualitative phenomenology and a general theory landscape classification, shown in Figure \ref{fig:QCD-general-PQ}.

Our results therefore highlight the multi-axion paradigm as a powerful framework for understanding fundamental physics and the ultraviolet completion of the SM. At the same time, they provide strong motivation for a combined effort from the experimental side, including present and new experimental techniques. 

The paper is organised as follows. In Section \ref{sec:MultiAxion}, we define the generic multi-axion setup. In Section \ref{sec:Twoaxion}, we study the structure of the simplest two-axion example and identify the different qualitatively distinct regimes, while also deriving a general sum rule that incorporates non-universal anomaly coefficients and general PQ breaking structure. Guided by these results, we turn to the analysis of a general multi-axion theory in Section \ref{sec:generalization}, and identify the relevant conditions as well as regimes under which axions can lie in different regions of parameter space. Similarly, we are able to identify the 
sum rule that governs the general $N$ axion system. Finally, we conclude and outline possible future directions in Section \ref{sec:conclusion}. 
In Appendix \ref{app:UVmodels} we include a simplified yet comprehensive introduction to well-motivated UV complete realisations of the phenomenological scenarios discussed in the main text. Appendix \ref{app:Hell} introduces an alternative parameterisation of the multi-axion setup to that of Section \ref{sec:MultiAxion}, which provides complementary perspectives on the structure of the axion theory. Appendix~\ref{app:Two-axion_app} provides additional details on the two-axion setup studied in \cref{sec:Twoaxion}. Although the main analysis of this paper focuses on the photon coupling, in Appendix~\ref{app:fermion_couplings} we comment on the structure of other phenomenologically relevant couplings in multi-axion systems, namely those to the lighter fermion families. Finally, in Appendix~\ref{app:N_0} we show that ALPs whose mixing with other axions is not induced by QCD instantons can also be mapped onto the phenomenological scenarios studied in the main part of this work.

\section{General structure of multi-axion systems} \label{sec:MultiAxion}
\subsection{Classification of solutions to strong CP and anomaly structure}
Let us define the generic setup of $\UPQ^N$ symmetries anomalous under QCD\footnote{In App.~\ref{app:N_0} we consider the presence of $\mathrm{U}(1)$ symmetries not anomalous under QCD in the interaction basis.}. In this case, we can write the low-energy Lagrangian as
\begin{equation}
\label{eq:gluon_photn_Lagrangian}
    \mathcal{L} = \frac{\alphas}{8\pi}\, (\bm{N}^T\cdot \bm{a}-\bar{\theta}) \, G\widetilde{G} +\frac{\alphaem}{8\pi}\, \bm{E}^T\cdot \bm{a} \, F\widetilde{F}  \, , \end{equation} 
where we define vectors in the $N-$axion space in bold letters,
\begin{align}   
    \bm{N}=\begin{pmatrix}
        \dfrac{1}{f_1} & \cdots &
        \dfrac{1}{f_N}
    \end{pmatrix}^T  , \, &\bm{E}=\begin{pmatrix}
        \dfrac{E_1}{N_1}\dfrac{1}{f_1} & \cdots &
        \dfrac{E_N}{N_N}\dfrac{1}{f_N}
    \end{pmatrix}^T\,, \nonumber \\  \bm{a} = &\begin{pmatrix}
        a_1 & \cdots &
        a_N
    \end{pmatrix}^T \, .
\end{align}
Here, the fields $a_i$ correspond to the Nambu-Goldstone Bosons (NGBs) associated to the $\UPQ^N$ symmetries. We have used the common redefinition $f_i=v_i/N_i$, where $v_i$ represent the corresponding Vacuum Expectation Values (VEVs), associated to the Spontaneous Symmetry Breakings (SSBs) of the abelian symmetries. Similarly, we allow for all different axion fields to have a generic anomalous coupling $N_i$ $(E_i)$ to QCD (QED). Additionally, we work in a basis with diagonal and canonically normalised axion kinetic terms, in which anomaly coefficients and $f_i$ are taken as generic effective parameters.

After QCD-confinement the gluon couplings generate a potential for the axions, which can be written as\footnote{For simplicity, we show only the leading cosine (harmonic) approximation of the QCD contribution. Our results and conclusions are unchanged by retaining the full form of the QCD potential.}
\begin{align}
    V(\bm{a})=-\LambdaQCD^4 \cos{\left(\bm{N}^T\cdot\bm{a}-\bar{\theta}\right)}+V_{\cancel{\rm{PQ}}}(\bm{a})\, \label{eq:general_potential}
\end{align}
where the QCD-potential breaks one direction of the PQ-symmetries, and, assuming all axions to be massive, the generic potential $V_{\cancel{\rm{PQ}}}(\bm{a})$ may generically break all the $\UPQ^N$ symmetries.

To solve the strong CP problem without spoiling the ``quality''
of the axion mechanism, it is \textit{sufficient} (though not \textit{necessary}) to impose that $V_{\cancel{{\rm {PQ}}}}$ preserves a PQ symmetry only broken by QCD. We refer to such cases as \textit{QCD-PQ systems}. More generally, however, there exist multi-axion systems where $V_{\cancel{{\rm {PQ}}}}$ breaks all PQ symmetries and yet the minimum of the complete potential is CP conserving. We refer to these as \textit{general-PQ systems}. An illustrative example of the latter in a two-axion system is provided by an axion potential of the form
\begin{align}
V= & -\Lambda_{\mathrm{QCD}}^{4}\cos\left(\frac{a_{1}}{f_{1}}+\frac{a_{2}}{f_{2}}-\overline{\theta}\right)\nonumber \\
 & -\Lambda_{1}^{4}\cos\left(\frac{a_{1}}{f_{1}}-\overline{\theta}_{1}\right)-\Lambda_{2}^{4}\cos\left(\frac{a_{2}}{f_{2}}-\overline{\theta}_{2}\right)\,.
\end{align}
If we impose $\Lambda_{1}=0$ or $\Lambda_{2}=0$, then one of the
two PQs $\UPQ^{1}\times \UPQ^{2}$ is only broken by QCD, the minimum is CP conserving and this is a QCD-PQ system. Otherwise, if both $\Lambda_{1}$ and $\Lambda_{2}$ are non-zero, the potential still has a CP conserving minimum if we impose $\bar{\theta}=\bar{\theta}_{1}+\bar{\theta}_{2}$. In this case both PQ symmetries are generally broken beyond QCD effects, the strong CP problem is solved and this is a general-PQ system.

A natural UV realisation of such a potential with the condition $\bar{\theta}=\bar{\theta}_{1}+\bar{\theta}_{2}$
is provided by models where QCD emerges as the diagonal subgroup of
a larger strong sector, e.g.~$\SU{N}_{1}\times\SU{N'}_{2}\to\QCD$
\cite{Agrawal:2017ksf,Gaillard:2018xgk,Fuentes-Martin:2019bue,Csaki:2019vte} (see also Appendix \ref{app:QCDdiag}). In this case, the effective QCD phase emerges as $\bar{\theta}=\bar{\theta}_{1}+\bar{\theta}_{2}$, where
the $\bar{\theta}_{i=1,2}$ are the CP-violating phases associated
with the individual gauge groups in the UV. Each of them introduces
an independent source of strong CP violation, but in the presence
of two axions the generalised PQ mechanism dynamically relaxes all
of them simultaneously, thereby also driving $\bar{\theta}$ to zero.
Importantly, the instanton effects of the additional confining sectors
generically break all PQ symmetries, and in some cases all of them
are larger than the QCD instantons~\cite{Agrawal:2017ksf,Csaki:2019vte}.
We note that this construction is not unique and in Appendix \ref{app:UVmodels} we
introduce more examples of general-PQ systems that originate from
well-motivated UV theories. Therefore, in order to capture the most
general multi-axion dynamics, we do not restrict our analysis to QCD-PQ systems, unlike Refs.~\cite{Agrawal:2022wyu,Gavela:2023tzu}.

Finally, the photon coupling at low energies is also modified by the mixing between the axion and the pion. Taking into account this universal contribution, the photon couplings at low energies can be defined 
\begin{equation}
\label{eq:photon_coupling}
\mathcal{L} = \frac{1}{4} \,\bm{g}_{a\gamma}^T\cdot \bm{a}\,F\widetilde{F} \, ,
\end{equation}
 with 
 \begin{align}
    \bm{g}_{a\gamma} &= \frac{\alphaem}{2\pi}\left(\bm{E}-C_{\chi}\bm{N}\right) \nonumber\\ &=  \frac{\alphaem}{2\pi}\begin{pmatrix}
        \left[\dfrac{E_1}{N_1}-C_{\chi}\right]\dfrac{1}{f_1} & \cdots & \left[\dfrac{E_N}{N_N}-C_{\chi}\right]\dfrac{1}{f_N}
    \end{pmatrix} \nonumber^T \\ &\equiv   \begin{pmatrix}
        \dfrac{C_{a_1\gamma}}{f_1} & \cdots & \dfrac{C_{a_N\gamma}}{f_N}
    \end{pmatrix}^T\, ,
\end{align}
where $C_{a_i \gamma}\equiv \dfrac{\alphaem}{2\pi}\left(\dfrac{E_i}{N_i}-C_\chi\right)$. 

In this notation the mass matrix of the axion system can be written as
\begin{equation}
    \mathcal{L} \supset \frac{1}{2}\bm{a}^T \bm{M}^2 \bm{a},
\end{equation}
with
\begin{align}
\label{eq:gen_mass_matrix}
 \bm{M}^2  &= \LambdaQCD^4 \bm{N}\bm{N}^T + \bm{M}_{\cancel{\rm{PQ}}}^2 \, ,\quad
    \bm{M}_{\cancel{\rm{PQ}}}^2 = \left.\pdv{V_{\cancel{\rm{PQ}}}(\bm{a})}{\bm{a}}{\bm{a}^T}\right|_{\bm{a}=\bm{0}} \, .
\end{align}
If we impose the existence of a PQ symmetry only broken by QCD dynamics, i.e.~we restrict to QCD-PQ systems, then
\begin{equation}
\label{eq:IR_sol}
      \det(\bm{M}_{\cancel{\rm{PQ}}}^2)=0 \, .
\end{equation}
However, this condition is \textit{sufficient} but not \textit{necessary} to solve the strong CP problem, as discussed before, and in particular it excludes well-motivated classes of multi-axion systems with non-trivial phenomenology.

Finally, since the mass matrix is real-symmetric, we can diagonalise it with orthonormal eigenvectors $\bm{u}_i$ such that $\bm{M}^2\bm{u}_{i}=  m_i^2 \bm{u}_i$ and define the corresponding mass-eigenstate axions $\tilde{a}_i= \bm{u}^T_i \cdot\bm{a}$. Using the completeness relation of the vector space $\sum_i \bm{u}_i \bm{u}_i^T =\mathbb{1}$ the photon coupling of the mass-eigenstate is
\begin{align}
    \mathcal{L} &= \frac{1}{4} \sum_i \bm{g}_{a\gamma}^T\cdot\bm{u}_i \bm{u}^T_i \cdot \bm{a} \widetilde{F} F = \frac{1}{4} \sum_i \bm{g}_{a\gamma}^T\cdot\bm{u}_i \, \tilde{a}_i \widetilde{F} F \nonumber \\
    \label{eq:photon_eigen_coupling}
    &\equiv \frac{1}{4} \sum_i \tilde{g}_{a_i\gamma} \, \tilde{a}_i \widetilde{F} F\, , \quad \textrm{with} \quad \tilde{g}_{a_i\gamma}=\bm{g}_{a\gamma}^T\cdot\bm{u}_i \, .
\end{align}
In Ref.~\cite{Gavela:2023tzu} it was shown that, 
if \cref{eq:IR_sol} is imposed, the QCD axion sector satisfies a sum rule. Concretely, \cref{eq:IR_sol} implies the existence of a vector $\bm{w}$ such that
\begin{equation}
    \bm{M}_{\cancel{\rm{PQ}}}^2 \cdot \bm{w}=0\,,
\end{equation}
which we can normalise as $\bm{N}^T\cdot \bm{w}=1$. Acting with the full mass matrix on this vector then gives
\begin{equation}
     \bm{M}^2 \cdot \bm{w} = \LambdaQCD^4\, \bm{N}\, .
\end{equation}
Assuming $\bm{M}^2$ has no zero eigenvalues so that it is invertible, we can write
\begin{align}
    1
    &=\bm{N}^T\cdot \bm{w}
    = \LambdaQCD^4\, \bm{N}^T \bm{M}^{-2} \bm{N}
    \nonumber\\
    \label{eq:QCD_sumrule}
    &\implies\;
    \LambdaQCD^4\sum_i \frac{\left(\bm{N}^T\cdot\bm{u}_i\right)^2}{m_i^2}=1\,,
\end{align}
which is precisely the sum rule derived in Ref.~\cite{Gavela:2023tzu}, with $\bm{u}_i$ and $m_i^2$ the eigenvectors and eigenvalues of $\bm{M}^2$, respectively.

A second sum rule was also obtained in Ref.~\cite{Gavela:2023tzu} under the additional assumption that all axions couple to photons with the same anomaly ratio, i.e.~that the photon anomaly vector is aligned with the gluon anomaly vector. It then follows immediately from \cref{eq:QCD_sumrule,eq:photon_eigen_coupling} that
\begin{equation}
\label{eq:photon_sumrule}
   \frac{\LambdaQCD^4}{C_{a\gamma}^2}
   \sum_i\frac{\tilde{g}_{a_i\gamma}^2}{m_i^2}
   = 1 \, .
\end{equation}
Equivalently, these sum rules imply a strict constraint on each individual axion eigenstate, 
\begin{equation}
    \frac{\tilde{g}_{a_i\gamma}^2}{m_i^2}
    \leq
    \frac{C_{a\gamma}^2}{\LambdaQCD^4}\, ,\label{eq:sumRule_bound}
\end{equation}
namely that all axions lie to the right of the QCD line as defined in \cref{eq:axion_window}. In Ref.~\cite{Gavela:2023tzu}, the authors also identified and studied the \textit{maxion} limit, in which all axions contribute equally to the sum rule, resulting in the maximal displacement of the QCD line by a factor of $\sqrt{N}$.

In this paper, we will explore the behaviour of axion masses and photon couplings beyond the maxion configuration. Moreover, we will explore the relaxation of the two assumptions of the QCD sum rule: the existence of a PQ symmetry broken exclusively by QCD, and the alignment of the photon coupling with the QCD anomaly vector. Once either of these assumptions is relaxed, an even richer multi-axion phenomenology emerges: first, the QCD axion line itself can be displaced by much more than the $\sqrt{N}$ bound implied by the sum rule. Second, once the alignment between the photon and gluon anomaly vectors is relaxed, the strict condition of Eq.~\eqref{eq:sumRule_bound} no longer applies, and axions can be naturally found to the left of the QCD line. Crucially, as we will show in the following sections, the phenomenology of multi-axion solutions to the strong CP problem is largely determined by those two aforementioned structural conditions:
\begin{itemize}
    \item PQ structure: $\text{det} (\bm M ^2_{\cancel{\mathrm{PQ}}})= 0$ (QCD-PQ) or $\text{det} (\bm M ^2_{\cancel{\mathrm{PQ}}})\neq 0$ (general-PQ),
    \item Anomaly structure: $\bm E \parallel\bm N$ or $\bm E \nparallel\bm N$.
\end{itemize}
The first condition is related to $\bm{M}^2_{\cancel{\rm PQ}}$: in QCD-PQ systems, one PQ direction remains broken only by QCD, while in general-PQ systems the additional potential may lift all PQ directions, without necessarily spoiling the CP-conserving minimum. The second condition depends on the alignment of $\bm{E}$ and $\bm{N}$. When $\bm{E}\parallel \bm{N}$, all axions share a universal value of the anomaly ratio $E_i/N_i$, so that the photon-coupling vector is aligned with the QCD direction. When $\bm{E}\nparallel \bm{N}$, the anomaly ratios are non-universal, and this will be shown to open qualitatively new possibilities for the axion mass--coupling relation.

These two ingredients connect the experimental phenomenology with the low-energy multi-axion system as well as the UV completion, allowing a general theory landscape classification as shown in Figure~\ref{fig:QCD-general-PQ} and a systematic exploration of all the qualitatively distinct phenomenological scenarios.

\subsection{From axion potential to mass matrix parameterisation}\label{sec:Lambda_bar_definition}
We begin from the general axion potential introduced in the previous section, \cref{eq:general_potential}. The $V_{{\cancel{{\rm PQ}}}}$ potential contains all sources of PQ breaking beyond QCD and is defined as a sum over generic periodic functions\footnote{For axion potentials generated by instantons or explicit PQ-breaking scalar couplings, it is common and typically accurate to perform harmonic expansions of the periodic functions, express them as cosines (up to phase shifts), and retain only the leading harmonic, $F_{I}(\bm{v}_I^T\cdot \bm{a}-\theta_{I})\approx -\Lambda^{4}_{I}\cos(\bm{v}_I^T\cdot \bm{a}-\theta'_{I})$. 
}
\begin{equation}
V_{{\cancel{{\rm PQ}}}}(a)\;=\sum_I V_I=\;\sum_{I}F_{I}\big(\bm{v}_I^T\cdot \bm{a}-\theta_{I}\big),\label{eq:VPQ_sum}
\end{equation}
where each $\bm{v}_{I}\in\mathbb{R}^{N}$ is an anomaly vector specifying a distinct PQ-breaking source. In order to efficiently study the different phenomenological scenarios arising in multi-axion systems, we split the set of vectors $\bm{v}_I$ into two non-overlapping subsets:
\begin{equation}
    \{ \bm{v}_I \} = \mathcal{C} \,\sqcup\, \mathcal{A} \,.
\end{equation}
We define the set of PQ-breaking sources \emph{fully aligned} with QCD as those whose component orthogonal to ${\bm{N}}$ vanishes, i.e. those that are exactly collinear with ${\bm{N}}$:
\begin{equation}
\mathcal{C}
\;\equiv\;
\left\{ \bm{v}_I\;\big|\; \bm{v}_I=\xi_{I}\,{\bm{N}}\;\text{ for }\xi_{I}\in\mathbb{R}\right\} \,.
\label{eq:subset_Lambda_bar}
\end{equation}
All remaining sources of PQ breaking are defined as \emph{non-fully-aligned} and belong to the set
\begin{equation}
\mathcal{A}
\;\equiv\;
\left\{ \bm{v}_I \;\big|\;\bm{v}_I \neq \xi_I\,{\bm{N}} \;\text{ for any } \xi_I \in \mathbb{R} \right\} \,.
\end{equation}
Crucially, this classification is defined at the level of the elementary
vectors that build the axion potential in Eq.~(\ref{eq:VPQ_sum}), directly related to the fundamental sources of PQ symmetry breaking, and is invariant under orthogonal field redefinitions of $\bm{a}$. The complete axion potential can be decomposed as
\begin{flalign}
&V(\bm{a})\;  
  = -\Lambda_{\mathrm{QCD}}^{4}\cos\!\big(\bm{N}^{T}\cdot \bm{a}-\overline{\theta}\big) \nonumber\\
  &-\sum_{I\in\mathcal{C}}F_{I}\big(\xi_{I}{\bm{N}}^T\cdot \bm{a}-\theta_{I}\big)-\sum_{I\in\mathcal{A}}F_{I}\big(\bm{v}_{I}^T\cdot \bm{a}-\theta_{I}\big)\,.
\end{flalign}
For any generic contribution, the mass matrix is extracted around
the minimum $\bm{a}_{\star}$ as
\begin{equation}
\left|\frac{\partial^{2}V_{I}}{\partial a\partial a^{T}}\right|_{a=a_{\star}}=F''_{I}\big(\bm{v}_{I}^T\cdot \bm{a}_{\star}-\theta_{I}\big)\bm{v}_{I}\bm{v}_{I}^{T}\equiv\kappa_{I}\Lambda_{I}^{4}\bm{v}_{I}\bm{v}_{I}^{T}\,,
\label{eq:formulaformass}
\end{equation}
where $F''_{I}(\bm{v}_{I}^T\cdot \bm{a}_{\star}-\theta_{I})\equiv \kappa_{I}\Lambda^4_{I}$ is a real number, which in general can be positive or negative. We use a convenient parameterisation where $\Lambda_I$ is a positive dimensionful scale while $\kappa_{I}=-1,1$ controls the sign. The total mass matrix therefore splits into two contributions: one fully aligned to the QCD anomaly vector
\begin{equation}
    \LambdaQCDbar^4\bm{N}\bm{N}^{T} \equiv \left(\Lambda_{\mathrm{QCD}}^{4}+\sum_{I\in\mathcal{C}}\kappa_{I}\Lambda_{I}^{4}\xi_{I}^{2}\right) \bm{N}\bm{N}^{T}\,,
     \label{eq:deflambdabar}
\end{equation}
while the remaining contributions containing the PQ-symmetry breaking effects that are not fully aligned
to QCD are collected in
\begin{equation}
     \bm{M}_A^2 \equiv \sum_{I\in\mathcal{A}}\kappa_{I}\Lambda_{I}^{4}\bm{v}_{I}\bm{v}_{I}^{T} \,.
\end{equation}
The total mass
matrix, given by 
\begin{equation}
\bm{M}^{2}=\LambdaQCDbar^4\bm{N} \bm{N}^{T}+\bm{M}_A^2\,,
\label{eq:mastereq}
\end{equation}
now presents a parameterisation that will be useful later to classify
the various phenomenological scenarios. Note that, in full generality, $\LambdaQCDbar$
is a real parameter that can be either positive or negative, while $\bm{M}_A^2$ is symmetric but not necessarily positive semi-definite. This is because in both cases, the coefficients $\kappa_{I}$ can be negative. However, the analysis of positivity performed in Appendix~\ref{app:Hell} reveals that when $\LambdaQCDbar<0$ or $\bm{M}_A^2$ has a negative eigenvalue, the phenomenology resembles other scenarios that are also present when $\barLambdaQCD>0$ and $\bm{M}_A^2$ is positive semi-definite. For simplicity, in the following we restrict to the latter case, which is a sufficient condition to ensure positivity of the full mass matrix. We refer to the corresponding axion mass–coupling relation as the generalised QCD line, or more informally simply the QCD line, defined by $\barLambdaQCD$. In QCD–PQ scenarios one automatically has $\barLambdaQCD=\LambdaQCD$, while in general-PQ scenarios this need not be the case.

Furthermore, it is always possible to perform a rotation that diagonalises $\bm{M}_A^2$ while preserving the rank-one structure of the $\Lambda_{\rm QCD}^4$ contribution. This is achieved by rotating the axion fields and the anomaly-coefficient vectors with orthogonal matrices, redefining the decay constants $f_i$ and $\Lambda_i$. In the following general analysis, we assume that we are working in this basis, where $\bm{M}_A^2=\mathrm{diag}(\Lambda^4_1/f^2_1,...,\Lambda^4_N/f^2_N)$. Since such a basis can always be found, the discussion remains fully general, and any phenomenological scenario can be mapped onto it (see Appendix \ref{app:explit22rotation} for an explicit example in a two-axion setup). 

This parameterisation isolates the subset of PQ-breaking effects that are fully aligned with QCD, treating the QCD direction as special and allowing for a direct comparison with the generic breaking scales encoded in $\bm M_A^2$. Generic non-fully-aligned sources can nevertheless have nonzero projection onto the QCD direction, and their
effect is not lost in the diagonal-$\bm M_A^2$ basis: it is encoded in the eigenvalues of $\bm M_A^2$ together with the orientation of the rotated anomaly vector $\bm N$.

While this basis is particularly useful to classify various phenomenological scenarios, it does not explicitly decompose PQ-breaking effects into components parallel, orthogonal, or mixed with respect to $\bm N$. Appendix~\ref{app:Hell} introduces a complementary and fully general parameterisation, in which the same mass matrix is expressed in terms of a parallel component, a mixing vector, and an orthogonal block. This makes the decomposition with respect to the QCD direction manifest and is particularly useful for interpreting generic PQ-breaking effects. In the two-axion case, this alternative parameterisation reproduces the same qualitatively distinct phenomenological scenarios discussed in the following sections. An analysis of the $N$-axion system within the framework of Appendix~\ref{app:Hell} is beyond the scope of this paper and is left for future work. Consequently, in the following we continue to use the present parameterisation and indicate the correspondence between the two parameterisations in Appendix~\ref{app:generaldecomposition}.

\section{Phenomenology of the two-axion system}\label{sec:Twoaxion}

The two-axion system illustrates the general features of multi-axion phenomenology while remaining analytically simple. In particular, it allows us to identify the different parametric regimes and the corresponding locations of axions in the ($m_a$, $g_{a\gamma\gamma}$) plane, which we later generalise to more complex multi-axion systems. In this section, we first characterise the distinct regimes of two-axion systems and study the mathematical structure that gives rise to them. Second, we observe that some of these cases are not captured by the sum rule in \cref{eq:photon_sumrule}, and we show that the discovery of one or more axions violating this rule could provide insight into different UV realisations. Finally, we comment on the phenomenological parameter space of two-axion systems.

\subsection{The structure of the two-axion system}
We consider a system of two axions associated with two broken PQ symmetries $\UPQ^1\times \UPQ^2$, both anomalous under QCD, leading to a generic axion potential as in \cref{eq:general_potential}. In the spirit of the general parameterisation detailed in \cref{sec:Lambda_bar_definition}, we decompose the mass matrix into a fully QCD-aligned contribution and a non-fully-aligned one, and work in a basis where the latter is diagonal. The former effectively redefines $\LambdaQCD \to \LambdaQCDbar$, as well as the scales $f_i$, allowing us to write, without loss of generality, the mass matrix as 
\begin{equation}
    \bm{M}^2=\begin{pmatrix}
        \dfrac{\barLambdaQCD^4+\Lambda_1^4}{f_1^2} & \dfrac{\barLambdaQCD^4}{f_1 f_2}\\
        \dfrac{\barLambdaQCD^4}{f_1 f_2}&\dfrac{\barLambdaQCD^4+\Lambda_2^4}{f_2^2} 
    \end{pmatrix} \, .
    \label{eq:2x2text}
\end{equation}

\noindent This includes contributions from the QCD potential as well as from a generic potential $V_{\cancel{\rm{PQ}}}$ that can break\footnote{As illustrated in \cref{sec:MultiAxion}, having a PQ symmetry only broken by QCD is a sufficient but not necessary condition to solve the strong CP problem.} both PQ symmetries non-trivially. The redefined $\barLambdaQCD$, as introduced in \cref{eq:deflambdabar}, accounts for PQ-breaking effects that fully align with QCD, and captures the fact that, in general-PQ systems, the QCD direction itself can be displaced, for example, in \textit{mirror world} constructions \cite{Berezhiani:2000gh,Hook:2018jle,DiLuzio:2021pxd} (see also Appendix~\ref{app:commuting_QCD}). In the absence of such effects, one has $\barLambdaQCD=\LambdaQCD$. The remaining PQ-breaking effects from $V_{\cancel{\rm{PQ}}}$ are captured by the diagonal entries $\Lambda_{i=1,2}$, with anomaly coefficients and $f_i$ treated as effective parameters.

In this setup, the eigenvectors can be written using the mixing angle $\theta$,
\begin{equation}
    \bm{u_+} = \begin{pmatrix}
        \cos\theta & \sin\theta
    \end{pmatrix}^T \, , \quad    \bm{u_-} = \begin{pmatrix}
        -\sin\theta & \cos\theta
    \end{pmatrix}^T \,, 
\end{equation}
associated with the eigenvalues $m_\pm^2$. The explicit formulae for eigenvalues and eigenvectors can be found in Appendix \ref{app:Two-axion_app}. We can study the different regimes in which the axions can be found by computing the ratios between mass and photon couplings,
\begin{widetext}
 \begin{align}
    \label{eq:heavy_line}
\frac{\tilde{g}_{a_+ \gamma}^2}{m_+^2}&=\frac{(C_{a_1\gamma} \cos\theta f_2+C_{a_2\gamma} \sin\theta f_1)^2}{f_1^2\Lambda_2^4\sin\theta^2+f_2^2\Lambda_1^4\cos\theta^2+(f_1\sin\theta+f_2\cos\theta)^2\barLambdaQCD^4}\, , \\
\label{eq:light_line}
     \frac{\tilde{g}_{a_- \gamma}^2}{m_-^2}&=\frac{(C_{a_1\gamma} \sin\theta f_2-C_{a_2\gamma} \cos\theta f_1)^2}{f_1^2\Lambda_2^4\cos\theta^2+f_2^2\Lambda_1^4\sin\theta^2+(f_1\cos\theta-f_2\sin\theta)^2\barLambdaQCD^4} \, .
     \end{align}
\end{widetext}
Note that in the cases where one axion originates mostly from the QCD direction, i.e.~$\tilde{g}_{a_ \gamma}^2/m_a^2\sim C_{a\gamma}^2 / \barLambdaQCD^4$, this QCD-like axion may already be significantly displaced from the canonical QCD band if $\barLambdaQCD\neq \LambdaQCD$, which already signals a general-PQ system. In the following, when characterising the different regimes of the two-axion system from \cref{eq:heavy_line,eq:light_line}, we refer to this modified line as the ``QCD line'', which would lie within the canonical QCD band in the absence of PQ-breaking effects fully aligned with QCD, or when such effects are negligible, so that \(\bar{\Lambda}_{\rm QCD}=\Lambda_{\rm QCD}\).

We begin by taking the limit $\Lambda_2\gg\barLambdaQCD$, for which we obtain
\begin{align}
    \frac{\tilde{g}_{a_+ \gamma}^2}{m_+^2}=\frac{C_{a_2\gamma}^2}{\Lambda_2^4}\ll \frac{C_{a_2\gamma}^2}{\barLambdaQCD}\,, \,\,
    \frac{\tilde{g}_{a_- \gamma}^2}{m_-^2}=\frac{C_{a_1\gamma}^2}{\Lambda_1^4+\barLambdaQCD^4}\, ,
\end{align}
provided that $f_1^2 \Lambda_2^4 - f_2^2 \Lambda_1^4 > 0$. In this limit, both axions are displaced to the right of the QCD line, unless $\Lambda_1=0$. If $\Lambda_1\gg\barLambdaQCD$, there is approximatively no mixing between the two original axions, and both axion mass eigenstates are aligned with one of the $\Lambda_i$ associated with a single PQ symmetry $\UPQ^i$. If, on the other hand, $\Lambda_1\lesssim \barLambdaQCD$, one axion is QCD-like. In both regimes, since the mixing is very small, we observe that each state retains its corresponding photon coupling.

At first sight, \cref{eq:heavy_line,eq:light_line} might suggest that both axions should always lie to the right of the QCD line, since the quantities in the denominator of both expressions are positive and therefore $[\cdots]>\barLambdaQCD^4$. While this is always true for the heavy axion in \cref{eq:heavy_line}, for the light state in \cref{eq:light_line} it holds only if $f_2 \sin\theta - f_1 \cos\theta \neq 0$. In the limit $\cos\theta \rightarrow (f_2/f_1)\sin\theta$, both expressions reduce to
\begin{align}
    \frac{\tilde{g}_{a_+ \gamma}^2}{m_+^2} &= \frac{(C_{a_2\gamma}f_1^2+C_{a_1\gamma}f_2^2)^2}{\barLambdaQCD^4\left(f_1^2 +f_2^2\right)^2+f_1^4\Lambda_2^4+f_2^4\Lambda_1^4} \, ,\\
\label{eq:lightaxion_limit}
    \frac{\tilde{g}_{a_- \gamma}^2}{m_-^2} &= \frac{(C_{a_2\gamma}-C_{a_1\gamma})^2}{\left(\Lambda_1^4 +\Lambda_2^4\right)} =  \left(\frac{\alphaem}{2\pi}\right)^2\frac{(E_1/N_1-E_2/N_2)^2}{\left(\Lambda_1^4 +\Lambda_2^4\right)}\, .
\end{align}
We see that, in this limit, the first expression is sensitive to $\barLambdaQCD$, whereas the second is not. This second expression exhibits two interesting features. First, for $E_1/N_1 = E_2/N_2$, it vanishes, implying that the axion moves rapidly to the right of the QCD line. If the two $E_i/N_i$ differ, the position relative to the QCD line is instead entirely determined by the combination $\Lambda_1^4+\Lambda_2^4$.

To identify the position of the light axion with respect to the QCD line, we need to find the limit in which $(f_2 \sin\theta- f_1 \cos\theta)\to 0$. As one can see in Appendix \ref{app:Two-axion_app}, there exist two ways to achieve this limit. The first one corresponds to the limit in which $\barLambdaQCD \gg \Lambda_{1,2}$, hence when the two $E_i/N_i$ differ, it is clear that the light axion state is to the left of the QCD line since 
\begin{equation}
\label{eq:left_of_theQCDline}
        \frac{4\pi^2}{\alphaem^2}\frac{\tilde{g}_{a_- \gamma}^2}{m_-^2} =\frac{(E_1/N_1-E_2/N_2)^2}{
        \left(\Lambda_1^4 +\Lambda_2^4\right)} \gg 
        \frac{(E_1/N_1-E_2/N_2)^2}{\barLambdaQCD^4} \, .
\end{equation}
We also note that this result is independent of the choice of $f_1$ and $f_2$, the only requirement being that $\LambdaQCDbar\gg\Lambda_{1,2}$. A second important point is that placing an axion to the left of the QCD line is only possible when the anomaly coefficient ratios are different. When $E_1/N_1 = E_2/N_2$, the expression is instead suppressed by $\barLambdaQCD$ as
\begin{equation}
\label{eq:lightaxion_series}
    \frac{\tilde{g}_{a_- \gamma}^2} {m_-^2}=\frac{C_{a\gamma}^{2}\left(f_{2}^{2}\Lambda_{1}^{4}-f_{1}^{2}\Lambda_{2}^{4}\right)^{2}}{\left(f_1^2+f_2^2\right)^2\left(\Lambda_{1}^4+\Lambda_2^4\right)\barLambdaQCD^{8}} +\mathcal{O}\left(\barLambdaQCD^{-12}\right)  \, ,
\end{equation}
with $C_{a_1\gamma}=C_{a_2\gamma}=C_{a\gamma}$. For $\barLambdaQCD=\LambdaQCD$, this contribution is always smaller than that of the traditional QCD axion; therefore, the axion lies to the right of the canonical QCD line and exhibits a suppressed photon coupling. Instead, if $\barLambdaQCD\ll\LambdaQCD$, one may still achieve an extra axion to the left of the canonical QCD line when $E_i/N_i$ are equal, but here the QCD-like state is itself to the left of the canonical QCD line.

The second way to achieve the limit of \cref{eq:lightaxion_limit} is by taking $f_1 \to \Lambda_1^2 f_2 /\Lambda_2^2$ (see \cref{eq:imrunningoutofnames}). In this case, note that with equal ratios of anomaly coefficients, the expression cancels again, and the axion moves below the QCD line, while in the case of different $E_i/N_i$, the position with respect to the QCD line is determined by the size of $\Lambda_{1,2}$.

We have identified the behaviour of the system, as well as the conditions under which an axion lies to the left or to the right of the QCD line. However, the underlying reason for this behaviour still remains to be explained. The first limit is realised when the mixing vanishes, $\theta\to 0$. In this case, each axion is sensitive to its respective photon coupling. The second limit found here depends on the hierarchy of scales, $\barLambdaQCD\gg \Lambda_{1,2}$, and the behaviour with respect to the QCD line is determined by the alignment, or misalignment, of the EM anomaly vector with the gluon anomaly vector. There also exists the degenerate limit $\Lambda_1^2/f_1=\Lambda_2^2/f_2$ and, similarly to the previous case, the numerator of \cref{eq:lightaxion_limit} approaches zero when the EM anomaly vector is again aligned with the gluon anomaly vector.

These limits are related to the particular structure of the mass matrix. First, turning our attention to the potential and applying the condition $\barLambdaQCD\gg\Lambda_1,\,\Lambda_2$, the mass matrix shows an interesting behaviour. In this limit, it is dominated by the QCD-like potential,
\begin{equation}
    \bm{M}^2 \simeq \barLambdaQCD^4 \bm{N}\bm{N}^T \, ,
\end{equation}
and is constructed entirely from a single vector. Therefore, it is manifestly rank one. As a result, only one linear combination of axions acquires mass, while the orthogonal combination remains approximately massless. In this notation, the corresponding eigenstates are immediately identified: the massive eigenstate is aligned with the vector that defines the rank-one mass matrix, whereas the orthogonal state is massless up to mass-corrections coming from the other sources of breaking $\Lambda_{1}$ and $\Lambda_{2}$:
\begin{align}
    &m_+^2 \simeq \frac{\LambdaQCDbar^4(f_1^2+f_2^2)}{f_1^2f_2^2},\, \bm{u}_+=\frac{\bm{N}}{|\bm{N}|}= \frac{1}{\sqrt{f_1^2+f_2^2}}\begin{pmatrix}
        f_2 \\ f_1
    \end{pmatrix} \label{eq:TwoAxionLimitplus} \\
     &m_-^2 \simeq \frac{\Lambda_{1}^4+\Lambda_{2}^4}{f_1^2+f_2^2},\,\quad\ \bm{u}_-=\frac{\bm{N}_\perp}{|\bm{N}_\perp|}= \frac{1}{\sqrt{f_1^2+f_2^2}}\begin{pmatrix}
        f_1 \\ -f_2
    \end{pmatrix}\, .
    \label{eq:TwoAxionLimitminus}
\end{align}
Using the eigenvectors we can compute the value of the photon coupling of the two mass-eigenstates
\begin{equation}
\label{eq:photon_coupling_large_LambdaQCD}
    \tilde{g}_{a_\pm\gamma} =\bm{g}_{a\gamma}\cdot\bm{u_\pm} =\begin{cases}
         \dfrac{C_{a_1\gamma} f_2^2+C_{a_2\gamma}f_1^2}{f_1 f_2 \sqrt{f_1^2+f_2^2}}\, ,\quad &\textrm{ (+)}\\[8pt]
        \dfrac{\alphaem}{2\pi}\dfrac{E_1/N_1-E_2/N_2}{\sqrt{f_1^2+f_2^2}}\, ,\quad &(-)
    \end{cases} 
\end{equation}
where in the case $E_1/N_1=E_2/N_2$ the result simplifies to $g_{a_+ \gamma}= C_{a\gamma} \sqrt{f_1^2+f_2^2}/f_1f_2$ and $g_{a_-\gamma}=\mathcal{O}(\barLambdaQCD^{-4})$. The reason behind these results can be traced back to the form of the anomalous vector coupling, namely
\begin{align}
    \bm{g_{a\gamma}}
    =& \frac{\alphaem}{2\pi}\left(\dfrac{\frac{E_1}{N_1} f_2^2+\frac{E_2}{N_2}f_1^2}{f_1^2 +f_2^2}-C_\chi \right)\bm{N} \nonumber\\ &+\frac{\alphaem}{2\pi} \frac{f_1 f_2}{f_1^2+f_2^2}\left(\frac{E_1}{N_1}-\frac{E_2}{N_2}\right) \bm{N}_\perp \, ,
\end{align}
where for the case $E_1/N_1=E_2/N_2$, the result simplifies further to $ \bm{g_{a\gamma}}=\frac{\alphaem}{2\pi}\left(\frac{E}{N}-C_\chi\right) \bm{N} \, , $ showing that the photon coupling is fully aligned to the gluon anomaly vector. This trivially leads to the result mentioned above, $\tilde{g}_{a_-\gamma} = \bm{g}_{a\gamma}^T\cdot \bm{N}_\perp = 0$. 

The second limit ($\Lambda_1^2/f_1 \to \Lambda_2^2/f_2$) is also interesting, although it can only happen in the presence of contributions almost fully aligned to QCD and tuning of PQ-breaking effects to cancel mixing with the orthogonal direction. In this case,  
the diagonal PQ-breaking potential leads to a universal contribution to the masses $\Lambda_1^4/f_1^2 \mathbf{1}$. In this case, one can immediately see that there exist two eigenvectors
\begin{align}
\label{eq:limit_universal_contribution}
    \bm{M}^2\frac{\bm{N}}{|\bm{N}|} &= \left(\LambdaQCD^4 |\bm{N}|^2+\frac{\Lambda_1^4}{f_1^2}\right)  \frac{\bm{N}}{|\bm{N}|} \, ,\\
    \bm{M}^2 \frac{\bm{N}_\perp}{|\bm{N}_\perp|}&=\frac{\Lambda_1^4}{f_1^2} \frac{\bm{N}_\perp}{|\bm{N}_\perp|} \, .
\end{align}
In both limits, we can see that the eigenvector associated to the second axion is orthogonal to the QCD anomaly vector. Therefore, in both limits, if the photon coupling is aligned to $\bm{N}$, the remaining axion is naturally ``photophobic''. We also note that, since the axion-pion contribution is aligned with the gluon-anomaly vector, it leaves a residual, though suppressed, contribution to $G\tilde{G}$. As a result, the corresponding contribution to the photon coupling remains suppressed as well. In both cases, one of the axion states approaches the QCD-anomaly vector while the other becomes orthogonal, hence a photon coupling aligned to the QCD-anomaly leads to a suppressed photon coupling.

\subsection{Modifications to the axion sum rule}
\label{sec:sumrule_2axions}
We can now turn to the photon coupling sum rule in \cref{eq:photon_coupling} and study how it is modified in this more general setup. Substituting \cref{eq:heavy_line,eq:light_line} into \cref{eq:photon_sumrule}, we achieve a general sum rule that takes into account general anomaly coefficients and PQ-breaking effects
\begin{equation}
    \sum_i \frac{\tilde{g}_{a_i\gamma}^2}{m_i^2}
    =
    \frac{
    C_{a_2 \gamma }^2\Lambda_{1}^4
    +C_{a_1 \gamma }^2\Lambda_{2}^4
    +\left(C_{a_2\gamma}-C_{a_1\gamma}\right)^2\barLambdaQCD^4
    }{
    \Lambda_{1}^4\barLambdaQCD^4+\Lambda_{2}^4\barLambdaQCD^4+\Lambda_{1}^4\Lambda_2^4
    } \, .
    \label{eq:sum_rule_photon_2axion}
\end{equation}
We can now explore a few limits in which the original relation in \cref{eq:photon_sumrule} is violated. Let us begin with the regime $\Lambda_1,\Lambda_2\gg\barLambdaQCD$, for which
\begin{equation}
    \sum_i \frac{\tilde{g}_{a_i\gamma}^2}{m_i^2}
    \simeq
    \frac{C_{a_1\gamma }^2}{\Lambda_1^4}
    +\frac{C_{a_2\gamma}^2}{\Lambda_2^4}
    \ll
    \frac{C_{a\gamma}^2}{\barLambdaQCD^4}\, .
    \label{eq:largeLambdai}
\end{equation}
In this case, independently of whether the anomaly coefficients are universal or not, the sum rule implies that both axion trajectories are displaced to the right of the QCD line. In this scenario there is no axion in the QCD line and all axions can be found to the right. This regime originates naturally in axion solutions of the strong CP problem where QCD arises as the diagonal subgroup of a larger strong sector \cite{Agrawal:2017ksf,Gaillard:2018xgk,Fuentes-Martin:2019bue,Csaki:2019vte}, which is a framework that naturally motivates multi-axion systems.

The second limit of interest is obtained by sending one of the extra scales to zero, e.g.~$\Lambda_2\to 0$ (and analogously for $\Lambda_1\to 0$), which yields
\begin{equation}
\label{eq:sum_rule_2axions_onescale}
       \sum_i \frac{\tilde{g}_{a_i\gamma}^2}{m_i^2}
       =
       \frac{C_{a_2 \gamma }^2}{\barLambdaQCD^4}
       +\frac{\left(C_{a_2\gamma}-C_{a_1\gamma}\right)^2}{\Lambda_{1}^4}
       \geq
       \frac{C_{a\gamma}^2}{\barLambdaQCD^4}\, .
\end{equation}
Here the inequality follows because both terms are manifestly positive. This expression shows that, if the photon coupling coefficients are non-universal, the position of one axion with respect to the QCD line is mostly determined by $\Lambda_{1}$. If $\Lambda_{1}\ll \barLambdaQCD$, then we have one axion to the left of the QCD line that dominates the sum rule. Otherwise, if $\Lambda_{1}\gg \barLambdaQCD$, then the sum rule is mostly dominated by the QCD-like contribution and the exotic axion is to the right of the generalised QCD line. Note that even restricting to QCD-PQ systems where $\barLambdaQCD=\LambdaQCD$, there are still corrections to the sum rule of \cref{eq:photon_sumrule} induced by the presence of non-universal anomaly coefficients. If they can be determined experimentally, these would rule out simple GUTs as UV completions of the SM, because they predict universal anomaly coefficients \cite{Agrawal:2022wyu}, even if both axions are in the canonical QCD line or to the right.  Note that if $\barLambdaQCD\ll\LambdaQCD$, both axions can be to the left of the canonical QCD-line without requiring non-universal anomaly coefficients.

\begin{figure*}[tbh]
    \centering
    \includegraphics[width=\linewidth]{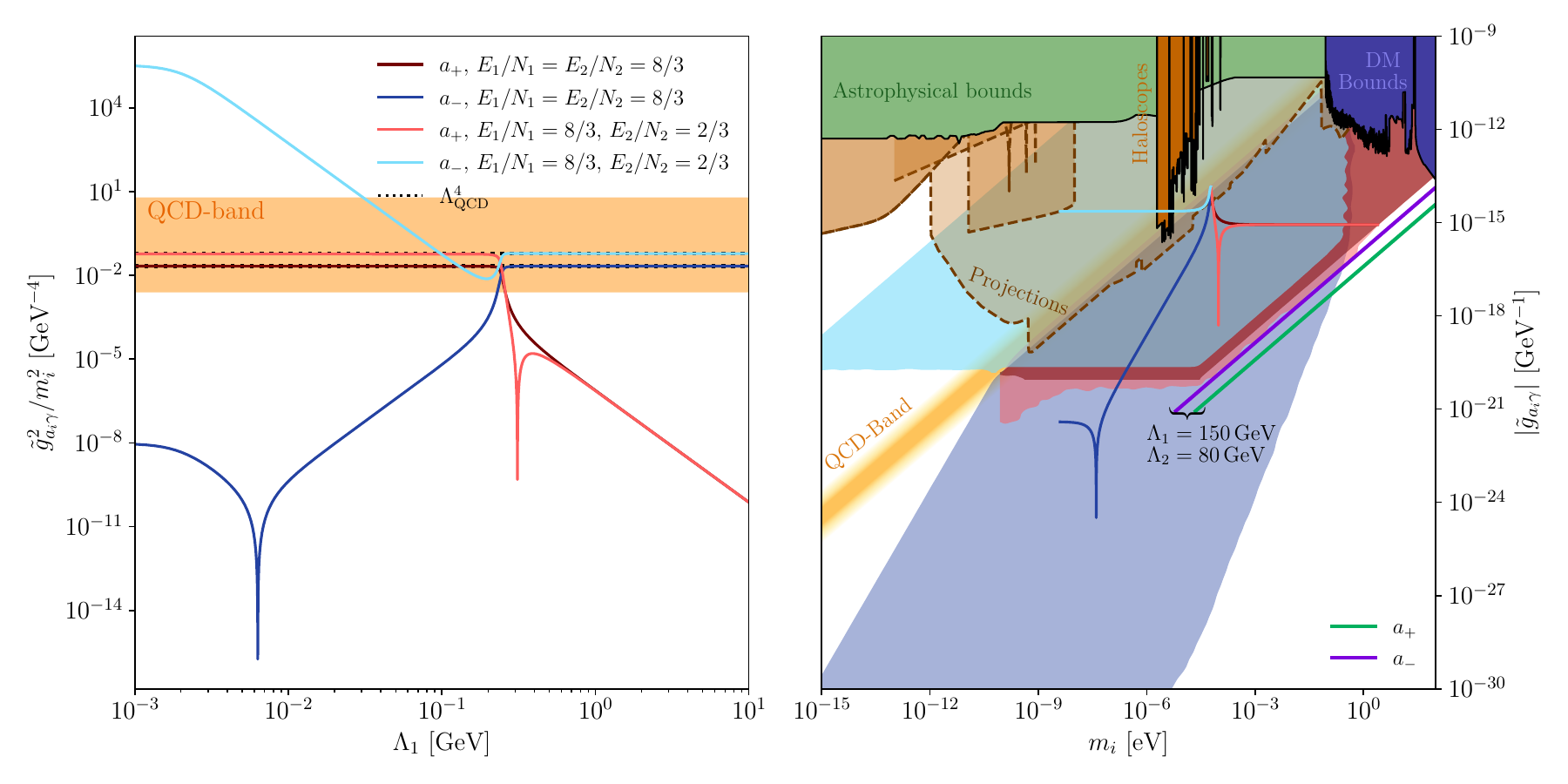}
    \caption{\textit{Left:} Behaviour of the two-axion model as a function of $\Lambda_1$, fixing $\Lambda_2=2\times 10^{-3}\,$GeV, $f_1=10^{12}\,$GeV, and $f_2=10^{11}\,$GeV, for two choices of $\left(E_1/N_1,\, E_2/N_2\right)$. \textit{Right:} Parameter scan in the $m_{a_i}$-$g_{a_i\gamma}$ plane, varying $f_i \in \left[10^7,\,10^{17}\right]\,$GeV and $\Lambda_1 \in \left[10^{-4},\,50\right]\,$GeV, with $\Lambda_2=0$. The colours correspond to the same choices as in the left plot, and the curves shown on the right correspond to the lines in the left plot. For this plot we used the code available in Ref.~\cite{AxionLimits}. 
    }
    \label{fig:Two_axions}
\end{figure*}

Finally, if both $\Lambda_1,\Lambda_2\ll \barLambdaQCD$ we obtain
\begin{align}
        \sum_i \frac{\tilde{g}_{a_i\gamma}^2}{m_i^2}
        \simeq&
        \frac{C_{a_2 \gamma }^2\Lambda_{1}^4+C_{a_1 \gamma }^2\Lambda_{2}^4}{\Lambda_{1}^4+\Lambda_{2}^4}\frac{1}{\barLambdaQCD^4}
        \nonumber\\ &+\frac{\left(C_{a_2\gamma}-C_{a_1\gamma}\right)^2}{\Lambda_{1}^4+\Lambda_{2}^4}+\mathcal{O}(\LambdaQCDbar^{-8})\, .
\end{align}
If the photon coupling coefficients are universal, $C_{a_1\gamma}=C_{a_2\gamma}$, the second term vanishes and for $\barLambdaQCD=\LambdaQCD$ the sum rule in \cref{eq:photon_sumrule} is approximately recovered, with only small corrections\footnote{Naively, one may interpret these corrections as a reintroduction of $\Delta\bar{\theta}$. Instead, if the naive $\Delta\bar{\theta}$ is in tension with nEDM experiments, one may interpret this as a general-PQ system where strong CP is solved. Then the specific pattern of data could motivate specific UV theories, as shown by the simplified examples of Appendix~\ref{app:UVmodels}.}. By contrast, for non-universal couplings the second term dominates, implying that a sizeable displacement of one axion to the left of the QCD line is possible. Note in particular that taking $\Lambda_1\to 0$ or $\Lambda_2\to 0$ with $\barLambdaQCD=\LambdaQCD$ does not restore the original sum rule; rather, it takes us back to \cref{eq:sum_rule_2axions_onescale}, where the non-universal term controls the result. 

A particular case presented in Ref.~\cite{Gavela:2023tzu} is the maximally deviating QCD axions, or \textit{maxions}. Defining a measure of distance with respect to the (generalised) QCD line
\begin{equation}
    \frac{1}{q_i}=\frac{\Lambda_{1}^4\barLambdaQCD^4+\Lambda_{2}^4\barLambdaQCD^4+\Lambda_{1}^4\Lambda_2^4
    }{C_{a_2 \gamma }^2\Lambda_{1}^4
    +C_{a_1 \gamma }^2\Lambda_{2}^4
    +\left(C_{a_2\gamma}-C_{a_1\gamma}\right)^2\barLambdaQCD^4
    } \cdot \frac{\widetilde{g}_{a_i\gamma}^2}{m_i^2}\,,
\end{equation}
then the sum rule can be read as
\begin{equation}
\sum_i \frac{1}{q_i}=1 \,.
\end{equation}
Therefore, the maxion configuration occurs when both $q_i=2$. If one of the $q_i$ were larger, the other would have to be smaller, and vice versa; consequently, the distance to the QCD line would be reduced. Assuming universal EM anomaly coefficients, the maximal contribution to the sum rule is
\begin{equation}
    \frac{\widetilde{g}_{a_i\gamma}^2}{m_i^2}\Bigg|_\text{Maxion}=\frac{C_{a\gamma}^2}{2\left(\barLambdaQCD^4+\frac{\Lambda_1^4\Lambda_2^4}{\Lambda_1^4+\Lambda_2^4}\right)}\,.
\end{equation}
In the limit of $\Lambda_1\rightarrow 0$ (or equivalently $\Lambda_2\rightarrow 0$) and $\barLambdaQCD=\LambdaQCD$, the maxion is displaced by a factor of $1/\sqrt{2}$ from the traditional QCD line, as found in Ref.~\cite{Gavela:2023tzu}. If $\barLambdaQCD\neq\LambdaQCD$, then the generalised maxions are displaced with respect to the QCD line defined by $\barLambdaQCD$. However, if both scales $\Lambda_{1,2}$ are non-zero, the above expression shows that the QCD line can indeed be displaced more than $\sqrt{2}$, even for $\barLambdaQCD=\LambdaQCD$. Note that the maxion limit requires imposing a very particular condition on the potential. Choosing $\Lambda_2=0$ (analogously for $\Lambda_1=0$), the maxion limit is satisfied when $\Lambda_1^4=(f_1^2+f_2^2)\barLambdaQCD^4/f_2^2$. Regardless of the sign of the coefficients and the hierarchy between scales $f_i$ and $\Lambda_i$, the maxion limit implies that the two axions are to the right of the QCD line. The maxion condition can be further generalised in the case of different photon couplings, see \cref{eq:Lambda1_2maxions} in Appendix~\ref{app:Two-axion_app}.

\subsection{The two-axion parameter space}

In \cref{fig:Two_axions} we numerically exemplify the analytical results derived in this section. For this analysis, we assume  
$\barLambdaQCD=\LambdaQCD$\footnote{If $\barLambdaQCD\neq\LambdaQCD$, then the reference QCD line changes and moves the entire axion contours in the right panel of \cref{fig:Two_axions} to the left or to the right.} and consider two benchmarks for the anomaly ratios $E_i/N_i$: in the first, we take $E_1/N_1 = E_2/N_2 = 8/3$, while in the second we fix $E_1/N_1 = 8/3$ and $E_2/N_2 = 2/3$. In the left plot of \cref{fig:Two_axions}, we fix the scales to $f_1 = 10^{12}\,$GeV, $f_2 = 10^{11}\,$GeV, and $\Lambda_2 = 2\times10^{-3}\,$GeV, while varying $\Lambda_1$. The peak at lower values of $\Lambda_1$ corresponds to the limit in which $f_1\to f_2 \Lambda_1^2/\Lambda_2^2$. Modifying the benchmark point of $\Lambda_2$ would shift the position of this peak. 

In the left plot of \cref{fig:Two_axions} we can distinguish two clear regimes. The first one (to the right) corresponds to $\Lambda_1 \gg \LambdaQCD$. In this case, for both scenarios, the heavier axions move towards smaller values of $\tilde{g}_{a_+\gamma}^2/m_+^2$ relative to the QCD axion reference, i.e.~it is displaced to the right of the QCD line. The second regime (to the left) is instead characterized by $\Lambda_1 \ll \LambdaQCD$, and here the behaviour differs significantly between the two benchmarks. For universal values of $E_i/N_i$, the lighter axion moves towards smaller values of $\tilde{g}_{a_-\gamma}^2/m_-^2$, and hence to the right of the QCD line. This behaviour is captured by the analytical expression in \cref{eq:lightaxion_series}, and originates from the vanishing of the photon coupling in the $\LambdaQCD$ limit, as can be seen from \cref{eq:photon_coupling_large_LambdaQCD}. By contrast, when the anomaly coefficients are different, \cref{fig:Two_axions} shows an axion moving towards larger values of $\tilde{g}_{a_-\gamma}^2/m_-^2$, i.e.~to the left of the QCD line, in agreement with the analytical expression in \cref{eq:left_of_theQCDline}. The intermediate regime is sensitive to the choice of scales $f_i's$, or the sign of $E_i/N_i$ but it leads to no qualitatively different results in the limits explained above.

The right plot of \cref{fig:Two_axions} shows the same benchmarks, now scanning over the three relevant scales, with $f_i \in \left[10^7,\,10^{17}\right]\,$GeV and $\Lambda_1 \in \left[10^{-4},\,50\right]\,$GeV, and displaying the region populated by the scan points. We also superimpose, as coloured lines, the same trajectories shown in the left plot in order to illustrate how they map onto the full parameter space.

We see that the region corresponding to $\Lambda_1 \gg \LambdaQCD$ is qualitatively similar in both benchmarks. One axion is in the QCD line while the other has a similar photon coupling but it is heavier. Testing this region would require a significantly enhanced sensitivity of haloscopes/helioscopes experiments, complemented by stronger axion tests via astrophysics and cosmology. Interestingly, in the regime $\Lambda_1,\,\Lambda_2 \gg \LambdaQCD$, both axions move to the right and there is no axion in the QCD line, yet they retain a photon coupling similar to the usual QCD axion. The purple-green lines show a benchmark in this regime ($\Lambda_1=150\,$GeV, $\Lambda_2=80\,$GeV)\footnote{We choose these values to exemplify this limiting case. In the particular models of Appendix \ref{app:UVmodels} based on $\SU{3}_1 \cross  \SU{3}_2\rightarrow \SU{3}_{\mathrm{QCD\equiv1+2}}$, one finds a maximum $\Lambda_i\sim1\,$GeV simultaneously.}. Discovering such a scenario may point to particular structures in the UV, such as a QCD emerging as the diagonal subgroup of a larger strongly coupled sector. 

Conversely, the opposite regime, $\Lambda_1 \ll \LambdaQCD$ shows qualitative differences between the two benchmarks for the anomaly coefficients. For the benchmark with universal $E_i/N_i$, the model populates only the region to the right of the QCD line, while in the non-universal case it instead extends to the left. The plot shows that the phenomenology of these axions changes dramatically depending only on whether the electromagnetic anomaly ratios are universal or non-universal. In the first case, one axion is light and ``photophobic'', and therefore practically invisible, whereas in the second case the axion lies in a region that could be probed by proposed future DM experiments \cite{Nagano:2019rbw,Berlin:2020vrk,Zhang:2021bpa,DMRadio:2022pkf}. Similarly, haloscopes targetting the QCD line~\cite{Stern:2016bbw,McAllister:2017lkb,Baryakhtar:2018doz,Beurthey:2020yuq,BREAD:2021tpx,Aja:2022csb,DeMiguel:2023nmz,Ahyoune:2023gfw,Alesini:2023qed,Fan:2024mhm} could also probe these axions if they constitute dark matter of the Universe, while helioscopes such as the proposed IAXO~\cite{IAXO:2019mpb} could detect them even if they do not account for dark matter. 

An important caveat for the phenomenology of these models is the production of dark matter, see Ref.~\cite{Baryakhtar:2026oun} for a related study in a multiple-axion setup. The sensitivity of the haloscopes shown in \cref{fig:Two_axions} depends on the local axion dark matter density, and the projected reach typically assumes that the axion makes up all of the local dark matter. The misalignment mechanism~\cite{Preskill:1982cy,Abbott:1982af,Dine:1982ah} may account for this abundance, and has been studied for two-axion systems in several works~\cite{Kitajima:2014xla,Ho:2018qur,Chen:2021wcf,Cyncynates:2023esj,Murai:2023xjn,Murai:2024nsp,Dunsky:2025sgz,Lee:2026umy}. Nevertheless, a fully reliable prediction usually requires UV input, because thermal corrections to the potential can be important and are generally related to the origin of the $\UPQ$-breaking terms. For this reason, the dark matter abundance must ultimately be determined on a model-by-model basis, and lies beyond the scope of the present generic low-energy analysis.

\textit{\textbf{Summary}} -- The main results of this section can be summarised as follows. If there exist PQ-breaking effects that fully align with the QCD direction, then one can have an effective scale $\barLambdaQCD$ that sets an effective ``QCD line''. This line may already be significantly displaced from the traditional QCD band to the left or to the right, while still solving the strong CP problem, see e.g.~Appendix \ref{app:commuting_QCD} for simplified models and references therein. In systems where such PQ-breaking effects fully aligned with QCD do not exist, then $\barLambdaQCD=\LambdaQCD$. In the limit $\barLambdaQCD \gg \Lambda_1, \Lambda_2$, the mass matrix is approximately rank-one and aligned with the gluon anomaly vector $\bm{N}$, yielding one eigenstate on the QCD line and a lighter state with mass controlled by $\Lambda_1$ and $\Lambda_2$. The photon coupling can be decomposed in the eigenbasis defined in terms of $\bm{N}$, implying two different patterns for the two-axion setup depending on the anomaly coefficients:
\begin{itemize}
\item If $E_1/N_1 = E_2/N_2$, the photon coupling vector is fully aligned with $\bm{N}$, suppressing the photon coupling of the lighter axion eigenstate. Although both the mass and the photon coupling vanish in this limit, the former scales as $\mathcal{O}(\Lambda_{1,2}^2)$ while the latter scales as $\mathcal{O}(\Lambda_{1,2}^4)$, such that the photon coupling approaches zero parametrically faster than the mass, implying that the lightest axion remains confined to the right of the QCD line. In this case the sum rule approximates to $1$, or is exactly $1$ if one of the two scales is exactly zero.
\item If $E_1/N_1 \neq E_2/N_2$, although the mass of the lighter axion is smaller than the QCD-like axion mass, the photon coupling does not vanish, and the corresponding state is therefore shifted to the left of the QCD line, in agreement with the analytical results of \cref{eq:light_line,eq:lightaxion_limit,eq:left_of_theQCDline,eq:lightaxion_series}. Similarly, in the regime $f_1 \simeq f_2 \Lambda_1^2/\Lambda_2^2$, the axion eigenstates approach the same limit, and one finds again that the photon coupling of the light axion vanishes when it is aligned with the gluon anomaly vector. For different $E_i/N_i$, the sum rule in \cref{eq:sum_rule_photon_2axion} is generally larger than $1$. 
\end{itemize}
In the regime in which one or both $\Lambda_i \gg \barLambdaQCD$, the mixing between the axion states is negligible, and each axion is therefore associated with its corresponding photon coupling and mass scale. If both scales are larger than $\barLambdaQCD$, both axions lie to the right of the QCD line. In this case, the generalised sum rule yields a value smaller than $1$ when the photon couplings are equal. If instead one of the scales is of order $\barLambdaQCD$ or smaller, one axion is QCD-like, and the sum rule approaches $1$ in the limit in which one of the two scales vanishes. 

Note that if $\barLambdaQCD\ll\LambdaQCD$, then one can have both axions to the left of the canonical QCD band even when the anomaly coefficients are universal. Moreover, even if we restrict to QCD-PQ systems (where $\barLambdaQCD=\LambdaQCD$) where the exotic axion is to the right of the QCD line, there are still corrections to the general axion sum rule induced in the presence of non-universal anomaly coefficients. If they can be identified experimentally, these would rule out simple GUTs as a UV completion of the SM, even if both axions are in the canonical QCD line or to the right (See discussion in \cref{sec:implications}).

\section{Phenomenology of \texorpdfstring{$N$}{N}-axion systems} \label{sec:generalization}

Having identified the origin of the different phenomenological scenarios in the structure of the mass matrix and the photon couplings of the two-axion system, in this section we generalise these results to a system with an arbitrary number of axions. We then show how the sum rule is extended to a general $N-$axion system, and illustrate the various scenarios through a numerical benchmark. Finally, we comment on their possible UV realisations.

\subsection{General regimes of \texorpdfstring{$N$}{N}-axion systems}

We will assume a general system of $N$ axions. As we did for the two-axion case, we work in a basis where we have diagonalised any contribution not fully aligned with the QCD direction. In this basis, the mass matrix of the fully general $N$ axion system is
\begin{equation}
    \bm{M}^2=\bm{M}_A^2+\barLambdaQCD^4\bm{N}\bm{N}^T\,,\label{eq:MassMatrix_basis}
\end{equation}
where $\bm{M}_A^2\equiv\mathrm{diag}(\Lambda^4_1/f_1^2,..,\Lambda^4_N/f_N^2)$. $\Lambda_i$ encode the breaking of the individual effective $\UPQ^{i}$ symmetries, and the off-diagonal contribution with the redefined $\barLambdaQCD$ accounts for QCD-breaking and for potential PQ-breaking effects beyond QCD which fully align with the QCD direction. As discussed in \cref{sec:Twoaxion} and Appendix \ref{app:commuting_QCD}, their interference with $\LambdaQCD$ could be positive or negative, such that both $\barLambdaQCD\leq\LambdaQCD$ and $\barLambdaQCD\geq\LambdaQCD$ are possible and compatible with multi-axion solutions of the strong CP problem. In absence of PQ-breaking effects fully aligned with the QCD direction, we have simply $\barLambdaQCD=\LambdaQCD$. We still denote ``QCD line" to the line defined with mass-coupling relation dictated by $\barLambdaQCD$. 

We begin by decomposing the matrix $\bm{M}_A^2$ into three contributions: larger ($\bm{M}_+^2$), similar order ($\bm{M}_0^2$) and smaller ($\bm{M}_-^2$) with respect to $\barLambdaQCD^2 \bm{N}\bm{N}^T$ 
\begin{equation}
    \bm{M}^2_A = \bm{M}^2_+ +\bm{M}^2_0 +\bm{M}^2_- \, ,
\end{equation}
in such a way that there are $K$ axions associated with $\bm{M}^2_+$, $M$ axions related to $\bm{M}^2_0$ and, finally, $N-M-K$ corresponding to $\bm{M}^2_-$. We can now define the corresponding gluon anomaly vectors associated with these scales such that the mass matrix can be written as 
\begin{widetext}
\begin{align}
    \bm{M}^2 &=  \LambdaQCDbar^4 \sum_{I,J} \bm{N}_I \bm{N}_J \;+\;\bm{M}^2_+ \;+\;\bm{M}^2_0\;+\;\bm{M}^2_-  \nonumber \\ &\simeq \barLambdaQCD^4 \sum_{I=+,0,-}\bm{N}_I\bm{N}^T_{I }+\barLambdaQCD^4 \bm{N}_-\bm{N}^T_0 +\barLambdaQCD^4 \bm{N}_0\bm{N}^T_-   \;+\;\bm{M}^2_+ \;+\;\bm{M}^2_0\;+\;\bm{M}^2_- \, .
\end{align}
By construction, we have $\Lambda_+ \gg \barLambdaQCD \sim \Lambda_0\gg \Lambda_-$. Therefore, we can approximate the mass matrix to a block diagonal form of the type
\begingroup
\newcommand{\TopStrut}{\rule{0pt}{4.2ex}}
\newcommand{\BotStrut}{\rule[-1.7ex]{0pt}{1.7ex}}

\begin{equation}
\bm{M}^2 \approx
\left(
\begin{array}{ccc|ccc|ccc}
\dfrac{\Lambda_{1}^4}{f_{1}^2} & \cdots & 0
& 0 & \cdots & 0
& 0 & \cdots & 0 \\
\vdots & \ddots & \vdots
& \vdots &  & \vdots
& \vdots &  & \vdots \\
0 & \cdots & \dfrac{\Lambda_{K}^4}{f_{K}^2}
& 0 & \cdots & 0
& 0 & \cdots & 0\BotStrut \\ \hline
0\TopStrut & \cdots & 0
& \dfrac{\barLambdaQCD^4+\Lambda_{K+1}^4}{f_{K+1}^2}
& \cdots
& \dfrac{\barLambdaQCD^4}{f_{K+1}f_{M+K}}
& \dfrac{\barLambdaQCD^4}{f_{K+1}f_{M+K+1}}
& \cdots
& \dfrac{\barLambdaQCD^4}{f_{K+1}f_{N}} \\
\vdots &  & \vdots
& \vdots & \ddots & \vdots
& \vdots &  & \vdots \\
0 & \cdots & 0
& \dfrac{\barLambdaQCD^4}{f_{M+K}f_{K+1}}
& \cdots
& \dfrac{\barLambdaQCD^4+\Lambda_{M+K}^4}{f_{M+K}^2}
& \dfrac{\barLambdaQCD^4}{f_{M+K}f_{M+K+1}}
& \cdots
& \dfrac{\barLambdaQCD^4}{f_{M+K}f_{N}}\BotStrut \\ \hline
0\TopStrut & \cdots & 0
& \dfrac{\barLambdaQCD^4}{f_{M+K+1}f_{K+1}}
& \cdots
& \dfrac{\barLambdaQCD^4}{f_{M+K+1}f_{M+K}}
& \dfrac{\barLambdaQCD^4}{f_{M+K+1}^2}
& \cdots
& \dfrac{\barLambdaQCD^4}{f_{M+K+1}f_{N}} \\
\vdots &  & \vdots
& \vdots &  & \vdots
& \vdots & \ddots & \vdots \\
0 & \cdots & 0
& \dfrac{\barLambdaQCD^4}{f_{N}f_{K+1}}
& \cdots
& \dfrac{\barLambdaQCD^4}{f_{M+K}f_{N}}
& \dfrac{\barLambdaQCD^4}{f_{N}f_{M+K+1}}
& \cdots
& \dfrac{\barLambdaQCD^4}{f_{N}^2}
\end{array}
\right) .
\label{eq:matrix_block}
\end{equation}

\endgroup
\end{widetext}
From the structure of \cref{eq:matrix_block}, we immediately see that the first block of the matrix is approximatively diagonal. Therefore, there exist $K$ axions whose masses are independent (in good approximation) of the QCD direction. Since the mixing of theses states is suppressed, the photon coupling of each axion $a_i$ is determined by its corresponding electromagnetic anomaly coefficient $E_i/N_i$ and decay constant $f_i$, for $i\in K$. These axions still receive a contribution from pion--axion mixing and therefore have a non-zero photon coupling even in the absence of an electromagnetic anomaly. In the case in which all scales are larger than $\barLambdaQCD$, so that $K=N$, all axions lie to the right of the (generalised) QCD line and there is no QCD-like axion. Beyond the maxion configuration of \cite{Gavela:2023tzu}, this is the only way to ``move'' the QCD line when $\barLambdaQCD=\LambdaQCD$, and it is only possible in general-PQ systems.

We can now reduce the dimensionality of the axion matrix\footnote{The zero-block off-diagonal entries can mix with the lighter axions, but their contributions are subleading.} and work with the $N-K$ axions left. This submatrix is approximately of rank $1+M$, where the one corresponds to the QCD direction while $M$ are the contributions of the scales close to $\barLambdaQCD$. We do not need to attempt the diagonalisation of this matrix, since we know that there exist eigenvectors with $0$-eigenvalue associated to the light eigenstates, in the limit $\bm{\Lambda}_- \to \bm{0}$. These eigenstates correspond to the orthogonal components of
\begin{equation}
\label{eq:light_eigenvectors}
    \bm{N}_0^T\cdot \bm{u}_{-,i}=0\, , \quad \bm{N}^T \cdot \bm{u}_{-,i}= \bm{N}_-^T \cdot\bm{u}_{-,i} =0 \, . 
\end{equation}
The first condition is easy to obtain, since we can assume that the eigenvector lives in the $\bm{V}_-$-subspace of the $\bm{V}_0\oplus \bm{V}_- -$vector space. From the second condition we also get that the masses correspond to the mass matrix projected onto the subspace orthogonal to $\bm{N}_-$
\begin{align}
    m_{-,i}^2 &= \bm{u}_{-,i}^T P_\perp^- \bm{M}_-^2 P_\perp^- \bm{u}_{-,i} \,\,\nonumber \\ &\textrm{with} \,\, P_\perp^- = \mathbf{1} - \hat{\bm{N}}_- \hat{\bm{N}}^T_- \, ,
\end{align} %
where 
$\hat{\bm{N}}_-=\bm{N}_-/|\bm{N}_-|$ stands for the normalised vector.
In our classification, we note that $1+M$ axions will in general have sizeable gluon couplings and live in the vicinity of the generalised QCD line. When the scales are close to $\barLambdaQCD$ we expect a quasi-maxion regime~\cite{Gavela:2023tzu}  characterised by a maximum displacement of $\sqrt{1+M}$ from the generalised QCD line. Finally we have $(N-M-K)-1$ axions with a mass-scale relation smaller than $\barLambdaQCD$. However, to identify the existence of states to the left of the QCD line we need to study the form of the photon-coupling vector
\begin{align}
    \tilde{g}_{a_i\gamma}^- &= \bm{g}_{a\gamma}^T\cdot \bm{u}_{-,i} = \frac{\alphaem}{2\pi}(\bm{E}-1.92\bm{N})^T\cdot\bm{u}_{-,i}  \nonumber \\ &=\frac{\alphaem}{2\pi} \bm{E}^T  \cdot \bm{P}_\perp^-  \cdot \bm{u}_{-,i} \equiv \frac{\alphaem}{2\pi}\bm{E}_-\cdot \bm{u}_{-,i}
\end{align}
where $\bm{E}_-$ is the projection of the $\bm{E}$ orthogonal to $\bm{N}$. We see that, naturally, the contribution from the pion-axion mixing is suppressed for these states, just from the orthogonality conditions in \cref{eq:light_eigenvectors}. Similarly, if $\bm{E}\propto \bm{N}$ with all $E_{i}/N_i$ equal, all states would be photophobic. 
In general, aside from accidental tuned cancellations, the existence of a subset of equal ratios $E_i/N_i$ is not by itself sufficient to guarantee a subset of vanishing photon couplings. The reason is that, in a generic light-sector mass matrix, the light eigenvectors $\bm u_{-,i}$ are linear combinations of all axion directions in the $\bm{V}_-$-subspace, so the corresponding photon couplings probe the full projected anomaly vector rather than individual entries. As a result, equal values of $E_i/N_i$ in a subset of components do not, in general, imply the existence of photophobic light eigenstates. However, it may happen that a subset of the axions has universal $E_i/N_i$. As mentioned above, mixing with the other light states can nevertheless induce a non-zero photon coupling. If this mixing is small, for instance due to a hierarchy among the $\Lambda_i $ scales, or if the coupling arises only through successive rotations before reaching a non-universal contribution, the resulting photon coupling is correspondingly suppressed.

A robust partial cancellation can only arise when the corresponding sector of the projected light mass matrix forms a degenerate block $M_{\rm light}^2 \supset m_I^2\,\mathbf{1}_I$, as we also saw in the two-axion section, in the limit of $\Lambda_1^2/f_1=\Lambda^2_2/f_2$. In this case, the light eigenspace associated with that block is well defined independently of the rest of the spectrum. If the restricted electromagnetic anomaly vector is aligned with the restricted QCD anomaly vector, then all light states but one, for the QCD direction, in that block have vanishing photon coupling.

The counting of photophobic states is particularly transparent when the projected light-sector mass matrix contains an exactly degenerate block $I$ of dimension $d$. In this case, the corresponding eigenspace is the full block, and the light states are selected only by orthogonality to the restricted QCD anomaly vector $\bm N_I$. Their photon couplings are controlled by the restricted electromagnetic anomaly vector $\bm E_I$, so photophobic directions are those orthogonal to both $\bm N_I$ and $\bm E_I$. Therefore, the counting depends only on whether $\bm E_I$ and $\bm N_I$ are aligned. If $\bm E_I\propto\bm N_I$, the two orthogonality conditions coincide, and all $d-1$ light states in the block are photophobic. If instead $\bm E_I\not\propto\bm N_I$, the two conditions are independent, leaving a $(d-2)$-dimensional photophobic subspace and one photon-coupled direction within the light subspace. Thus, the detailed pattern of $E_i/N_i$ entries inside the block is irrelevant for the counting; only the distinction between alignment and misalignment matters.

\subsection{A general axion sum rule } \label{sec:general_sumrule}

As we did for the two axion case, we turn our attention to computing the modifications to the sum rule in \cref{eq:photon_sumrule}. It is useful to note that this sum rule can be written in a manifestly invariant form,
\begin{align}
    \sum_i \frac{\tilde{g}_{a_i\gamma}^2}{m_i^2} &= \tilde{\bm{g}}_{a\gamma}^T\cdot\bm{M}_{\textrm{diag}}^{-2} \cdot \tilde{\bm{g}}_{a\gamma} = \tilde{\bm{g}}_{a\gamma}^T\cdot \bm{U}^T \cdot\bm{M}^{-2} \cdot \bm{U} \cdot \tilde{\bm{g}}_{a\gamma} \nonumber \\
    &= \bm{g}_{a\gamma}^T\cdot \bm{M}^{-2} \cdot\bm{g}_{a\gamma} \, .
    \label{eq:sum_rule_basis_dep}
\end{align}
Although the computation of the inverse may be challenging, using the fact that the sum rule is basis-independent we can work out this equation in the most adequate form. Concretely, in \cref{sec:MultiAxion}, we defined the full mass matrix as $\bm{M}^2=\bm{M}_{\cancel{\mathrm{PQ}}}^2+\LambdaQCD^4\bm{N}\bm{N}^T$, i.e. it is composed of the generic mass matrix $\bm{M}_{\cancel{\mathrm{PQ}}}^2$ corrected by the rank-1 contribution of the QCD direction. Since we assume all axions are massive\footnote{Our results can also be applied if some axions were to be massless. If $M$ out of the $N$ eigenstates are massless, one can block diagonalize those states first, and then apply our results to the $(N-M)\times (N-M)$ dimensional matrix.}, then $\bm{M}^2$ is invertible and the inverse can be calculated by using the Sherman-Morrison formula~\cite{Sherman_Morrison}\footnote{The Sherman-Morrison formula is a particular form of the Woodbury formula, which extends this equation to rank-$k$ corrections to matrices.}, which reads 
\begin{equation}
\label{eq:sherman}
    (\bm{Q}+\bm{u}\bm{v}^T)^{-1}= \bm{Q}^{-1}-\frac{\bm{Q}^{-1}\bm{u}\bm{v}^T\bm{Q}^{-1}}{1+\bm{v}^T\bm{Q}^{-1}\bm{u}}\,,
\end{equation}
provided that $1+\bm{v}^T\bm{Q}^{-1}\bm{u}\neq 0$, where $\bm{Q}$ is a square matrix and $\bm{u}, \bm{v}$ column vectors. Identifying $\bm{Q}=\bm{M}^2_{\cancel{\mathrm{PQ}}}$ and $\bm{u}=\bm{v}=\LambdaQCD^2 \bm{N}$ and using \cref{eq:sherman} for our case, we obtain
\begin{widetext}
\begin{equation}
\label{eq:THE_sumRule}
\begin{aligned}
    \sum_i \frac{\tilde{g}_{a_i\gamma}^2}{m_i^2} &= \tilde{\bm{g}}_{a\gamma}^T\cdot\bm{M}_{\textrm{diag}}^{-2} \cdot \tilde{\bm{g}}_{a\gamma}=\left(\frac{\alpha_\textrm{em}}{2\pi}\right)^2 (\bm{E}^{T}-C_\chi \bm{N}^{T})\left(\bm{M}_{\cancel{\mathrm{PQ}}}^{-2}-\LambdaQCD^4\frac{\bm{M}_{\cancel{\mathrm{PQ}}}^{-2}\bm{N}\bm{N}^{T}\bm{M}_{\cancel{\mathrm{PQ}}}^{-2}}{1+\LambdaQCD^4 \bm{N}^{T}\bm{M}_{\cancel{\mathrm{PQ}}}^{-2}\bm{N}}\right)(\bm{E}-C_\chi \bm{N}) \\
    &=
    \left(\frac{\alpha_\textrm{em}}{2\pi}\right)^2\left[\frac{1}{1+r}\left(\bm{E}-C_\chi\bm{N}\right)^T\bm{M}_{\cancel{\mathrm{PQ}}}^{-2}\left(\bm{E}-C_\chi\bm{N}\right)
    +
    \frac{r}{1+r}\left(\bm{E}^T\bm{M}_{\cancel{\mathrm{PQ}}}^{-2}\bm{E}-\LambdaQCD^4\frac{\left(\bm{E}^T\bm{M}_{\cancel{\mathrm{PQ}}}^{-2}\bm{N}\right)^2}{r}\right)\right] \, ,
\end{aligned} 
\end{equation}
\end{widetext}
where we have defined $r=\LambdaQCD^4 \bm{N}^{T}\bm{M}_{\cancel{\mathrm{PQ}}}^{-2}\bm{N}$. The equation above is explicitly basis independent and uses a generic parameterisation that only isolates the QCD contribution. We can immediately note that the second piece cancels exactly, whenever $\bm{E}$ is parallel to $\bm{N}$, and, since this expression can be used in any basis, if originally the vectors are aligned, the second piece is in any basis zero. Deviations from the assumptions $\textrm{det}(\bm{M}_{\cancel{\textrm{PQ}}}^2)=0$, as we will see below, are captured by the $r$ factor. In the basis of Appendix~\ref{app:generaldecomposition}, the sum rule can also be obtained (see~\cref{eq:newbasissumrule}), which makes the interplay between the aligned and orthogonal contributions with respect to QCD particularly transparent.

To identify the different phenomenological scenarios, we follow the procedure outlined in \cref{sec:Lambda_bar_definition} and extract from $\bm{M}_{\cancel{\mathrm{PQ}}}^{2}$ any contribution fully aligned with QCD. The full mass matrix is decomposed as
\begin{equation}
    \bm{M}^2=\bm{M}_A^2+\barLambdaQCD^4\bm{N}\bm{N}^T\,.
\end{equation}
Using this parameterisation, we now write explicitly \cref{eq:THE_sumRule} in the basis of \cref{eq:MassMatrix_basis} where $\bm{M}_A^2=\mathrm{diag}(\Lambda^4_1/f^2_1,...,\Lambda^4_N/f^2_N)$ and $r=\sum_{i=1}^N\barLambdaQCD^4/\Lambda^4_{i}$,

\begin{widetext}
\begin{align}
     \LambdaQCD^4 &\left(\frac{2\pi}{\alpha_\textrm{em}}\right)^2\sum_i \frac{\tilde{g}_{a_i\gamma}^2}{m_i^2}= \nonumber\\ &=\frac{1}{1+r}\left[\sum_i \left(\frac{E_i}{N_i}-C_\chi\right)^2\frac{\LambdaQCD^4}{\Lambda_i^4} + \left\{\left(\sum_i\frac{E_i^{^2}}{N_i^{2}}\frac{\LambdaQCD^4}{\Lambda_i^4}\right)\left( \sum_j\frac{\LambdaQCDbar^4}{\Lambda_j^4}\right)-\left(\sum_i\frac{E_i}{N_i}\frac{\LambdaQCD^2\LambdaQCDbar^2}{\Lambda_i^4}\right)^2\right\}\right]
\nonumber \\
&=
\underbrace{\frac{1}{1+r}}_{\det(\bm{M}_A^2)\neq 0}\left[\sum_i \left(\frac{E_i}{N_i}-C_\chi\right)^2
\frac{\LambdaQCD^4}{\Lambda_i^4}
+
\sum_{i<j}
\frac{\LambdaQCD^4\LambdaQCDbar^4}{\Lambda_i^4 \Lambda_j^4}
\underbrace{\left(\frac{E_i}{N_i}-\frac{E_j}{N_j}\right)^2}_{\text{non-universal }\bm{E}} \right]. \label{eq:THE_sumRule_basis}
\end{align}
\end{widetext}
We see that this expression exhibits two departures from the sum rule of \cref{eq:photon_sumrule}. The first is associated with general-PQ solutions to the strong CP problem and is encoded in the factor $1/(1+r)$ and in the presence of $\barLambdaQCD$. Indeed, assuming all axions to be massive, the existence of a PQ symmetry only broken by QCD translates to the following equivalence
\begin{flalign}
\exists\text{ QCD-PQ} & \Leftrightarrow\mathrm{det}\left(\bm{M}_{\cancel{\mathrm{PQ}}}^{2}\right)=0\\
 & \Leftrightarrow[\mathrm{det}\left(\bm{M}_A^2\right)=0]\land[\barLambdaQCD=\LambdaQCD]\,.
\end{flalign} The second departure arises from the non-universality of the photon anomaly vector. Restoring the two assumptions underlying the QCD sum rule of \cref{eq:QCD_sumrule}, one recovers the result of \cref{eq:photon_sumrule}. Explicitly, taking $\bm{E}=\frac{E}{N}\bm{N}$, \cref{eq:THE_sumRule_basis} further simplifies to
\begin{equation}
\begin{aligned}
    \LambdaQCD^4 \left(\frac{2\pi}{\alpha_\textrm{em}}\right)^2\sum_i \frac{\tilde{g}_{a_i\gamma}^2}{m_i^2} =& \left(\frac{E}{N}-C_{\chi}\right)^2\frac{\LambdaQCD^4}{\barLambdaQCD^4}\frac{r}{1+r} \\
    &\xrightarrow{\substack{\bar\Lambda_{\rm QCD}=\Lambda_{\rm QCD} \\ r\to\infty}} \left(\frac{E}{N}-C_{\chi}\right)^2 \, .
\end{aligned}
\end{equation}
Here, the limit $r\to\infty$ corresponds to sending the determinant of $\bm{M}_A^2$ to zero, and together with $\barLambdaQCD=\LambdaQCD$ imposes $\mathrm{det}(\bm{M}_{\cancel{\rm{PQ}}}^2)=0$. We thus recover precisely the sum rule of \cite{Gavela:2023tzu}, showing that \cref{eq:THE_sumRule} provides its natural form for a general multi-axion system, while using the basis of \cref{eq:MassMatrix_basis} gives a particularly descriptive form in \cref{eq:THE_sumRule_basis}. We also note that, in the limit $\bm{E}=(E/N)\bm{N}$, the modification to the sum rule of \cref{eq:photon_sumrule} is given by
\begin{equation}
\label{eq:photon_sumrule_gen_eqcoef}
     \frac{\LambdaQCD^4}{C_{a\gamma}^2} \sum_i \frac{\tilde{g}_{a_i\gamma}^2}{m_i^2} =\frac{\LambdaQCD^4}{\barLambdaQCD^4}\frac{r}{1+r} \, .
\end{equation}
Therefore, general-PQ systems can give values of the generalised sum rule larger or smaller than one. The PQ-breaking effects that do not fully align with QCD, contained in $\bm{M}_A^2$ and described in our particular basis by $\Lambda_i$, tend to make the sum rule smaller than 1, since $r/(1+r)\leq1$, i.e.~they can only move the QCD line to the right\footnote{Note that the fully aligned contribution feeds only into the QCD direction, while a partially aligned contribution also feeds in the orthogonal. For a partially aligned contribution to dominate the QCD direction (i.e.~to exceed $\LambdaQCDbar$), a much larger $\Lambda_i$ in the interaction basis is needed, which will then raise as well the orthogonal direction and make other eigenvalues heavy. If there exists a dominant almost fully aligned contribution that turns $\bm{N}$ into an approximate eigenvector of $\bm{M}^2_A$, this regime simply resembles qualitatively the $\barLambdaQCD\neq \LambdaQCD$ scenario.}. In contrast, the presence of PQ-breaking effects that fully align with QCD can make the sum rule larger or smaller than 1, since both $\barLambdaQCD\leq\LambdaQCD$ and $\barLambdaQCD\geq\LambdaQCD$ are possible and generated in consistent UV theories (an illustrative example is mirror world constructions, see e.g.~\cite{Berezhiani:2000gh,Hook:2018jle,DiLuzio:2021pxd} and Appendix~\ref{app:commuting_QCD}).

In the event that several axions are observed, a sufficiently precise experimental determination of the sum rule yielding a value below one would point to two possibilities (or to a combination of both): 1) additional axions remain to be discovered or 2) the underlying theory does not contain an exact QCD-PQ direction, i.e.~no PQ combination is broken exclusively by QCD. One could still attempt to reconstruct an ansatz for the multi-axion potential that describes data, and identify the right conditions that lead to a CP-conserving minimum. This would help identify candidate UV theories that naturally generate such a multi-axion potential. Similarly, the sum rule yielding a value larger than one, would point to two possibilities (or to a combination of both): 1) the presence of specific dynamics in the UV that can generate PQ-breaking effects that interfere destructively with QCD. To interpret such effects in the context of axion potentials, we again refer the reader to the aforementioned literature or to Appendix~\ref{app:commuting_QCD}. In this case, one can have axions to the left of the QCD line with universal anomaly coefficients. 2) Non-universal anomaly coefficients, as we explore in the following.

For simplicity, and to explore the deviations from the sum rule in \cref{eq:photon_sumrule}, let us now consider a QCD-PQ system, where the solution to strong CP is ensured by imposing $\det \bm{M}_{\cancel{\rm{PQ}}}^2=0$. This implies $\barLambdaQCD=\LambdaQCD$ and $\det\bm{M}_A^2=0$, meaning that at least one eigenvalue of the $\bm{M}_A^2$ matrix must vanish. Assuming that only one eigenvalue is zero\footnote{The Sherman-Morrison formula cannot be applied if more than one eigenvalue of $\bm{M}^2_A$ vanishes, because a rank-one addition is not sufficient to render the matrix invertible.}, one has $\LambdaQCD^4/\Lambda_1^4 \to \infty$, and therefore $1/(1+r)\to \Lambda_1^4/\LambdaQCD^4$. The expression then simplifies to
\begin{widetext}

\begin{align}
    \LambdaQCD^4\left(\frac{2\pi}{\alpha_\textrm{em}}\right)^2 \sum_i \frac{\tilde{g}_{a_i\gamma}^2}{m_i^2}\xrightarrow{\substack{\bar\Lambda_{\rm QCD}=\Lambda_{\rm QCD} \\ \Lambda_1\to 0}}& \Lambda_1^4\left[\frac{1}{\Lambda_1^4}\left(\frac{E_1}{N_1}-C_{\chi}\right)^2+\frac{1}{\Lambda_1^4}\sum_{j=2}\frac{\LambdaQCD^4}{\Lambda_j^4}\left(\frac{E_1}{N_1}-\frac{E_j}{N_j}\right)^2\right] \nonumber\\
    =& \left(\frac{E_1}{N_1}-C_{\chi}\right)^2+\sum_{j=2}\frac{\LambdaQCD^4}{\Lambda_j^4}\left(\frac{E_1}{N_1}-\frac{E_j}{N_j}\right)^2 \, .
\end{align}
\end{widetext}
In this limit, it is clear that the second term contributes only for non-universal anomaly coefficients. It is possible to rewrite the expression, simplifying the comparison to \cref{eq:photon_sumrule}:
\begin{equation}
     \frac{\LambdaQCD^4}{C_{a_1\gamma}^2} \sum_i \frac{\tilde{g}_{a_i\gamma}^2}{m_i^2}
     = 1+\sum_{j=2}\frac{\LambdaQCD^4}{\Lambda_j^4}\,\Delta E_j \geq 1 \, ,
\end{equation}
where
\begin{equation}
   \Delta E_j=\left(\frac{\alphaem}{2\pi \, C_{a_1\gamma}}\right)^2\left(\frac{E_1}{N_1}-\frac{E_j}{N_j}\right)^2 \, .
\end{equation}
Hence, a measurement of the sum rule yielding a value greater than one could point to a non-aligned photon anomaly vector. This may be accompanied by axions to the left of the QCD line or not. As discussed in \cref{sec:implications}, the former has been shown to have important implications for the UV completion of the SM, such as ruling out simple GUTs \cite{Agrawal:2022wyu}. However, the latter is also interesting. In principle, one may have a perfectly valid system of axions to the right of the QCD line with distinct anomaly coefficients. That would also lead to a sum rule larger than one and would similarly rule out simple GUTs. However, the deviation will be small if all $\Lambda_j\gg\LambdaQCD$, while more pronounced if at least one of the scales is much smaller, so the axions to the left of the QCD line or to the right but close to it are likely the most accessible scenarios.

For completeness, we also study the maxion case. However, for simplicity we will limit ourselves to the case of equal photon couplings to show the departure from \cref{eq:photon_sumrule}, once general PQ breaking is considered. In \cref{eq:photon_sumrule_gen_eqcoef}, we obtain the structure of the sum rule in the case when the QED and QCD anomalies are aligned. All axions will contribute equally to the photon sum rule when 
\begin{align}
    \frac{\LambdaQCD^4}{C_{a\gamma}^2} 
    \frac{\tilde{g}_{a_i\gamma}^2}{m_i^2} =\frac{1}{N}\frac{\LambdaQCD^4}{\barLambdaQCD^4}\frac{r}{1+r}\,,
\end{align}
where the specific value of $\tilde{g}_{a_i\gamma}^2/m_i^2$ will depend on the specific theory at hand through $r$. With respect to the sum rule in \cref{eq:photon_sumrule}, here the (generalised) maxions can displace a generalised QCD line that depends on $\barLambdaQCD$ and $r$ (controlled by the sum of all $\Lambda_i$ scales). This generalised QCD line may be to the left or to the right of the canonical QCD line. If this generalised QCD line is found to the right of the traditional QCD line, it could be either because the PQ-breaking effects fully aligned with QCD interfere constructively, i.e. $\barLambdaQCD>\LambdaQCD$, or because PQ-breaking effects non-fully aligned with the QCD direction are much larger than $\LambdaQCD$. If it moves to the left, it means that we have $\Lambda_i\ll\LambdaQCD$ for at least one $i$, and the PQ-breaking effects fully aligned with QCD interfere negatively. The generalised maxions can further displace this generalised QCD line by a maximum factor of $\sqrt{M}$, with $M$ being the number of (generalised) maxions.

We recover the maxion condition from Ref.~\cite{Gavela:2023tzu}, valid only for QCD-PQ systems with fully universal anomaly coefficients, if we take the limit $r\to \infty$, which corresponds to at least one $\Lambda_i\to 0$, and impose $\barLambdaQCD=\LambdaQCD$. 

In summary, it is important to emphasise that the logic of this sum rule stems from the fact that, while the left-hand side can in principle be accessed experimentally—through measurements of photon couplings and masses, appropriately combined and normalised to $\Lambda_{\rm QCD}$—the right-hand side is diagnostic and interpretative rather than uniquely reconstructible. It can be used to guide the interpretation of data and to motivate candidate axion potentials or UV completions, but it does not uniquely determine the underlying axion system. In the event of detecting a multi-axion system, one can confront our sum rule with the one of \cite{Gavela:2023tzu}, that we recover as a special limit of our general sum rule. For a given system, if the sum rule $< 1$ then in principle one may invoke the sum rule of \cite{Gavela:2023tzu} to predict the existence of undiscovered axions and to identify the regions of parameter space where they can be detected. Instead, if those predicted axions are not found, one may drop the assumptions of \cite{Gavela:2023tzu} to use our more general sum rule in \cref{eq:THE_sumRule}, and identify alternative theories that describe the main data. The details of the axion spectrum may be able to disentangle between both approaches. In contrast, if the sum rule $>1$, the system can only be described with our more general sum rule. Here our sum rule would guide candidate theories that agree well with data, since it connects the experimental data directly with the structure of PQ breaking and alignment of anomaly coefficients. But one could still go beyond our assumptions, i.e.~consider the presence of ALPs from $\mathrm{U}(1)$ symmetries not anomalous under QCD (see Appendix \ref{app:N_0}), and propose theories that may also describe data well. Again, the specific details of the axion spectrum may (or may not) be able to disentangle.

\subsection{A numerical example}

Finally, as we did for the previous section, we will verify numerically the results of this section. We will choose for this a setup of six axions, assuming $\barLambdaQCD=\LambdaQCD$ with the rest of scales and couplings shown in \cref{tab:benchmark_axion}.

Following our arguments, we see that in this benchmark one scale is larger than QCD, and hence we expect an axion widely separated to the right of the QCD line with unsuppressed photon coupling. Next we have a scale close to QCD and four scales much smaller with two of them being degenerate. For the photon coupling, we have chosen coefficients that typically appear in axion models, and fixed $E_i/N_i$ arbitrarily, except for the two scale-degenerate axions, since we want to show numerically that a universal subset of photon couplings gives rise to a photophobic light axion. Hence, following the results of this section we expect two axions close to the QCD line, one axion to the right of the QCD line with unsuppressed photon coupling, and another with a light mass and suppressed coupling, and finally two axions to the left of the QCD line.

\begin{figure*}[tbh]
    \centering
    \includegraphics[width=\linewidth]{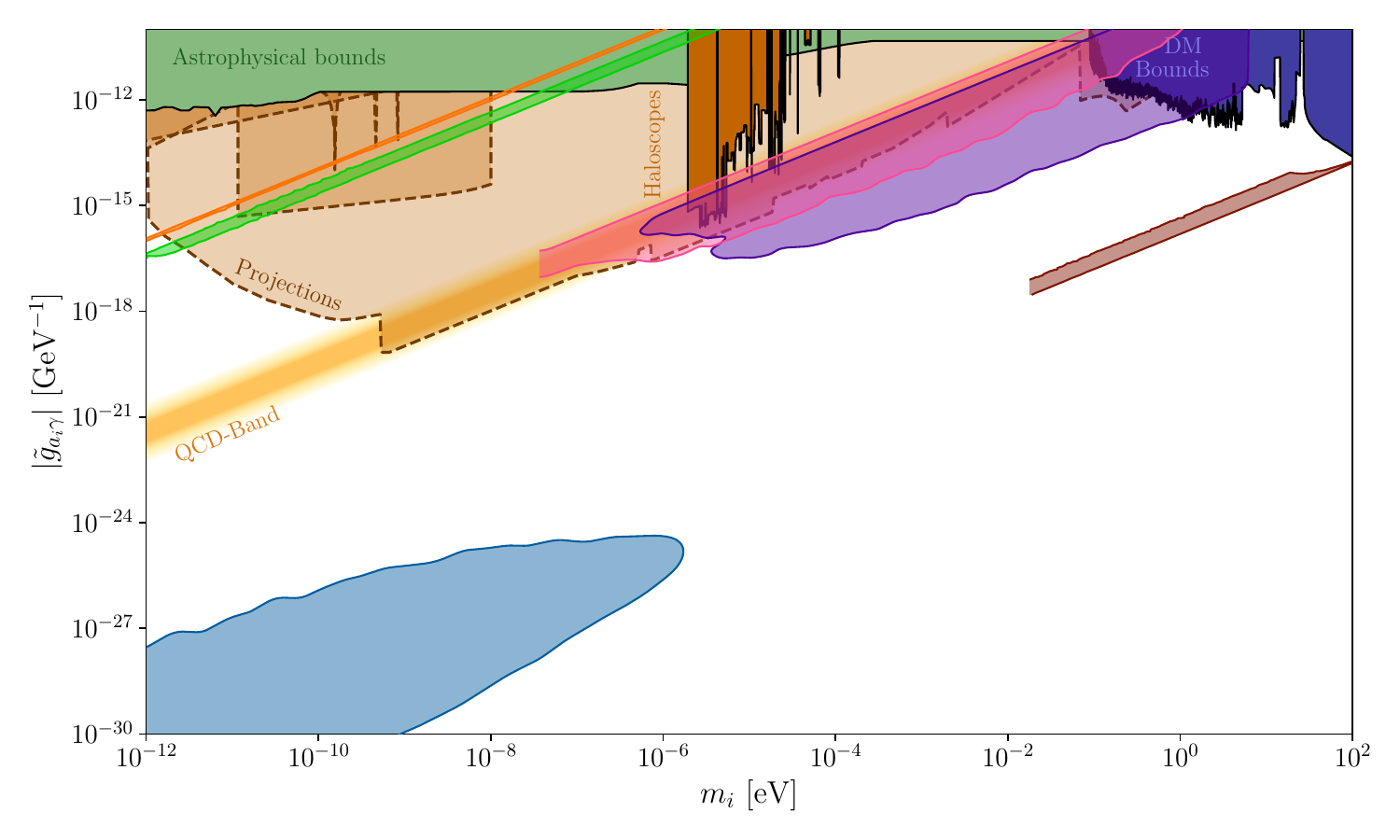}
    \caption{Numerical scan of the six-axion benchmark of \cref{tab:benchmark_axion}, with the six regions corresponding to the six different axion mass eigenstates. More details can be found in the main text.}
    \label{fig:6axions}
\end{figure*}

The numerical result of this model is presented in \cref{fig:6axions}, where we scan over all $f_i\in\left[10^6,\, 10^{14}\right]\,$GeV, with the only constraint of $f_5=f_6$ to show this degenerate limit. We also discard points that may break the hierarchy of the matrix. In \cref{fig:6axions} we observe the predictions given above, with one clear axion state far to the right of the QCD line (brown), two QCD-like axions (pink and purple), one photophobic light axion (blue) and two axions to the left of the QCD line (green and orange). We note that we have forced the degeneracy of two of the states to show this feature. However, the bottom line of this study should be that barring this highly tuned degeneracy, we expect all light states to be to the left of the QCD band, unless the EM anomaly vector is universal. The remaining states are then found to the right of the QCD line. Finally, although there exist points in between which mix some of the regions, the distinction of axions to the left and to the right is fully captured by the arguments above.

\begin{table}
    \centering
    \begin{tabular}{c|c|c|c|c|c|c}
         & $a_1$ &$a_2$ & $a_3$&$a_4$ &$a_5$ & $a_6$  \\
         \hline
         $\Lambda_i$ [GeV]&  42 & $61\times10^{-3}$& $3\times10^{-4}$ &$2.1\times10^{-4}$ &$10^{-4}$ &$10^{-4}$  \\[2pt]
         \hline 
         $E_i/N_i$ & 5/3 & -5/3 & 2 &-5/3 &8/3 &8/3
    \end{tabular}
    \caption{Benchmark point used for \cref{fig:6axions}. More details can be found in the main text.}
    \label{tab:benchmark_axion}
\end{table}

\subsection{Implications for UV theories} \label{sec:implications}
Before concluding this section, we comment on the theoretical plausibility of the scenarios discussed here and on their consistency within natural UV-complete theories. A simplified yet comprehensive introduction to multi-axion model-building can be found in Appendix \ref{app:UVmodels}. This includes simplified models that capture the core mechanisms underlying well-motivated UV constructions in the literature. A general conclusion from this section is that PQ structure and anomaly structure not only govern the low-energy phenomenology of the multi-axion system, but also connect this with potential UV completions and allow a systematic classification of the strong CP axion theory landscape, as shown in Figure \ref{fig:QCD-general-PQ}. 

First, we note that, while it is well known that a generic ALP that is not anomalous under QCD but has an anomalous coupling to photons can broadly populate the $(m_a,g_{a\gamma})$ plane, our analysis shows that multi-axion systems exhibit a non-trivial interplay between axion masses and photon couplings. The way in which this behaviour maps onto regions of parameter space lying to the left or to the right of the QCD line is, in itself, an interesting result of our study. If such a particle is ever discovered, it would be theoretically possible (even if experimentally difficult) to discriminate whether it is a generic ALP or a member of a multi-axion system if the photon coupling can be measured with enough precision to disentangle a potential $G\tilde{G}$ (axion-pion mixing) component. 

On the other hand, the general multi-axion system explored here does not need to originate from an ad-hoc set of PQ symmetries, since well-motivated UV constructions have been proposed in which more than one axion is naturally predicted, for instance to relax additional sources of CP violation in BSM theories, as illustrated in Appendix \ref{app:new_strong_sectors} with simple examples. Another motivated class of multi-axion systems arises in theories with extra dimensions or in string theory, where a tower of 4D pseudo-scalars coupled to $G\tilde{G}$ may originate from a single bulk axion, and multiple bulk or brane axions may arise from higher-dimensional gauge fields (see Appendix \ref{app:extra_dimensions} for simple examples). In Appendix \ref{app:Multi-Peccei-Quinn-model-building}, we provide a brief introduction to multi-PQ systems and general model-building guidelines, while in Appendix \ref{app:scalar_mixing} we discuss how axion mixing may originate from explicit PQ breaking in the scalar sector.

In particular, finding extra axion states to the right of the QCD line, while there exists a QCD axion, is a prediction of very generic multi-axion models and compatible with generic UV completions of the SM. On the other hand, if no QCD axion is found but several axions are found away of the QCD band, this would suggest the existence of a general-PQ axion system. Depending on the specific pattern of axions, and considering the value of the generalised axion sum rule, this could motivate specific UV completions. For example, finding axions to the left of the QCD line but no QCD axion, could imply that $\barLambdaQCD\neq \LambdaQCD$, some $\Lambda_i$ are smaller than $\barLambdaQCD$ and no $\Lambda_i$ is accidentally on the QCD band. This delivers a sum rule larger than one, and would suggest a mirror world construction \cite{Berezhiani:2000gh,Hook:2018jle,DiLuzio:2021pxd} or non-trivial embedding of QCD in a parent-colour group \cite{Valenti:2022tsc} (see also Appendix \ref{app:commuting_QCD}). In contrast, finding all axions to the right of the QCD band would more likely suggest that all $\Lambda_i$ are much larger than $\LambdaQCD$, which can happen for example if QCD emerges as the diagonal subgroup of a larger strong sector \cite{Agrawal:2017ksf,Gaillard:2018xgk,Fuentes-Martin:2019bue,Csaki:2019vte} (see also Appendix \ref{app:QCDdiag}). This delivers a sum rule smaller than one.

Moreover, by discovering a multi-axion system with non-universal anomaly coefficients, one may extract crucial information about the UV completion of the SM. This could be verified by finding an axion to the left of the QCD line, as discussed in \cite{Agrawal:2022wyu}, but also by measuring a generalised sum rule larger than one in a system with all axions to the right of the QCD line, as discussed in this paper. In principle, different ratios of anomaly coefficients for different axions can be achieved in very generic scenarios, not requiring fine-tuning nor convoluted UV theories. For example, one may consider a generalisation of the KSVZ model~\cite{Kim:1979if,Shifman:1979if} with a different fermion representation for each PQ symmetry, e.g.~$U_{1}\sim(\boldsymbol{3},\boldsymbol{1},2/3)$ for $\UPQ^1$ and $D_{2}\sim(\boldsymbol{3},\boldsymbol{1},-1/3)$ for $\UPQ^2$ in a two-axion system. This would naturally induce $E_1/N_1\neq E_2/N_2$. Similarly, one could envisage implementations of the DFSZ model~\cite{Zhitnitsky:1980tq,Dine:1981rt} with different choices of fermion PQ charges or non-minimal DFSZ models. In fact, as long as the QCD or QED anomalies are induced by different fermion representations, this scenario generically arises, which from the point of view of the SM seems more natural than having degenerate PQ sectors with exactly the same fermion representations.

Nevertheless, when QCD and QED are embedded in the same simple gauge group, the ratio of anomaly coefficients depends only on the index of embedding  \cite{Wise:1981ry,Srednicki:1985xd,DiLuzio:2018gqe,Ernst:2018bib,Agrawal:2022wyu}, and is therefore independent of the fermion representations inducing the anomaly. For instance, embedding the SM in $\mathrm{SU}(5)$ or $\mathrm{SO}(10)$ yields $E/N = 8/3$ irrespective of the representation. Here, one may still achieve different $E_i/N_i$ in non-minimal scenarios involving an additional unbroken hidden $\mathrm{U}(1)$ with large kinetic mixing with hypercharge, generalising the mechanism of \cite{Daido:2018dmu}. It may also be possible to achieve axions to the left of the QCD line in mirror world constructions with negative instanton interference \cite{Hook:2018jle,DiLuzio:2021pxd}. Instead, if QCD and QED are not embedded within a single simple group, then different $E_i/N_i$ can be generically obtained. Explicit examples include the Pati-Salam \cite{Pati:1974yy}, flipped $\SU{5}$/$\mathrm{SO}(10)$ \cite{Barr:1981qv,Bertolini:2010yz} and trinification \cite{DeRujula:1984mnu} models\footnote{In principle, one may still achieve unification of gauge couplings by incorporating cyclic symmetries to these groups. In fact, to achieve different $E_i/N_i$ in the trinification setup, the simplest possibility suggests a multi-axion system.}.

On the other hand, it is possible, and in fact relatively generic, to realise axions with different $E_i/N_i$ factors from string theory, for instance in intersecting D-brane models from Type II compactifications \cite{Halverson:2019cmy,Gendler:2023kjt}, where QCD and QED originate from different sectors. Another example could be heterotic string theory with gauge symmetry $E_8\times E_8$, where QCD and QED can be embedded in different $E_8$ factors \cite{Agrawal:2024ejr,Reig:2025dqb}. Therefore, the discovery of such an axion would rule out well-motivated UV scenarios and point to specific embeddings of the SM in the UV. We believe this to be a strong theoretical motivation to search for these states in axion experiments, taking advantage of the general formalism and sum rule developed in this paper.

\section{Conclusions and outlook}\label{sec:conclusion}

Multi-axion solutions to the strong CP problem lead to non-generic phenomenology and motivate axion searches beyond the traditional QCD axion band. While this framework is well motivated and has attracted increasing attention in recent years, a systematic and fully general study has been missing from the literature, where the relevant phenomenological details are often diluted because most works focus on specific models or scenarios. In this paper, by contrast, we have carried out a systematic and general analytical study of these systems, highlighting the different phenomenological opportunities to probe multi-axion scenarios in axion experiments. In doing so, we have also uncovered new scenarios that remain under-explored in the literature, despite their strong theoretical and experimental motivation.

In this work, we have considered a general multi-axion system with $N$ Peccei-Quinn symmetries anomalous under QCD. In addition to the QCD contribution to the axion potential, we allow for a generic PQ-breaking potential that generically breaks all PQ symmetries, assuming no massless axions. In contrast to much of the existing literature, we do not restrict our analysis to potentials that preserve a PQ symmetry broken only by QCD. While this is a sufficient condition to solve the strong CP problem, it is not a necessary one; in particular, it is not realised in some UV-complete models that naturally motivate multi-axion systems as a way to relax new sources of strong CP violation, see e.g.~\cite{Agrawal:2017ksf,Gaillard:2018xgk,Fuentes-Martin:2019bue,Csaki:2019vte} and Appendix~\ref{app:new_strong_sectors}. Moreover, we allow for the possibility that the different axions have different $E/N$ anomaly coefficient ratios, rather than assuming them to be universal, as is often done in the literature. This more general treatment uncovers a wide range of phenomenological possibilities, some of which remain under-explored in the literature despite providing strong motivation for new axion searches.

\begin{table*}[ht]
    \centering
    \renewcommand{\arraystretch}{1.15}
    \begin{tabular}{|c|c|c|c|c|}
    \hline
\textbf{PQ structure} &\begin{tabular}{@{}c@{}}\textbf{Anomaly} \\ \textbf{structure}\end{tabular}  &\textbf{Sum rule} &\textbf{Spectrum} &\textbf{UV implications} \\
\hline
 \begin{tabular}{@{}c@{}}QCD-PQ:\\$\Lambda_i \ll \Lambda_{\mathrm{QCD}}$\end{tabular}&$\mathbf{E}\parallel \mathbf{N}$&$=1$&Photophobic + QCD axion&Canonical multi-axion\\
\hline
\begin{tabular}{@{}c@{}}QCD-PQ:\\ $\Lambda_i \gg \Lambda_{\mathrm{QCD}}$\end{tabular}&$\mathbf{E}\parallel \mathbf{N}$& $=1$&Heavy gluon-coupled + QCD axion&Canonical multi-axion\\
\hline
\begin{tabular}{@{}c@{}}QCD-PQ:\\ $\Lambda_i \simeq \Lambda_{\mathrm{QCD}}$\end{tabular}&$\mathbf{E}\parallel \mathbf{N}$& $=1$&Maxions&Coincidence of scales\\
\hline
\begin{tabular}{@{}c@{}}QCD-PQ:\\ $\Lambda_i \ll \Lambda_{\mathrm{QCD}}$\\ $\Lambda_i > \Lambda_{\mathrm{QCD}}$\end{tabular}&$\mathbf{E}\nparallel \mathbf{N}$&$>1$
&Axions left, or right but close&\begin{tabular}{@{}c@{}} Simple GUTs ruled out $\rightarrow$\\  non-GUT / string / \\intersecting branes\end{tabular}
\\
\hline
\begin{tabular}{@{}c@{}}General-PQ: \\ $\bar{\Lambda}_{\mathrm{QCD}} < \Lambda_{\mathrm{QCD}}$\end{tabular}
&$\mathbf{E}\parallel \mathbf{N}$&$>1$&Axions left but GUT-compatible& \begin{tabular}{@{}c@{}}Negative alignment $\rightarrow$\\ mirror-world-like\\ (axion non-linear)\end{tabular} 
\\
\hline
\begin{tabular}{@{}c@{}}General-PQ: \\ $\bar{\Lambda}_{\mathrm{QCD}} > \Lambda_{\mathrm{QCD}}$\end{tabular}&$\mathbf{E}\parallel \mathbf{N}$&
$<1$&\begin{tabular}{@{}c@{}}No axion in QCD band;\\  left requires $\mathbf{E}\nparallel \mathbf{N}$\end{tabular} &\begin{tabular}{@{}c@{}}Positive alignment $\rightarrow$\\   mirror-world-like\\(axion linear), parent-colour\end{tabular} \\
\hline
\begin{tabular}{@{}c@{}}General-PQ: \\ $\Lambda_{i=1,\ldots,N} \gg \Lambda_{\mathrm{QCD}}$\end{tabular} &\begin{tabular}{@{}c@{}}$\mathbf{E}\parallel \mathbf{N}$ or\\ $\mathbf{E}\nparallel \mathbf{N}$\end{tabular}&
$<1$&No axion in QCD band nor left&Embedded / diagonal QCD\\
\hline
    \end{tabular}
    \caption{Structural classification of multi-axion solutions to the strong CP problem.
    The PQ-breaking structure and anomaly alignment determine the qualitatively distinct spectra and the implications for UV completions. The sum rule acts as a diagnostic tool that helps to discriminate experimentally between the different scenarios.}
    \label{tab:pq_classification}
\end{table*} 

We have classified systems with a PQ symmetry only broken by QCD as \textbf{QCD-PQ} systems, while we denote the more general case as \textbf{general-PQ} systems. Within the latter, there is the possibility that there exist PQ-breaking effects that fully align with QCD, interfering positively or negatively, while still solving strong CP. A well-known example of this are mirror world constructions, see e.g.~Appendix \ref{app:commuting_QCD}. This defines a generalised QCD line controlled by an effective parameter $\barLambdaQCD$, which could be to the right or to the left of the traditional QCD band. 

A fundamental result of our work is that the aforementioned PQ structure (whether the system is QCD-PQ or general-PQ) and the (non-)universality of the anomaly coefficient ratios $E_i/N_i$ for all axions govern the multi-axion system. They connect the low-energy phenomenology with the underlying system and the ultraviolet completion, allowing a simple yet general and systematic classification of qualitatively distinct scenarios.

For a system of $N$ axions we have identified an illustrative basis choice where $N$ diagonal sources of PQ breaking beyond QCD, characterized by scales $\Lambda_i$ ($i=1,\ldots,N$), allow to describe qualitatively all the following phenomenological scenarios, together with the conditions under which they arise. We have highlighted how these different scenarios occur in a general two-axion system, and then we have generalised the system to $N$ axions and extracted the conditions under which the various scenarios arise, using a generalised vector formalism that allows to classify the various scenarios in a very simple way in terms of sub-matrix rank and anomaly vector alignment in the $N$-axion vector space. This allowed us to derive a \textit{general axion sum rule} that controls the behaviour of mass and photon couplings in the general multi-axion system. Under the discovery of a multi-axion system, our sum rule and the more restricted sum rule of \cite{Gavela:2023tzu}, which only applies to QCD-PQ systems with universal anomaly coefficients, can be contrasted to identify the structure of PQ symmetry breaking in the multi-axion system, hence helping in the reconstruction of the axion potential and pointing to specific constructions in the UV.

The various phenomenological scenarios and the conditions under which they arise are summarised in Tab.~\ref{tab:pq_classification} and are the following:
\begin{itemize}
    \item If $\Lambda_i \gg \barLambdaQCD=\LambdaQCD$ for $i=1,\ldots,K$, then there are $K$ axions that are generally heavier than the usual QCD axion, but have a similar photon coupling induced by pion mixing or, in addition, UV contributions. This scenario may be probed through a combined effort involving significantly improved haloscope and helioscope experiments, together with cosmological and astrophysical tests.
    \item If $K=N$, then the non-QCD PQ-breaking potential provides the leading breaking of all PQ symmetries and there is \textit{no} axion in the QCD band. The signal of the axion solution to the strong CP problem would be here the presence of multiple axions to the right of the QCD band, perhaps suggesting a non-trivial embedding of QCD in the UV. This may be tested by a complemented effort of significantly enhanced haloscopes/helioscopes experiments and cosmology/astrophysics tests. This scenario predicts a sum rule $< 1$.
    \item If $\Lambda_i \ll \Lambda_{\mathrm{QCD}}$ with $i=1,...,M$, then there are $M$ axions generally lighter than the usual QCD axion. Their coupling to photons depends crucially on the $E/N$ ratios. If these are universal, meaning $E_i/N_i=E_j/N_j$ for $i,j=1,...,M$, then the photon coupling is further suppressed than the axion mass, and all these axions are photophobic and very difficult to discover in axion experiments.
    \item Conversely, if some of the $E_i/N_i$ differ, so that the photon-coupling vector is not fully aligned with the gluon-anomaly vector, then these states generically lie to the left of the QCD band, and may fall within the projected sensitivity of haloscopes and helioscopes. From the theoretical point of view, this appears to be a rather generic PQ scenario from the perspective of the SM, since it can be realised simply by introducing different fermion representations that generate the PQ anomalies. However, the detection of such an axion would rule out the possibility of embedding QCD and QED into the same simple group, including well-motivated unification frameworks such as $\SU{5}$ and $\textrm{SO}(10)$, which predict a universal value $E/N=8/3$ independently of the fermion representations \cite{Agrawal:2022wyu}. Instead, such an observation would point towards more specific embeddings of the SM, such as Pati-Salam, trinification or string theories where QCD and QED originate from different sectors \cite{Halverson:2019cmy,Gendler:2023kjt}.
    \item Similarly, one may have a system with a generic QCD axion and other axions to the right of the QCD line that have non-universal anomaly coefficients. This would also rule out simple GUTs, and may be identified by measuring a value of our general axion sum rule larger than one, but would likely require very high experimental precision if all the extra axions are very far to the right of the QCD line.
    \item If $\barLambdaQCD\ll\LambdaQCD$ then one could measure axions to the left of the QCD line that have universal anomaly coefficients, meaning they are compatible with gauge unification of QCD and QED in the same simple group, but also suggesting more non-trivial dynamics in the UV such as the aforementioned mirror-world constructions. Crucially, this scenario predicts a sum rule $>$ 1. Conversely, if $\barLambdaQCD\gg\LambdaQCD$, then one could see axions to the left of the QCD line but no axion in the QCD line. These two scenarios allow to distinguish whether the PQ-breaking effects that fully align with QCD interfere positively or negatively. Together with the case $\Lambda_{i=1,...,N} \gg \barLambdaQCD=\LambdaQCD$ and the maxion case of Ref. \cite{Gavela:2023tzu}, these are the only ways to avoid the QCD band in a multi-axion system, since otherwise at least one axion is expected to live in the traditional QCD band.
\end{itemize}

Taken together, our results establish a systematic link between experimental axion searches, multi-axion phenomenology, and the ultraviolet structure of solutions to the strong CP problem. They also show that axions away from the canonical QCD band should not be interpreted solely as generic ALPs: they can arise as characteristic signatures of multi-axion solutions to the strong CP problem. Their discovery could have implications for our understanding of fundamental physics and the ultraviolet completion of the Standard Model.

\section*{Acknowledgements }
The authors thank Jason Aebischer, Arturo de Giorgi, Luca Di Luzio, Luca Merlo and Pablo Qu\'ilez for comments on a preliminary version of the draft. MFN is grateful to Pablo Qu\'ilez for very enjoyable discussions on axions during the La Thuile 2026 conference. 
This research was supported by the Swiss National Science Foundation, project No. 2000-1-240011. 
The work of MFZ is supported by the Spanish MIU through the National Program FPU (grant number FPU22/\hspace{0pt}03625), by the European Union's Horizon 2020 research and innovation programme under the Marie Sk\l odowska-Curie grant agreement No.~101086085-ASYMMETRY and by the Spanish Research Agency (Agencia Estatal de Investigaci\'on) through the grant IFT Centro de Excelencia Severo Ochoa No CEX2020-001007-S and by the grant PID2022-137127NB-I00 funded by MCIN/\hspace{0pt}AEI/\hspace{0pt}10.13039/501100011033.
MP is grateful for the funding from the Swiss National Science Foundation (SNSF) through grant TMSGI2-225951. 
This article is based upon work from COST Action COSMIC WISPers CA21106, supported by COST (European Cooperation in Science and Technology).
%-----------------------------------------------------------------------------
\appendix

\section{UV complete models} \label{app:UVmodels}

In this section, we present explicit UV-complete models that give rise to two-axion systems with massive axions that naturally mix. We show how these constructions realise the different phenomenological regimes discussed in the main text. All models can be straightforwardly generalised to the case of $N$ axions. Rather than attempting a comprehensive review of the literature, we instead provide general guidelines for multi-axion model building and highlight simple setups that naturally motivate the emergence of multi-axion systems.

\subsection{ Multi-Peccei-Quinn model building}\label{app:Multi-Peccei-Quinn-model-building}

Let us consider a renormalisable model with two SM singlet complex
scalars $S_{1,2}\sim(\boldsymbol{1},\boldsymbol{1},0)$,
\begin{equation}
\mathcal{L}\supset|\partial_{\mu}S_{1}|^{2}+|\partial_{\mu}S_{2}|^{2}-V(S_{1,2})\,.\label{eq:UVmodel_potential}
\end{equation}
If the scalar potential only contains trivial self-conjugate quadratic
terms, i.e.~terms of the form $|S_{1}|^{n}|S_{2}|^{m}$ with $n,m=0,2,4$
and $n+m\leq4$, then the Lagrangian is invariant under two independent
global $\Uone$ symmetries associated with independent re-phasing of
the two singlets. If both fields develop VEVs, $\langle S_{i}\rangle=v_{i}/\sqrt{2}$,
then both $\Uone$ symmetries are spontaneously broken and the scalar
fields can be decomposed in terms of two radial and two angular (axion)
modes,
\begin{equation}
S_{1,2}=\frac{1}{\sqrt{2}}\left(v_{1,2}+\rho_{1,2}\right)e^{ia_{1,2}/v_{1,2}}\,.\label{eq:polar_decomposition}
\end{equation}
Assuming both scalars couple to arbitrary coloured fermions such that
an anomaly under QCD is induced, then the following anomalous couplings
arise at low energies
\begin{equation}
\mathcal{L}\supset\frac{\alpha_{s}}{8\pi}\left(\frac{a_{1}}{f_{1}}+\frac{a_{2}}{f_{2}}\right)G\widetilde{G}+\frac{\alpha_{\mathrm{em}}}{8\pi}\left(\frac{E_{1}}{N_{1}}\frac{a_{1}}{f_{1}}+\frac{E_{2}}{N_{1}}\frac{a_{2}}{f_{2}}\right)F\widetilde{F}\,,
\end{equation}
where we have absorbed the QCD anomaly coefficients into a redefinition
of the VEVs as $f_{i}=v_{i}/N_{i}$. The anomaly coefficients depend on the representations of the coloured fermions, where the most popular classes of models include:
\begin{itemize}
\item KSVZ models: the anomalies are induced by new heavy fermions that
are vector-like under the SM but chiral under the PQ symmetries, getting
their mass from $\langle S_{i}\rangle$. For example, one may consider
one heavy quark for each PQ symmetry, like the set $U_{1}\sim(\mathbf{3},\mathbf{1},2/3)$
and $D_{2}\sim(\mathbf{3},\mathbf{1},-1/3)$, which gives $E_{1}/N_{1}=8/3$,
$E_{2}/N_{2}=2/3$. Choosing SM-like representations is desirable
since these mix and decay into SM fermions, while more exotic representations
like $\Psi\sim(\mathbf{3},\mathbf{1},0)$ generically lead to stable
coloured relics and fractionally charged hadrons unless these are
diluted by inflation\footnote{For a catalogue of consistent KSVZ models see Ref.~\cite{DiLuzio:2024xnt}.}. Similarly, smaller sets of coloured representations
are preferable to avoid the formation of dangerous domain walls from
PQ spontaneous breaking. This is classified in terms of the domain
wall number $N_{\mathrm{\mathrm{DW}}}=N/k$, where $k$ accounts for
the largest true discrete symmetry $Z_{k}\subset \UPQ$
that remains unbroken after PQ breaking. In simple models typically
$k=1$ and $N_{\mathrm{\mathrm{DW}}}=N$. For a single heavy quark
in the fundamental of QCD we have $N=1$, while for a quark doublet
$Q\sim(\mathbf{3},\mathbf{2},1/6)$ we have $N=2$ as this contains
two fundamentals. In general, $N_{\mathrm{DW}}>1$ leads to a domain
wall problem unless PQ breaking happens before inflation.
\item DFSZ models: the anomalies are induced by SM fermions. To be consistent
with the structure of Yukawa couplings of the SM, one needs to add
a second Higgs doublet, in such a way that $H_{u}$ couples to up-quarks
and $H_{d}$ to down-quarks, both charged under PQ along with the
PQ breaking scalar(s) $S_{i}$. Minimal DFSZ models involve charging
$\SU{2}$ singlets with flavour universal PQ charges, and the only
freedom remaining is whether charged leptons couple to $H_{u}$ or
$H_{d}$, resulting in $E/N=8/3$ or $2/3$ respectively. In either
case, $N_{\mathrm{DW}}\geq3$. To address this without invoking inflation,
one must go beyond minimal DFSZ, e.g.~considering flavour dependent
PQ charges~\cite{Davidson:1983tp,Davidson:1984ik,DiLuzio:2023ndz,Cox:2023squ}, extra Higgs doublets, extra fermions...
\end{itemize}
In principle, when building multi-PQ models one needs to take into account
the considerations above for each PQ symmetry. Generically, each PQ
sector could be realised by its own KSVZ or DFSZ mechanism.

Regardless of the fermion content inducing the anomalies, after QCD
confinement the resulting low-energy potential contains the QCD contribution
generated by the axion-gluon couplings, plus contributions from generic
sources of PQ breaking,
\begin{widetext}
    \begin{equation}
V\approx-\Lambda_{\mathrm{QCD}}^{4}\cos\left(\frac{a_{1}}{f_{1}}+\frac{a_{2}}{f_{2}}-\overline{\theta}\right)-\sum_{i}\Lambda_{i}^{4}\cos\left(c_{i1}\frac{a_{1}}{f_{1}}+c_{i2}\frac{a_{2}}{f_{2}}-\delta_i\right)\,,
\end{equation}
with $\delta_i$ being new generic CP-violating phases that may originate from the non-QCD sector. Note that here we approximate sources of PQs breaking to cosine functions in axion fields thanks to the periodicity of the axion potential. In some cases, one needs to go beyond this approximation, as discussed later. The mass matrix is then
\begin{equation}
\bm{M}^{2}=\left.\frac{\partial^{2}V}{\partial a_{i}\partial a_{j}}\right|_{a_{1},a_{2}=0}=\Lambda_{\mathrm{QCD}}^{4}\left(\begin{array}{cc}
{\displaystyle \frac{1}{f_{1}^{2}}} & {\displaystyle \frac{1}{f_{1}f_{2}}}\\
{\displaystyle \frac{1}{f_{1}f_{2}}} & {\displaystyle \frac{1}{f_{2}^{2}}}
\end{array}\right)+\sum_{i}\Lambda_{i}^{4}\left(\begin{array}{cc}
{\displaystyle \frac{c_{i1}^{2}}{f_{1}^{2}}} & {\displaystyle \frac{c_{i1}c_{i2}}{f_{1}f_{2}}}\\
{\displaystyle \frac{c_{i1}c_{i2}}{f_{1}f_{2}}} & {\displaystyle \frac{c_{i2}^{2}}{f_{2}^{2}}}
\end{array}\right)
\end{equation}
\end{widetext}
In the absence of PQs breaking beyond QCD, the linear combination
$\tilde{a}_{2}=\frac{1}{\sqrt{f_{1}^{2}+f_{2}^{2}}}(f_{2}a_{1}+f_{1}a_{2})$
gets mass from QCD while $\tilde{a}_{1}=\frac{1}{\sqrt{f_{1}^{2}+f_{2}^{2}}}(-f_{1}a_{1}+f_{2}a_{2})$
remains massless. The presence of PQ breaking sources beyond
QCD leads to the rich phenomenology discussed in the main text. In the following, we discuss well-motivated explicit constructions that give rise to PQs breaking sources beyond QCD and without introducing a quality problem in the solution of strong CP. We also distinguish between scenarios were a PQ symmetry is only broken by QCD (QCD-PQ), and scenarios were all PQ symmetries are generally broken beyond QCD (general-PQ), showing well-motivated UV scenarios that realise each PQ-breaking structure while achieving a CP conserving minimum.

\subsection{Explicit breaking from scalar sector}\label{app:scalar_mixing}

If the scalar potential in \cref{eq:UVmodel_potential} contains
a non-self-conjugate coupling of the form
\begin{equation}
V(S_{1,2})\supset\lambda \, S_{1}^{n_{1}}(S^{(\dagger)}_{2})^{n_{2}}+\mathrm{h.c.}\,,\label{eq:scalar_breaking}
\end{equation}
then the above coupling breaks the two PQs down to a subgroup
$\UPQ^{1}\times \UPQ^{2}\rightarrow \UPQ^{n_{1}\alpha_{1}\mp n_{2}\alpha_{2}}$.
By substituting the polar decomposition of $S_{1,2}$, \cref{eq:polar_decomposition}, and taking into account the hermitian conjugate term, the cosine potential
emerges as
\begin{widetext}
\begin{equation}
V_{\mathrm{\cancel{\textrm{PQ}}}}(a_{1},a_{2})\approx-|\lambda|\frac{(N_{1}f_{1})^{n_{1}}(N_{2}f_{2})^{n_{2}}}{2^{(n_{1}+n_{2})/2-1}}\cos\left(\frac{n_{1}}{N_{1}}\frac{a_{1}}{f_{1}}\pm\frac{n_{2}}{N_{2}}\frac{a_{2}}{f_{2}}-\arg(\lambda)\right)\,.
\end{equation}
where the (+) sign corresponds to the case $S_{1}^{n_{1}}S_{2}^{n_{2}}$ in \cref{eq:scalar_breaking} and the (-) sign to $S_{1}^{n_{1}}(S^{(\dagger)}_{2})^{n_{2}}$. Any other case is equivalent by an overall minus sign that does not affect the cosine. The resulting mass matrix is
\begin{equation}
\bm{M}^{2}=\Lambda_{\mathrm{QCD}}^{4}\left(\begin{array}{cc}
{\displaystyle \frac{1}{f_{1}^{2}}} & {\displaystyle \frac{1}{f_{1}f_{2}}}\\
{\displaystyle \frac{1}{f_{1}f_{2}}} & {\displaystyle \frac{1}{f_{2}^{2}}}
\end{array}\right)+|\lambda|\frac{(N_{1}f_{1})^{n_{1}}(N_{2}f_{2})^{n_{2}}}{2^{(n_{1}+n_{2})/2-1}}\left(\begin{array}{cc}
{\displaystyle \frac{n_{1}^{2}/N_{1}^{2}}{f_{1}^{2}}} & {\displaystyle \pm \frac{(n_{1}/N_{1})(n_{2}/N_{2})}{f_{1}f_{2}}}\\
{\displaystyle \pm\frac{(n_{1}/N_{1})(n_{2}/N_{2})}{f_{1}f_{2}}} & {\displaystyle \frac{n_{2}^{2}/N_{2}^{2}}{f_{2}^{2}}}
\end{array}\right)\,,\label{eq:matrix_coupling}
\end{equation}
\end{widetext}
where $n_{1}+n_{2}=4$ if $\lambda$ is a dimensionless coupling,
otherwise $\lambda$ will account for the Wilson coefficient of a
non-renormalisable operator. In the (+) case, if $N_{1}n_{1}\neq N_{2}n_{2}$ the residual PQ is only broken by QCD, therefore the system reabsorbs $\overline{\theta}$ at the minimum and this is a QCD-PQ system. In this case, the mass matrix reveals that, in general, QCD gives mass to a linear combination while the orthogonal state is either heavier or lighter depending on how the $\lambda$-dependent term compares to $\Lambda_{\mathrm{QCD}}$, as studied in the main text.

In principle, if the coupling in \cref{eq:scalar_breaking}
is renormalisable, we need extremely small values of $\lambda$ so
that one of the axions is not extremely heavy and completely decoupled from the $(m_{a},\, g_{a\gamma\gamma})$ plane. In this sense, we note that $\lambda$ being small is technically natural (i.e. radiatively stable). Alternatively, the coupling $\lambda$ may originate from non-renormalisable operators or at loop-level
if the PQ symmetries are broken in the fermion sector, see for instance~\cite{Hill:1988bu,Frigerio:2011in,deGiorgi:2023tvn,deGiorgi:2024str}, while the residual PQ is preserved. In the QCD-PQ case, one needs to make sure that other couplings breaking the residual PQ are either not present or extremely suppressed. This corresponds to the familiar PQ quality problem arising in our simplified two-axion setup.

A possibility to protect the quality of the model while still generating a scalar coupling that breaks the 2 PQ symmetries down to a residual PQ is incorporating an extended gauge symmetry.\footnote{This idea is similar to the one studied in the context of one axion/ALP, cf.~\cite{Barr:1992qq,Rothstein:1992rh,Greljo:2025suh}.} For example, the tri-hypercharge models of Refs.~\cite{FernandezNavarro:2023rhv,FernandezNavarro:2024hnv,FernandezNavarro:2025zmb} impose the flavour deconstruction of SM hypercharge in the
UV, $\Uone_{Y_{1}+Y_{2}}\times \Uone_{Y_{3}}\rightarrow \Uone_{Y}$.
This is spontaneously broken by scalars with hypercharges $\phi_{q}\sim(-1/6,\, 1/6)$ and $\phi_{\ell}\sim(1/2,\, -1/2)$. This choice of hypercharges is not ad-hoc, but it is chosen so that these scalars couple to SM fermions, dynamically generating the flavour structure of the SM. Remarkably, at the renormalisable level, these allow only the non-self-conjugate coupling $\phi_{q}^{3}\phi_{\ell}$ in the scalar potential, i.e.~corresponding to the $n_{1}=3$ and $n_{2}=1$ case of our general setup. Any operator of the form $\phi_{q}^{n}\phi_{\ell}^{m}$ is forbidden by the gauge symmetry, including non-renormalisable ones, except for $(\phi_{q}^{3}\phi_{\ell})^{n}$, all of them leaving unbroken the residual PQ that is a good candidate for the QCD axion. This is just an example of how one may still achieve accidental PQ symmetries and a high quality solution to strong CP in multi-axion models with
explicit multi-PQ breaking.

Alternatively, the system may contain several couplings of the form of \cref{eq:scalar_breaking} that break all PQ symmetries, including the QCD direction. In this case, to achieve a CP conserving minimum, one may enforce conditions over the CP phases that could be motivated by a symmetry in the UV. An example is the (+) case of \cref{eq:matrix_coupling} when $N_{1}n_{1}= N_{2}n_{2}$, then one may enforce $\bar{\theta}=\arg(\lambda)$ to achieve a CP conserving minimum, and introduce another scalar coupling to give mass to the orthogonal axion. This would constitute an example of general-PQ system as studied in the main text.

\subsection{New strong sectors} \label{app:new_strong_sectors}

\subsubsection{Commuting with QCD}\label{app:commuting_QCD}

In general, new confining sectors introduce new sources of strong
CP violation (see e.g.~\cite{Strassler:2006im,Berezhiani:2000gh,Hook:2014cda,DiLuzio:2021pxd,Chen:2021jcb,Kondo:2025hdc,Lee:2026umy}). For example, new strong sectors under which QCD-PQ is anomalous can generate new CP violating contributions that destabilise the single axion solution to the strong CP problem. In these cases, the effective axion potential gains another term
\begin{equation}
V\approx-\Lambda_{\mathrm{QCD}}^{4}\cos\left(\frac{a}{f}-\overline{\theta}\right)-\Lambda_{d}^{4}\cos\left(\frac{N_{d}}{N}\frac{a}{f}-\overline{\theta}_{d}\right)\,,
\end{equation}
where the additional $\overline{\theta}_{d}$ may originate generically
from a new confining sector that commutes with QCD, or even by coloured gravitational instantons, which are a generic prediction of the SM with General Relativity \cite{Chen:2021jcb}. This motivates the introduction of a second PQ symmetry to relax the new source of CP violation, implying the existence of a ``companion axion'' with the potential
\begin{align}
V\approx&-\Lambda_{\mathrm{QCD}}^{4}\cos\left(\frac{a_{1}}{f_{1}}+\frac{a_{2}}{f_{2}}-\overline{\theta}\right) \nonumber \\ &-\Lambda_{d}^{4}\cos\left(c_{1}\frac{a_{1}}{f_{1}}+c_{2}\frac{a_{2}}{f_{2}}-\overline{\theta}_{d}\right)\:,\label{eq:new_strong_sector}
\end{align}
where $c_{1}=N_{1}^{d}/N_{1}$ and $c_{2}=N_{2}^{d}/N_{2}$.

Imposing $c_1\neq c_2$ solves strong CP and leads to a QCD-PQ system. This leads to the following mass matrix 
\begin{equation}
\bm{M}^{2}=\Lambda_{\mathrm{QCD}}^{4}\left(\begin{array}{cc}
{\displaystyle \frac{1}{f_{1}^{2}}} & {\displaystyle \frac{1}{f_{1}f_{2}}}\\
{\displaystyle \frac{1}{f_{1}f_{2}}} & {\displaystyle \frac{1}{f_{2}^{2}}}
\end{array}\right)+\Lambda_{d}^{4}\left(\begin{array}{cc}
{\displaystyle \frac{c_{1}^{2}}{f_{1}^{2}}} & {\displaystyle \frac{c_{1}c_{2}}{f_{1}f_{2}}}\\
{\displaystyle \frac{c_{1}c_{2}}{f_{1}f_{2}}} & {\displaystyle \frac{c_{2}^{2}}{f_{2}^{2}}}
\end{array}\right)\,.
\end{equation}
If $\Lambda_{d}$ is generated by coloured gravitational instantons,
then it is estimated to be $\Lambda^{4}_{d}\sim(0.04-0.6)\Lambda^{4}_{\mathrm{QCD}}$ \cite{Chen:2021jcb}. Therefore, we expect a QCD-like axion as well as a lighter companion, whose coupling to photons crucially depends on $E_{i}/N_{i}$. Alternatively, dark strong sectors can generate $\Lambda_{d}\gg\Lambda_{\mathrm{QCD}}$. Here, we expect again a QCD axion and a heavier companion. The $G\tilde{G}$ coupling of the QCD-like axion is controlled by $c_1-c_2$, which typically is a $\mathcal{O}(1)$ number.

Instead, in the case $c_1=c_2=c$, one can still achieve a CP conserving minimum in the potential of \cref{eq:new_strong_sector} by imposing $\bar{\theta}=c\,\bar{\theta}$. A natural realisation of this condition originates if the new strong sector resembles a mirror QCD,
\begin{equation}
\QCD\times \SU{3}_{\mathrm{mirror}}\,,
\end{equation}
where a discrete symmetry $\mathbb{Z}_2$ exchanges both strong sectors $\QCD\leftrightarrow \SU{3}_\mathrm{mirror}$. This naturally imposes $\overline{\theta}=\overline{\theta}_{\mathrm{mirror}}$, and a degenerate PQ sector for both strong sectors, i.e. naturally realising $c_{1}=c_{2}=1$. In this case, the QCD axion is $a_1+a_2$ while the orthogonal combination is massless. We may lift the mass of the orthogonal combination by adding a suitable scalar coupling of the form of \cref{eq:scalar_breaking}. For example, we choose $\lambda (S_1S_2^{\dagger})^2$. The resulting mass matrix is (assuming $N_{1,2}=1$ for simplicity)
\begin{align}
\bm{M}^{2}= & (\Lambda_{\mathrm{QCD}}^{4}+\Lambda_{\mathrm{mirror}}^{4})\left(\begin{array}{cc}
{\displaystyle \frac{1}{f_{1}^{2}}} & {\displaystyle \frac{1}{f_{1}f_{2}}}\\
{\displaystyle \frac{1}{f_{1}f_{2}}} & {\displaystyle \frac{1}{f_{2}^{2}}}
\end{array}\right)\nonumber \\
 & +2|\lambda|f_{1}^{2}f_{2}^{2}\left(\begin{array}{cc}
{\displaystyle \frac{1}{f_{1}^{2}}} & {\displaystyle -\frac{1}{f_{1}f_{2}}}\\
{\displaystyle -\frac{1}{f_{1}f_{2}}} & {\displaystyle \frac{1}{f_{2}^{2}}}
\end{array}\right)\,.
\end{align}
If $\SU{3}_{\mathrm{mirror}}$ is spontaneously broken at a very high scale, the small-size instantons generating $\Lambda_{\mathrm{mirror}}$ can be calculated reliably using the dilute instanton gas approximation (DIGA) \cite{Callan:1977gz,tHooft:1976snw,Affleck:1980mp,Csaki:2019vte}, and in general one finds $\Lambda_{\mathrm{mirror}}\gg\LambdaQCD$, such that the axion solving strong CP in this system can be much heavier than the traditional QCD axion. This is another example of general-PQ system, where all PQ symmetries are broken beyond QCD effects, and yet the minimum is CP conserving. This simplified model can be consistently extended to full mirror world constructions \cite{Berezhiani:2000gh}.

Another possibility would be to impose $\bar{\theta}_d=\bar{\theta}+\pi$ while $c_1=c_2=1$ in the potential of \cref{eq:new_strong_sector}. Such a potential would be naturally generated in a mirror world construction where $\mathbb{Z}_2$ is non-linearly realised by one axion field as $a_i \rightarrow a_i+\pi f_{i}$ (in the UV theory, this could be realised e.g.~if a PQ scalar is odd under $\mathbb{Z}_2$), as in \cite{Hook:2018jle,DiLuzio:2021pxd}. Then the two cosines interfere negatively as
\begin{flalign}
V\supset & -\Lambda_{\mathrm{QCD}}^{4}\cos\left(\frac{a_{1}}{f_{1}}+\frac{a_{2}}{f_{2}}-\overline{\theta}\right) \nonumber\\
 & -\Lambda_{\mathrm{mirror}}^{4}\cos\left(\frac{a_{1}}{f_{1}}+\frac{a_{2}}{f_{2}}-\overline{\theta}-\pi\right)\nonumber\\
 & =-\left[\Lambda_{\mathrm{QCD}}^{4}+\Lambda_{\mathrm{mirror}}^{4}\cos(\pi)\right]\cos\left(\frac{a_{1}}{f_{1}}+\frac{a_{2}}{f_{2}}-\overline{\theta}\right)\,.
\end{flalign}
Expanding near the vaccuum reveals that the mass-decay constant relation of the QCD-like state goes like $m^2_a f^2_a\sim |\Lambda^{4}_{\mathrm{QCD}}-\Lambda^{4}_{\mathrm{mirror}}|$. If $\Lambda_{\mathrm{mirror}} \approx \LambdaQCD$, then the leading cosine approximation is not reliable, and a full treatment of the axion potential reveals that the CP conserving extremum is actually not a minimum but a maximum. However, we find that if $\Lambda_{\mathrm{mirror}}/\LambdaQCD\lesssim 0.77$, the minimum is CP conserving, and by saturating the inequality the QCD-like axion may be found significantly to the left of the canonical QCD band. This is another example of general-PQ system with non-trivial axion phenomenology, and yet the minimum of the axion potential is CP conserving, and the strong CP problem is solved.

Note that in models with new confining sectors, the criteria to identify a domain wall problem depends on the determinant of the anomaly coefficients matrix,
\begin{equation}
\mathcal{\bm N}=\left(\begin{array}{cc}
N_{1} & N_{2}\\
N_{1}^d & N_{2}^{d}
\end{array}\right)\,,
\end{equation}
such that $N_{\mathrm{DW}}=\det(\mathcal{\bm N})$. Therefore, multi-axion models of this kind allow for large anomaly coefficients while avoiding a domain wall problem if $\det(\mathcal{\bm N})=1$ \cite{Kondo:2025hdc}.

\subsubsection{Embedding of QCD} \label{app:QCDdiag}

Let us now assume that QCD emerges as the diagonal subgroup of a larger
strong sector (see e.g.~\cite{Agrawal:2017ksf,Gaillard:2018xgk,Fuentes-Martin:2019bue,Csaki:2019vte}). For simplicity, let us assume two fundamental $\SU{3}$ factors in the UV which are spontaneously broken down to QCD by the
VEV of a bifundamental scalar $\Phi\sim(\mathbf{3},\mathbf{\bar{3}})$,

\begin{equation}
\SU{3}_{\mathrm{1}}\times \SU{3}_{\mathrm{2}}\overset{\langle\Phi\rangle}{\longrightarrow}\SU{3}_{\mathrm{QCD}}\,.
\end{equation}
This implies a matching condition between the gauge couplings of the
larger strong sector and the QCD gauge coupling,
\begin{equation}
g_{s}=\frac{g_{1}g_{2}}{\sqrt{g_{1}^{2}+g_{2}^{2}}}\,.
\end{equation}
For each separate coupling, this implies $g_{1,2}>g_{s}$ at the matching
scale, meaning that both new sectors are necessarily strongly coupled. In
the unbroken phase, the model allows for two independent sources of
strong CP violation,
\begin{equation}
\mathcal{L}=\frac{\alpha_{1}}{8\pi}\bar{\theta}_{1}G_{1\mu\nu}^{a}\tilde{G}_{1}^{a\,\mu\nu}+\frac{\alpha_{2}}{8\pi}\bar{\theta}_{2}G_{2\mu\nu}^{a}\tilde{G}_{2}^{a\,\mu\nu}\,.
\end{equation}
In the broken phase, one gets the usual QCD term as $\bar{\theta}=\bar{\theta}_{1}+\bar{\theta}_{2}$.
One may introduce one axion to relax only $\bar{\theta}$,
but then one still has dangerous CP violation from the broken direction,
which couples to all SM fermions as well regardless of the fermion embedding in the full symmetry. Moreover, new instantons induced by $\SU{3}_{1}$ and $\SU{3}_{2}$ introduce new contributions to the axion potential that destabilise the single-axion solution to the strong CP problem. Therefore, this motivates the introduction
of two axions to relax $\bar{\theta}_{1}$ and $\bar{\theta}_{2}$ simultaneously. The two-axion potential is then given by
\begin{align}
V\approx &-\Lambda_{\mathrm{QCD}}^{4}\cos\left(\frac{a_{1}}{f_{1}}+\frac{a_{2}}{f_{2}}-\overline{\theta}\right)\nonumber \\ &-\Lambda_{1}^{4}\cos\left(c_{1}\frac{a_{1}}{f_{1}}-\overline{\theta}_{1}\right)-\Lambda_{2}^{4}\cos\left(c_{2}\frac{a_{2}}{f_{2}}-\overline{\theta}_{2}\right)\,,\label{eq:FD_potential}
\end{align}
where $c_{1}=N'_{1}/N_{1}$ and $c_{2}=N'_{2}/N_{2}$ take into account
potentially different anomaly coefficients. If $c_1=c_2=1$, the minimum of this potential $\langle a_{1}/f_{1}-\overline{\theta}_{1}\rangle=\langle a_{2}/f_{2}-\overline{\theta}_{2}\rangle=0$
is CP conserving and solves the strong CP problem by a generalisation of the Vafa-Witten theorem~\cite{Vafa:1984xg}. It also removes CP violation from the broken direction and protects against $\SU{3}_{1}$ and $\SU{3}_{2}$ instantons breaking
the single axion solution to strong CP. Therefore, this class of models naturally motivates the existence of a second axion. The model is naturally UV-completed by having two PQ symmetries, one anomalous under each $\SU{3}_{i}$. Then we expect the fermions inducing these anomalies to transform simultaneously under one single $\SU{3}_{i}$, naturally generating $c_{1,2}=1$. One can see that in the axion potential of \cref{eq:FD_potential}, there is no PQ symmetry only broken by QCD, but the minimum is CP conserving. This constitutes another example of general-PQ system. The resulting mass matrix is
\begin{equation}
\bm{M}^{2}=\Lambda_{\mathrm{QCD}}^{4}\left(\begin{array}{cc}
{\displaystyle \frac{1}{f_{1}^{2}}} & {\displaystyle \frac{1}{f_{1}f_{2}}}\\
{\displaystyle \frac{1}{f_{1}f_{2}}} & {\displaystyle \frac{1}{f_{2}^{2}}}
\end{array}\right)+\left(\begin{array}{cc}
{\displaystyle \frac{\Lambda_{1}^{4}}{f_{1}^{2}}} & {\displaystyle 0}\\
0 & {\displaystyle \frac{\Lambda_{2}^{4}}{f_{2}^{2}}}
\end{array}\right)\,.
\end{equation}
The phenomenology is crucially determined by how the $\Lambda_{1,2}$
compare to $\Lambda_{\mathrm{QCD}}$. As in the mirror QCD example, the small-size instantons generating $\Lambda_{i}$ can be calculated reliably using again the DIGA \cite{Callan:1977gz,tHooft:1976snw,Affleck:1980mp,Csaki:2019vte}. It is beyond the scope of this paper to do a quantitative analysis, but we will highlight the qualitative behaviour and give some indicative values of what can be achieved. Both $\Lambda_{1,2}$ grow with larger VEVs $\langle\Phi\rangle$, and separately with their corresponding gauge couplings. Taking into account the matching condition, $g_{1}\gg g_{s}\simeq g_{2}$ enhances $\Lambda_{1}$, while $g_{2}\gg g_{s}\simeq g_{1}$ enhances $\Lambda_2$. Moreover, fermions
transforming under a given $\SU{3}_{i}$ suppress the corresponding $\Lambda_{i}$. For example, assuming for simplicity all SM quarks embedded under $\SU{3}_{1}$, then typically $\Lambda_{1}\ll\Lambda_{2}$ for most values of the gauge couplings. The suppression of fermions becomes less relevant as the VEV increases, e.g.~for $\langle\Phi\rangle\sim10^{17}\,\mathrm{GeV}$
we find the largest simultaneous value of $\Lambda_{i}$ to be
\begin{equation}
\Lambda_{1}\sim\Lambda_{2}\sim1\,\mathrm{GeV}\gg\Lambda_{\mathrm{QCD}}\,,
\end{equation}
showing that they can be simultaneously larger than $\Lambda_{\mathrm{QCD}}$,
leaving no axion in the QCD line. Indeed this is not only an example of a general-PQ system, but one in which the breaking effects of the QCD-PQ direction can be large enough to have important phenomenological consequences, while still solving strong CP.

For smaller VEVs, typically one
finds $\Lambda_{2}\gg\Lambda_{\mathrm{QCD}}\gg\Lambda_{1}$, meaning
that one axion is in the QCD line and the other one is heavier. However, for very small VEVs such as $\langle\Phi\rangle=1\,\mathrm{TeV}$ and $g_{2}\ll g_{1}$ we find configurations were 
\begin{equation}
\Lambda_{1}\sim\Lambda_{2}\sim10^{-3}\,\mathrm{GeV}\ll\Lambda_{\mathrm{QCD}}\,,
\end{equation}
where we have one axion in the QCD line and the other one is much lighter, with the position in the ($m_{a},g_{a\gamma\gamma}$)  plane crucially depending on the $E_{i}/N_{i}$ ratios. We comment that if the third family of quarks is embedded in a different $\SU{3}$ with respect to the light families, one can generate dynamically small CKM quark mixing. Such a model will be explored in great detail in an upcoming publication~\cite{FernandezNavarro:wip}. More generally, this construction is a fundamental component in many models of flavour deconstruction, a non-exhaustive list includes~\cite{Salam:1979p,Craig:2011yk,Greljo:2018tuh,Bordone:2017bld,FernandezNavarro:2023hrf}. All of them lead in general to large strong CP violation effects and would benefit from a multi-axion implementation.

Another well-motivated class of models that generates general-PQ systems is that of non-trivial embeddings of QCD into a parent-colour group, see e.g ~\cite{Dimopoulos:1979pp,Valenti:2022tsc}. The breaking of the parent group provides a new strong sector that commutes with QCD but whose strong CP-violating phase is aligned with $\bar{\theta}$ due to the gauge structure, and an axion solution still solves strong CP with the PQ symmetry broken by QCD and by non-perturbative effects of the commuting subgroup. The qualitative effect is similar to linearly realised $\mathbb{Z}_2$ in mirror world constructions, discussed in the previous section.

We conclude that the presence of new strong sectors beyond QCD naturally motivates the existence of multi-axion systems and allow for general-PQ systems. Well-motivated UV models can realise the complete phenomenology described in the main text.

\subsection{Axions from extra dimensions}\label{app:extra_dimensions}

For simplicity, we will just consider the case of a single bulk axion model in flat 5D spacetime (to go beyond our simplified scenario, see e.g.~\cite{deGiorgi:2024elx,Flacke:2006ad}). The metric is then $ds^{2}=\eta_{\mu\nu}dx_{\mu}dx_{\nu}-dy^{2}$, with $\eta_{\mu\nu}=\mathrm{diag(1,-1,-1,-1)}$ being the Minkowski metric in the mostly minus convention, with $x^{\mu}$ the 4D coordinates and $y$ representing the extra dimension. We assume the extra dimension is compactified on a $S^{1}/\mathbb{Z}_2$ orbifold with $y\,\in\,[0,\pi R]$, imposing the identification $y\rightarrow-y$. Therefore, the model is controlled by a single geometric parameter, the size of the extra dimension $R$, and the fixed points at $y=0$ and $y=\pi R$ define 4D branes. We assume the PQ symmetry is implemented by a $\mathbb{Z}_2$-even complex scalar field in the bulk, $\langle\Phi\rangle=f_{5}/\sqrt{2}$, such that the axion is the corresponding angular mode $a(x,y)$. The fermion content will induce a single anomaly coefficient ratio $E/N$ for the 5D axion. Now we assume that QCD lives on the $y=0$ brane, such that the action is
\begin{align}
S=&\int d^{4}x\int_{0}^{\pi R}dy\frac{1}{2}\left((\partial_{\mu}a)^{2}-(\partial_{y}a)^{2}\right) \nonumber \\ &+\int d^{4}x\frac{a(x,0)}{f_{5}}\frac{\alpha_{s}}{8\pi}G\widetilde{G}\,.\label{eq:5D_axion}
\end{align}
The strong CP problem is automatically solved by the PQ mechanism, now realised in 5D. Since the extra dimension is compact, to obtain the effective 4D theory we expand the axion field in Kaluza-Klein (KK) modes
\begin{equation}
a(x,y)=\sum_{n=0}^{\infty}a_{n}(x)\psi_{n}(y)\,,\label{eq:KK_modes}
\end{equation}
where $a_{n}(x)$ represents the 4D degrees of freedom and $\psi_{n}(y)$ the
wave function profiles in the extra dimension. To find the correct $\psi_{n}$ we require that the action becomes diagonal in KK modes,
then the $y$-derivative term returns the eigenvalue equation
\begin{equation}
-\frac{d^{2}\psi_{n}(y)}{dy^{2}}=m_{n}^{2}\psi_{n}(y)\,.
\end{equation}
Imposing orbifold boundary conditions, only modes even under $y\rightarrow-y$ are retained. Then we obtain $m_{n}=n/R$ with $n=0,1,2,...$, $\psi_{0}=C$ and $\psi_{n}=A\cos(m_{n}y)$. We finally fix the constants by requiring canonically normalised 4D kinetic terms to obtain 
\begin{equation}
a(x,y)=\frac{1}{\sqrt{\pi R}}a_{0}(x)+\sqrt{\frac{2}{\pi R}}\sum_{n=1}^{\infty}a_{n}(x)\cos\left(\frac{n}{R}y\right)\,.
\end{equation}
Reinserting this expression into \cref{eq:5D_axion} we obtain the effective 4D action,
\begin{align}
S=&\int d^{4}x\frac{1}{2}\left[\sum_{n=0}^{\infty}(\partial_{\mu}a_{n})^{2}-\sum_{n=1}^{\infty}m_{n}^{2}a_{n}^{2}\right]\nonumber \\ &+\int d^{4}x\frac{\alpha_{s}}{8\pi}\left(\frac{a_{0}}{f_{0}}+\sum_{n=1}^{\infty}\frac{a_{n}}{f_{n}}\right)G\widetilde{G}\,,
\end{align}
where $f_{0}=\sqrt{\pi R}f_{5}$, $f_{n}=f_{0}/\sqrt{2}$
and the 5D decay constant has mass dimension 3/2. The resulting 4D theory therefore contains a single axion zero mode together with a tower of massive KK pseudoscalars. Since QCD couples to the axion field evaluated on the brane, the instanton potential depends on the linear combination of all KK modes appearing in $a(x,0)$, leading
to a mixed multi-field potential for the axion and its KK excitations,
\begin{equation}
V\approx-\Lambda^4_{\mathrm{QCD}}\cos\left(\frac{a_{0}}{f_{0}}+\sum_{n=1}^{\infty}\frac{a_{n}}{f_{n}}\right)+\frac{1}{2}\sum_{n=1}^{\infty}m_{n}^{2}a_{n}^{2}\,.
\end{equation}
Regardless of the fact that the KK modes cannot be interpreted as pseudo-Goldstone bosons (their shift symmetries are explicitly broken by the KK masses), one can still interpret the system as an (infinite) set of approximate $\Uone$ symmetries, inherited from the higher-dimensional PQ symmetry, of which a single linear combination is broken by QCD while the explicit KK masses break the remaining ones. Indeed, the mass matrix of the system is fully captured by the general formalism of the main text, and therefore also the phenomenology. As a simplified example, let us now truncate the KK sum for the first mode to achieve an effective two-axion system,
\begin{equation}
\bm{M}^{2}=\frac{\Lambda_{\mathrm{QCD}}^{4}}{f_{0}^{2}}\left(\begin{array}{cc}
1 & \sqrt{2}\\
\sqrt{2} & {\displaystyle 2}
\end{array}\right)+\frac{1}{R^{2}}\left(\begin{array}{cc}
0 & {\displaystyle 0}\\
0 & 1
\end{array}\right)\,.
\end{equation}
Therefore, we generally expect a QCD-like axion plus a lighter or heavier companion depending on the size of $R^{-1}$. If only the 5D axion lives in the bulk, then $R^{-1}$ is almost unconstrained and the exotic companion can be light or heavy. Instead, if gravity also propagates in the bulk then $R^{-1}\gtrsim10^{-3}\mathrm{eV}$ \cite{Kapner:2006si}, which is typically
larger or at most comparable to the QCD contribution for the typical values of $f_{0}$ (which can still be independently controlled by $f_{5}$). When including higher modes, the mass matrix is
\begin{widetext}
    \begin{equation}
\bm{M}^{2}=\frac{\Lambda_{\mathrm{QCD}}^{4}}{f_{0}^{2}}\left(\begin{array}{ccccc}
1 & \sqrt{2} & \sqrt{2} & \cdots & \sqrt{2}\\
\sqrt{2} & {\displaystyle 2} & 2 & \cdots & 2\\
\sqrt{2} & 2 & 2 & \cdots & 2\\
\vdots & \vdots & \vdots & \ddots & \vdots\\
\sqrt{2} & 2 & 2 & \cdots & 2
\end{array}\right)+\frac{1}{R^{2}}\left(\begin{array}{ccccc}
0 & {\displaystyle 0} & 0 & \cdots & 0\\
0 & 1 & 0 & \cdots & 0\\
0 & 0 & 4 & \cdots & 0\\
\vdots & \vdots & \vdots & \ddots & \vdots\\
0 & 0 & 0 & \cdots & n^{2}
\end{array}\right)\,.
\end{equation}
\end{widetext}
In this form, the connection with our general multi-axion formalism in \cref{sec:MultiAxion} is manifest: the brane-localised QCD contribution gives a rank-one matrix \(\Lambda_{\rm QCD}^4\bm{N}\bm{N}^T\), while the KK masses provide a positive diagonal contribution to \(\bm{M}_{\cancel{\mathrm{PQ}}}^2\). The $n$-th KK mode gets a heavier mass with respect to the first mode by a factor $n$. If we allow for small $R^{-1}$, the model predicts a full multi-axion spectrum that spans some of the regimes studied in \cref{sec:generalization}.

Finally, if QED is localised in the $y=0$ brane like QCD, the photon coupling in the 5D theory is
\begin{equation}
S\supset\int d^{4}x\frac{E}{N}\frac{a(x,0)}{f_{5}}\frac{\alpha_{\mathrm{em}}}{8\pi}F\tilde{F}\,,
\end{equation}
so all modes effectively share the same $E/N$ and inherit the same pion-mixing factor after QCD confinement. This could change by doing simple modifications to the 5D axion model:
\begin{itemize}
\item Assume QCD and QED propagate in the bulk, but heavy PQ fermions that induce anomalies are localised on different branes, e.g.~one fermion in $y=0$ and another one in $y=\pi R$. When we insert the KK expansion, each KK mode couples with different strength
\begin{equation}
\left(\frac{E}{N}\right)_{n}=\frac{E_{1}\psi_{n}(0)+E_{2}\psi_{n}(\pi R)}{N_{1}\psi_{n}(0)+N_{2}\psi_{n}(\pi R)}\,.
\end{equation}
\item Introduce multiple bulk axions. This is essentially a 5D generalisation
of the 4D multi-PQ model studied in \cref{app:Multi-Peccei-Quinn-model-building}.
Each PQ may carry its own $E/N$ ratio as usual, and the different sets of 4D KK modes inherit these coefficients.
\item Localise QCD and QED in different branes\footnote{In more realistic constructions QCD and QED can indeed live in different brane stacks, as in intersecting D-brane constructions in type II string theory. This provides different $E/N$ factors for different axions and KK modes \cite{Halverson:2019cmy,Gendler:2023kjt}, as illustrated in our simplified example. However, the two branes must intersect at some point for a realistic SM.}, e.g.~in $y=0$ and $y=\pi R$
respectively, such that their overlap with the bulk axion differs. Then 
\begin{equation}
\left(\frac{E}{N}\right)_{n}=\frac{\psi_{n}(\pi R)}{\psi_{n}(0)}\,.
\end{equation}
Since in our case $\psi_{n}(\pi R)/\psi_n(0)=(-1)^n$, the $E/N$ ratio alternates sign for the KK tower.
\end{itemize}

Interestingly, the presence of 5D gauge fields propagating in the bulk may naturally generate an effective 5D bulk axion model such as the one we just described. For simplicity, let us consider a 5D abelian gauge field $A_{M}$ with $M=\mu,5$, with gauge symmetry
acting as
\begin{equation}
A_{M}\rightarrow A_{M}+\partial_{M}\Lambda(x,y)
\end{equation}
The $A_{\mu}$ components describe a 4D gauge field, $A_{\mu}\rightarrow A_{\mu}+\partial_{\mu}\Lambda(x,y)$,
while the 5th component $A_{y}\equiv A_{5}$ describes a 4D scalar and transforms as $A_{5}\rightarrow A_{5}+\partial_{y}\Lambda(x,y)$. We assume $A_\mu$ is odd under $\mathbb{Z}_2$ while $A_5$ is even, so that only the latter has a zero mode in the $y=0$ brane. For gauge parameters compatible with the boundary conditions, we have $\Lambda(x,-y)=-\Lambda(x,y)$. In particular, the transformation $\Lambda(x,y)=\alpha \, y$, with constant $\alpha$, induces $A_{5}\rightarrow A_{5}+\alpha$. This corresponds to a constant shift that defines a residual global symmetry, identified with the Peccei–Quinn symmetry.

If there exist massive fermions in the QCD brane $y=0$ chiral under the residual PQ, then they can induce a Chern-Simons-like term in the brane. The action of the $A_{5}(x,y)$ field then contains 
\begin{align}
S\supset & \int d^{4}x\int dy\frac{1}{2}\left((\partial_{\mu}A_{5})^{2}-(\partial_{y}A_{5})^{2}\right)\nonumber \\
 & +\int d^{4}x\,\frac{A_{5}(x,0)}{f_5}\frac{\alpha_{s}}{8\pi}G\widetilde{G}\,.\label{eq:5D_gauge_axion}
\end{align}
where $f_5=1/(Ng^2_5\sqrt{\pi R})$ in our toy model, with $g_5$ being the gauge coupling of the bulk $\Uone$ with mass dimension -1/2 and $N$ the QCD anomaly coefficient. We can see that the fifth component of the gauge field $A_5(x,y)$ behaves like the bulk axion in Eq. \eqref{eq:5D_axion}, leading to a similar multi-axion model\footnote{Strictly speaking, in our toy $S^1/\mathbb{Z}_2$ gauge-theory realisation, the non-zero KK modes of $A_5(x,y)$ are eaten by the massive KK vectors via the Stückelberg mechanism, so only the zero mode survives as a physical state. More general boundary conditions, boundary symmetry breaking, or enlarged gauge sectors can instead leave a physical tower of KK pseudoscalars from $A_5(x,y)$, thereby reproducing the genuine bulk axion phenomenology.} in the effective 4D theory.

This toy model captures the core mechanism by which multiple axions emerge from higher-dimensional theories such as string theory. The extra degrees of freedom from higher-dimensional gauge fields can provide 4D scalars exhibiting a shift symmetry. In realistic string constructions, additional axions arise from higher-form gauge fields propagating in non-trivial compact manifolds, such as Calabi–Yau manifolds. In particular, in intersecting D-brane constructions in type II string theory, QCD and QED are commonly localised in distinct brane stacks that intersect. From the point of view of multi-axion systems, this naturally leads to non-universal $E/N$ factors for the various axions \cite{Halverson:2019cmy,Gendler:2023kjt}, as illustrated by our toy 5D model with QCD and QED localised on separate branes.

In conclusion, extra-dimensional theories naturally generate multi-axion systems in the 4D effective theory, which motivate the phenomenological scenarios described in the main text. Here we have developed a toy 5D model that illustrates the basic mechanism at work in more realistic constructions. An important message is that some classes of extra dimensional and stringy theories generically predict non-universal $E/N$.

\section{Alternative parameterisation and general sum rule}
\label{app:Hell}
\subsection{General results}
\label{app:generaldecomposition}
\textit{\textbf{Definitions: }}In this Appendix, we provide further details on the decomposition of the mass matrix $\bm{M}^2$ introduced in \cref{eq:gen_mass_matrix}, and connect it to an alternative parameterisation that illustrates, in full generality, how PQ-breaking sources decompose along the QCD direction.
Consider a generic $n\times n$ system
\begin{equation}
\bm{M}^2=\Lambda_{\mathrm{QCD}}^4\bm{N}\bm{N}^{T}+\bm{M}^2_{\cancel{\rm PQ}}\,.
\end{equation}
First, we normalise $\bm{N}$ and define $\hat{\bm{N}}=\bm{N} / \left| \bm{N} \right|$. Then, we choose an orthonormal basis $\left\{\bm{N}_1^{\perp},\ldots,\bm{N}_{n-1}^{\perp}\right\}$ of the orthogonal subspace, satisfying
\begin{equation}
\hat{\bm{N}}^{T}\cdot\hat{\bm{N}}_i^{\perp}=0,
\quad
\hat{\bm{N}}_i^{\perp T}\cdot\hat{\bm{N}}_j^{\perp}=\delta_{ij}\,,
\end{equation}
and we collect these orthogonal vectors into a matrix
\begin{equation}
\bm{N}^{\perp}
=
\left(
\begin{array}{lll}
\hat{\bm{N}}_1^{\perp} & \cdots & \hat{\bm{N}}_{n-1}^{\perp}
\end{array}
\right)\,.
\end{equation}
The decomposition of $\bm{M}^2$ becomes
\begin{equation}
\bm{M}^2
=
\beta\,\hat{\bm{N}}\hat{\bm{N}}^{T}
+\hat{\bm{N}} \bm{\delta}^{\,T}\bm{N}^{\perp T}
+\bm{N}^{\perp}\bm{\delta}\,\hat{\bm{N}}^{T}
+\bm{N}^{\perp}\bm{\Sigma} \bm{N}^{\perp T}\,.
\label{eq:decMmybasis}
\end{equation}
The coefficient $\beta$ is given by
\begin{equation}
\beta=\hat{\bm{N}}^{T}\bm{M}_{\cancel{\rm PQ}}^2\hat{\bm{N}}+\Lambda_{\mathrm{QCD}}^4\,\left| \bm{N} \right|^2\,.
\end{equation}
The mixing between the QCD direction and the orthogonal subspace is parameterised by $\bm{\delta}$, whose components are
\begin{equation}
\bm{\delta}_i=\hat{\bm{N}}^{T} \bm{M}^2\hat{\bm{N}}^{\perp}_i \,,
\end{equation}
while the orthogonal subspace is given by a symmetric $(n-1) \times (n-1)$ matrix, whose entries are:
\begin{equation}
\bm{\Sigma}_{i j}=\hat{\bm{N}}^{\perp T}_i \bm{M}^2 \hat{\bm{N}}^{\perp}_j \,.
\end{equation}
\textit{\textbf{Dictionary between parameterisations: }}The correspondence with the parameterisation in the main text, \cref{eq:mastereq}, is given by
\begin{equation}
\beta=\LambdaQCDbar^{4}\left| \bm{N} \right|^2+ \hat{\bm{N}}^T \bm{M}_A^2 \hat{\bm{N}} = \LambdaQCDbar^{4}\left| \bm{N} \right|^2+ \beta_{A}\,,
\label{eq:correspondancebeta}
\end{equation}
while $\bm{\Sigma}$ and $\bm{\delta}$ are the same i.e.
\begin{equation}
    \begin{aligned}
    &\left(\bm{\Sigma}_A\right)_{ij} \equiv \hat{\bm{N}}_i^{\perp T} \bm{M}_A^2 \hat{\bm{N}}_j^{\perp} = \hat{\bm{N}}^{\perp T}_i \bm{M}^2\hat{\bm{N}}^{\perp}_j = \bm{\Sigma}_{ij} \,, \\
         &\left(\bm{\delta}_A\right)_i \equiv \hat{\bm{N}}^T \bm{M}_A^2 \hat{\bm{N}}_i^{\perp} = \hat{\bm{N}}^{T} \bm{M}^2\hat{\bm{N}}^{\perp}_i = \bm{\delta}_i \,,
    \end{aligned}
\label{eq:correspondancesigmadelta}
\end{equation}
because the $\LambdaQCDbar$-term has vanishing overlap with the vector space spanned by $\{\hat{\bm{N}}^\perp_i \}$. Equivalently, we see that, while the $\LambdaQCDbar$-term contains only contributions from $\bm{v}_I$ vectors exactly parallel to $\bm{N}$, the matrix $\bm{M}_A^2$ contains the orthogonal and mixed components, as well as the parallel projections of the non-fully-aligned vectors (i.e.~$\beta_A$):
\begin{equation}
\begin{aligned}
        &\LambdaQCDbar^4 \bm{N} \bm{N}^T\equiv \frac{\left(\beta -\beta_A\right)}{\left| \bm{N} \right|^2} \bm{N} \bm{N}^T  \,, \\
        & \bm{M}_A^2 \equiv \beta_A\hat{\bm{N}}\hat{\bm{N}}^{T}
+\hat{\bm{N}} \bm{\delta}^{\,T}\bm{N}^{\perp T}
+\bm{N}^{\perp}\bm{\delta}\hat{\bm{N}}^{T}
+\bm{N}^{\perp}\bm{\Sigma} \bm{N}^{\perp T}.
\end{aligned}
\end{equation}
Equivalently, we can also relate $\beta$, $\bm \delta$ and $\bm \Sigma$ to the explicit decomposition of the axion potential in \cref{sec:Lambda_bar_definition}:
\begin{equation}
\begin{aligned}
&\beta = \nonumber \\ &\left(\Lambda_{\mathrm{QCD}}^{4}+\sum_{I\in\mathcal{C}}\kappa_{I}\Lambda_{I}^{4}\frac{\xi_{I}^{2}}{\left| \bm{N} \right|^2}\right)\left|\bm{N}\right|^2+\sum_{I\in\mathcal A}\kappa_I\Lambda_I^4\,(\hat{\bm N}^T\cdot\bm v_I)^2 \\&=\LambdaQCDbar^4 \left|\bm{N}\right|^2+\sum_{I\in\mathcal A}\kappa_I\Lambda_I^4\,(\hat{\bm N}^T\cdot\bm v_I)^2, \\
&\bm{\delta}_i  = \sum_{I\in\mathcal A}\kappa_I\Lambda_I^4\,(\hat{\bm N}^T\cdot\bm v_I)( \bm v_I^T\cdot  \hat{\bm{N}}^{\perp }_i), \\
&\bm{\Sigma}_{ij}  = \sum_{I\in\mathcal A}\kappa_I\Lambda_I^4\,(\hat{\bm{N}}^{\perp T}_i\bm v_I )(\bm v_I^T\cdot \hat{\bm{N}}^{\perp }_j)\,.
\end{aligned}
\label{eq:correspondancemariotomarko}
\end{equation}
\textit{\textbf{Positivity condition: }}The positivity condition on the full mass matrix $\bm{M}^2$ expressed in the parameterisation of \cref{eq:mastereq} is given by 
\begin{equation}
\LambdaQCDbar^{4}\left| \bm{N} \right|^2+\beta_{A}-\bm{\delta}_{A}^{T}\bm{\Sigma}_{A}^{-1}\bm{\delta}_{A} \overset{!}{>} 0 \text{ and } \bm{\Sigma}_A \overset{!}{\succ} 0 \,,
\label{eq:positivityA}
\end{equation}
where the symbol $\succ 0$ denotes positive-definiteness. Note that $\bm{M}_A^2$ is not required to be positive definite, hence $\beta_A$ could be negative, whereas positivity of the full mass matrix forces $\beta=\LambdaQCDbar^4 \left| \bm{N} \right|^2 + \beta_A > 0$. In particular, two interesting regimes can be seen from the positivity conditions:
\begin{itemize}
\item If $\LambdaQCDbar^4$ is negative, the potentially dangerous direction that may lead to a negative eigenvalue, and hence to a violation of positivity, is aligned with $\hat{\bm N}$. Positivity of the full mass matrix then requires the non-aligned sector $\bm{M}_A^2$ to compensate along this direction. In particular, while positivity already requires $\bm{\Sigma}_A \succ 0$ (and hence $\bm{\delta}_{A}^{T}\bm{\Sigma}_{A}^{-1}\bm{\delta}_{A} > 0$) in the orthogonal subspace, one further needs a sufficiently large \emph{positive} projection along the QCD direction,
\begin{equation}
\beta_A > \bm{\delta}_A^{T}\bm{\Sigma}_A^{-1}\bm{\delta}_A 
+ |\LambdaQCDbar^{4}|\left| \bm{N} \right|^2 >0\,,
\end{equation}
which in particular implies $\beta_A>0$ and therefore that $\bm{M}_A^2$ is positive definite. In the basis where $\bm{M}_A^2$ is diagonal, this means that all its eigenvalues are positive, and that the spectrum is dominated by the $\mathcal{A}$ sector. The system then reproduces qualitatively the regime $\Lambda_{i}\gg\bar{\Lambda}_{\mathrm{QCD}}$ explored in \cref{sec:Twoaxion,sec:generalization}.
\item Alternatively, $\bm{M}_A^2$ can have only one negative eigenvalue only if
\begin{equation}
\beta_A - \bm{\delta}_A^{T}\bm{\Sigma}_A^{-1}\bm{\delta}_A <0\,,
\end{equation}
in which case the instability lies again along the QCD-aligned direction. Positivity of the full mass matrix then requires a positive aligned contribution, so that necessarily $\LambdaQCDbar^{4}>0$, with
\begin{equation}
\LambdaQCDbar^4\,\left| \bm{N} \right|^2
>
\bm{\delta}_A^{T}\bm{\Sigma}_A^{-1}\bm{\delta}_A - \beta_A >0\,.
\end{equation}
In this case, if the mixing term dominates, $\bm{\delta}_A^{T}\bm{\Sigma}_A^{-1}\bm{\delta}_A \gg \LambdaQCDbar^{4}\,\left| \bm{N} \right|^2$, the system is effectively $\bm{M}_A^2$-dominated and reduces qualitatively to the regime $\Lambda_i \gg \bar{\Lambda}_{\mathrm{QCD}}$. Conversely, if $\LambdaQCDbar^{4}\,\left| \bm{N} \right|^2 \gg \bm{\delta}_A^{T}\bm{\Sigma}_A^{-1}\bm{\delta}_A$, the light mode aligns with the (generalised) QCD direction.
\end{itemize}
The parameterisation adopted in the main text makes explicit the non-trivial interplay, highlighted at the end of \cref{sec:Lambda_bar_definition}, between $\bar{\Lambda}_{\rm QCD}^4$ and $\beta_A$: while their sum is required to be positive, the individual contributions need not be.
\noindent \newline
\textit{\textbf{Details on the $\{\hat{\bm{N}}, \bm{N}^\perp \}$ basis: }}Coming back to the parameterisation given in \cref{eq:decMmybasis}, we can define the matrix $\bm{V} \equiv (\hat{\bm{N}} \,\,\hat{\bm{N}}_1^{\perp} \ldots \hat{\bm{N}}_{n-1}^{\perp})$, and rotate the mass matrix in the new basis to get
\begin{equation}
\bm{V}^T \boldsymbol{M}^2 \bm{V}=\left(\begin{array}{cc}
\beta & \boldsymbol{\delta}^T \\
\boldsymbol{\delta} & \bm{\Sigma}
\end{array}\right) \,,
\end{equation}
which is positive-definite provided that
\begin{equation}
\bm{\Sigma} \overset{!}{\succ} 0 \quad \text { and } \quad \beta-\boldsymbol{\delta}^T \bm{\Sigma}^{-1} \boldsymbol{\delta}\overset{!}{>}0 \,.
\label{eq:positivityappendix}
\end{equation}
In particular, this also implies $\beta >0$. The inverse, necessary to compute the sum rule in this basis, is given by
\begin{equation}
\boldsymbol{M}^{-2}=\left(\begin{array}{cc}
\left(\beta-\boldsymbol{\delta}^T \bm{\Sigma}^{-1} \boldsymbol{\delta}\right)^{-1} & -\dfrac{\boldsymbol{\delta}^T \bm{\Sigma}^{-1}}{\beta-\boldsymbol{\delta}^T \bm{\Sigma}^{-1} \boldsymbol{\delta}} \\
-\dfrac{\bm{\Sigma}^{-1} \boldsymbol{\delta}}{\beta-\boldsymbol{\delta}^T \bm{\Sigma}^{-1} \boldsymbol{\delta}} & \bm{\Sigma}^{-1}+\dfrac{\bm{\Sigma}^{-1} \boldsymbol{\delta} \boldsymbol{\delta}^T \bm{\Sigma}^{-1}}{\beta-\boldsymbol{\delta}^T \bm{\Sigma}^{-1} \boldsymbol{\delta}}
\end{array}\right) \,.
\end{equation}
The Schur complement $\beta - \boldsymbol{\delta}^T \bm{\Sigma}^{-1} \boldsymbol{\delta} \equiv \det \bm{M}^2/\det \bm{\Sigma}$ corresponds to the effective curvature along the QCD direction $\hat{\bm N}$ after accounting for mixing with the orthogonal subspace. In other words, it represents the component of the mass matrix aligned with $\hat{\bm N}$ once the mixing with the orthogonal sector has been removed. Physically, it can be interpreted as the reduction of the naive parameter $\beta$ due to the ``leaking" of the system along orthogonal directions, and it vanishes precisely when a flat direction develops in the spectrum i.e.~when there is at least one massless axion. 

Note that $\bm{\Sigma}$ corresponds to the restriction of $\bm M^2$ to the subspace orthogonal to $ \hat{\bm{N}}$. Since $\bm M^2$ is assumed to be positive definite (all axions are massive i.e.~no zero eigenvalues), it is invertible and all its principal submatrices are also positive definite. In particular, $\bm{\Sigma}$ is positive definite and therefore invertible, with full rank $n-1$. If $ \hat{\bm{N}}$ is an eigenvector of $\bm M^2$ (with eigenvalue $\beta$), the mixing vector $\bm\delta$ vanishes and $\bm{\Sigma}$ simply coincides with the block of $\bm M^2$ acting on the orthogonal subspace. Otherwise, when $ \hat{\bm{N}}$ is not an eigenvector, $\bm\delta \neq \bm{0}$ but $\bm{\Sigma}$ still describes the dynamics in the $(n-1)$-dimensional orthogonal subspace.
\noindent \newline
\textit{\textbf{General sum rule: }}In this basis, similarly to \cref{sec:sumrule_2axions}, we can  compute the general sum rule. We start by decomposing the photon coupling in our new basis:
\begin{align}
\boldsymbol{g}_{a\gamma}&\equiv\begin{pmatrix}
    g_\parallel & \bm{g}_{\perp}
\end{pmatrix}^T =g_\parallel \hat{\bm{N}}+\sum_i g_{a_i\gamma}^\perp \hat{\bm{N}}_i^{\perp} \nonumber \\ &=\frac{\alpha_{{\rm em}}}{2\pi}\left[(E^{\parallel}-C_{\chi}|\bm{N}|)\hat{\bm{N}}+E^{\perp}_i\hat{\bm{N}}_i^{\perp}\right].
\label{eq:photon_coupl_decomp}
\end{align}
This makes the computation of the sum rule straightforward, as we simply contract $\bm{M}^{-2}$ on both sides with the anomaly vector to obtain:
\begin{equation}
\boxed{\sum_i \frac{\tilde{g}_{a_i\gamma}^2}{m_i^2}=\boldsymbol{g}_{\perp}^T \bm{\Sigma}^{-1} \boldsymbol{g}_{\perp}+\frac{\left(g_\parallel-\boldsymbol{\delta}^T \bm{\Sigma}^{-1} \boldsymbol{g}_{\perp}\right)^2}{\LambdaQCD^4|\bm{N}|^2+\beta_{\cancel{\mathrm{PQ}}}-\boldsymbol{\delta}^T \bm{\Sigma}^{-1} \boldsymbol{\delta}}}
\label{eq:newbasissumrule}
\end{equation}
where $\beta_{\cancel{\mathrm{PQ}}} \equiv \hat{\bm{N}}^{T}\bm{M}_{\cancel{\rm PQ}}^2\hat{\bm{N}}$. The second term of the sum rule shows that the photon coupling of the QCD direction can be modified by mixing with the orthogonal sector. The first term shows that the orthogonal sector always contributes if $\bm{E}$ is not parallel to $\bm{N}$. This general sum rule shows that the system distinguishes between two qualitative regimes:
\begin{itemize}
\item \emph{\underline{Non-universal anomaly coefficients:}} In this regime, the first term can dominate the sum rule. This only happens when $\bm{E}$ is not parallel to $\bm{N}$, and  the best example is when there are light axions. The second term of \cref{eq:newbasissumrule} is typically pulled down by the leaking of parallel photon coupling into the orthogonal direction via mixing, and by the presence of parallel contributions inside $\bm{M}^2_{\cancel{\mathrm{PQ}}}$. Then, if there are orthogonal states that are light and $\bm{E}$ is not parallel to $\bm{N}$, the first term dominates the sum rule, and there are light axions either to the left of the QCD line (these states become orthogonal automatically) or to the right but close to it. The sum rule can be larger than 1.

\item \emph{\underline{Universal anomaly coefficients}: } This regime corresponds to the situation in which anomaly coefficients are universal,
\begin{equation}
\quad E_{1}/N_1=\cdots=E_{n}/N_n
\quad \Longleftrightarrow \quad
\boldsymbol{g}_{\perp} = \bm{0}\,.
\end{equation}
This means that $\bm{E}$ is parallel to $\bm{N}$ and the sum rule reduces to
\begin{equation}
\quad\quad\sum_i \frac{\tilde{g}_{a_i\gamma}^2}{m_i^2}
=
\frac{\left(\frac{\alpha_{{\rm em}}}{2\pi}|\bm{N}|(E/N-C_{\chi})\right)^2}{\LambdaQCD^4|\bm{N}|^2+\beta_{\cancel{\mathrm{PQ}}}-\boldsymbol{\delta}^T \bm{\Sigma}^{-1} \boldsymbol{\delta}}\,,
\end{equation}
Here the light states are orthogonal and photophobic. If there are large PQ-breaking effects very well aligned with QCD, these translate into a large $\beta_{\cancel{\mathrm{PQ}}}$ that could be positive or negative, moving a dominant QCD-like state to the left or to the right of the QCD line. This is the effect captured by $\barLambdaQCD$ in the main text, and the sum rule can be larger or smaller than 1. For generic PQ-breaking that has non-zero parallel projection (even if they are not very well aligned), if these effects are large they will lead to very heavy states that pull the $\bm{\Sigma}$ term down and feed into $\beta_{\cancel{\mathrm{PQ}}}$ until they overcome QCD. This is the $\Lambda_i\gg \LambdaQCD$ regime in the main text (or $\Lambda_i\gg \barLambdaQCD$ if one were to extract the fully aligned effects). The sum rule is then equal to one or smaller for the latter case.
\end{itemize}
\textit{\textbf{Axion coupling to $G\tilde{G}$: }}Additionally, using our general parameterisation, we can derive the axion coupling to $G\tilde{G}$ in the mass basis. Starting from
\begin{equation}
\mathcal L \supset \frac{1}{4}\,\bm g_{ag}^T\cdot\bm a\,G\tilde G \,,
\end{equation}
we rotate the axion fields to the basis $\{\hat{\bm{N}}, \bm{N}^\perp \}$. We denote by $\bm u_i$ the $i$-th normalised eigenvector written in the $\{\hat{\bm{N}}, \bm{N}^\perp \}$ basis as $\bm u_i =
(
u_i^\parallel \,\,\,\,
\bm u_i^\perp
)^T
$. The gluon coupling of the mass eigenstate $\tilde{a}_i$ is
\begin{equation}
\tilde{g}_{\tilde{a}_ig}
=
|\bm{\tilde{g}}_{\tilde{a}g}|\,u_i^\parallel
=
\frac{\alpha_s}{2\pi}|\bm N|\,u_i^\parallel .
\end{equation}
The eigenvalue equation reads
\begin{equation}
\begin{pmatrix}
\beta & \bm\delta^T \\
\bm\delta & \bm\Sigma
\end{pmatrix}
\begin{pmatrix}
u_i^\parallel \\
\bm{u}_i^\perp
\end{pmatrix}
=
m_i^2
\begin{pmatrix}
u_i^\parallel \\
\bm u_i^\perp
\end{pmatrix}.
\end{equation}
Solving the equation in the orthogonal subspace, one finds
\begin{equation}
\bm{u}_i^\perp
=
-(\bm\Sigma-m_i^2\mathbf 1)^{-1}\bm\delta\,u_i^\parallel \,.
\end{equation}
This shows\footnote{In the non-degenerate case, where all eigenvalues are distinct, $(\bm\Sigma-m_i^2\mathbf 1)$ is invertible for any eigenstate with non-vanishing overlap with the QCD direction i.e. $u_i^\parallel\neq0$. If $u_i^\parallel=0$, the corresponding state is entirely contained in the orthogonal subspace and its gluon coupling vanishes. In the presence of degenerate eigenvalues, $(\bm\Sigma-m_i^2\mathbf 1)$ may be non-invertible even for a QCD-coupled eigenspace. Since the QCD direction is one-dimensional, one may choose an orthonormal basis of the degenerate eigenspace such that at most one state has non-vanishing projection
along $\bm N$, while the remaining states are fully orthogonal to $\bm N$ and therefore decouple from gluons. One may then project out the QCD-decoupled directions and work in the complementary subspace, where the inverse of $(\bm\Sigma-m_i^2\mathbf 1)$ , understood as the inverse restricted to that subspace, is well defined.} that the orthogonal component is entirely induced by mixing
and is proportional to $u_i^\parallel$. Imposing the normalization
condition on the eigenvectors and substituting the expression above, one obtains
\begin{equation}
u_i^\parallel
=
\left[
1+
\bm\delta^T
(\bm\Sigma-m_i^2\mathbf 1)^{-2}
\bm\delta
\right]^{-1/2}.
\end{equation}
Therefore, the explicit gluon coupling in the mass basis is
\begin{equation}
\tilde{g}_{\tilde{a}_ig}
=
\frac{\alpha_s}{2\pi}
\frac{\left|\bm{N} \right|}{
\sqrt{
1+
\bm\delta^T
(\bm\Sigma-m_i^2\mathbf 1)^{-2}
\bm\delta}
}\,.
\end{equation}

\subsection{Example: two-axion system}
Let us apply our generic derivation of \cref{app:generaldecomposition} to the simple two-axion system. Consider a mass matrix given by
\begin{equation}
\bm{M}^2=\Lambda_{\mathrm{QCD}}^4\bm{N}\bm{N}^{ T}+\begin{pmatrix}
\dfrac{\Lambda_1^{4 \prime}}{f_1^2} & \dfrac{\Lambda_{12}^{4 \prime}}{f_1f_2} \\
\dfrac{\Lambda_{12}^{4 \prime}}{f_1f_2} & \dfrac{\Lambda_2^{4 \prime}}{f_2^2} 
\end{pmatrix}\,,
\end{equation}
where $\bm{N}=\left( 1/f_1\,, 1/f_2 \right)^T$. We apply the procedure describe in the previous section and rotate $\bm{M}^2$ to a new basis,
\begin{equation}
\bm{V}^T \bm{M}^2 \bm{V}=\left(\begin{array}{ll}
\beta & \delta \\
\delta & \Sigmatwo
\end{array}\right) \,,
\end{equation}
with the rotation defined as
\begin{equation}
   \bm{V} = \frac{1}{\sqrt{f_1^2+f_2^2}}\begin{pmatrix}
        f_2 & f_1 \\ f_1 & -f_2
    \end{pmatrix} \,.
\end{equation}
The matching between $\{\beta, \Sigmatwo, \delta\} \leftrightarrow\left\{\Lambda_1^{\prime}, \Lambda_{12}^{\prime}, \Lambda_2^{\prime}\right\}$ is given by
\begin{equation}
\begin{aligned}
\beta
&= \dfrac{f_2^4\Lambda_1^{\prime 4}+2f_1^2f_2^2\Lambda_{12}^{\prime 4} +f_1^4\Lambda_2^{\prime 4}}{f_1^2f_2^2\left(f_1^2+f_2^2\right)}  \\ &+\LambdaQCD^4 \frac{f_1^2+f_2^2}{f_1^2f_2^2}\,, \\
\Sigmatwo
&= \dfrac{\Lambda_1^{\prime 4}-2\Lambda_{12}^{\prime 4} +\Lambda_2^{\prime 4}}{f_1^2+f_2^2}\,, \\
\delta
&= \dfrac{f_2^2\left(\Lambda_1^{\prime 4}-\Lambda_{12}^{\prime 4} \right)+f_1^2\left(\Lambda_{12}^{\prime 4}-\Lambda_{2}^{\prime 4} \right)}{f_1f_2\left(f_1^2+f_2^2\right)}\,.
\label{eq:beta2x2}
\end{aligned}
\end{equation}
We can now define 
\begin{equation}
        \bm{R}^\prime = \begin{pmatrix}
        \cos \theta^\prime & -\sin \theta^\prime \\ \sin \theta^\prime & \cos \theta^\prime
    \end{pmatrix} \,, \quad \theta^\prime \equiv \frac{1}{2} \arctan\frac{2\delta}{\beta-\Sigmatwo} \,,
\end{equation}
and rotate the mass matrix into diagonal form:
\begin{equation}
\bm{R}^{\prime T}\left(\bm{V}^T \bm{M}^2 \bm{V}\right) \bm{R}^\prime=\begin{pmatrix}
m_+^2 & 0\\
0 & m_-^2
\end{pmatrix} \,.
\end{equation}
Equivalently, the full rotation from the original basis to the mass basis can be written as 
\begin{equation}
    \bm{U}^T \bm{M}^2 \bm{U} = \begin{pmatrix}
m_+^2 & 0\\
0 & m_-^2
\end{pmatrix} \,,
\end{equation}
where we defined\footnote{Notice that, in the limit $\theta'=\phi$, one finds that the total rotation angle $\theta=\phi-\theta'$ vanishes, so that $\bm{U}=\mathbb{1}$. This implies that the mass matrix is already diagonal in the original $(a_1,a_2)$ basis, and no further rotation is required to reach the mass eigenstates. Equivalently, the mixing induced by the mass matrix exactly compensates the rotation to the $\{\hat{\bm N},\bm N^\perp\}$ basis. In this situation, the QCD-aligned direction coincides with one of the mass eigenstates, and the system exhibits no mixing already in the original basis. }
\begin{equation}
    \bm{U} = \begin{pmatrix}
        \cos \theta & -\sin \theta \\ \sin \theta & \cos \theta
    \end{pmatrix}\,, \quad \theta \equiv \phi-\theta^\prime\,, \quad \phi \equiv \arctan \frac{f_1}{f_2}\,.
\end{equation}
The eigenvalues are 
\begin{equation}
m_{ \pm}^2=\frac{\beta+\Sigmatwo}{2} \pm \frac{1}{2} \sqrt{(\beta-\Sigmatwo)^2+4 \delta^2} \,.
\end{equation}
Positive-definiteness translates to
\begin{equation}
    \det \bm{M}^2 > 0 \Longleftrightarrow \beta \Sigmatwo - \delta^2 > 0 \,\text { and } \Sigmatwo >0\,,
    \label{eq:positivity2x2case}
\end{equation}
which implies $\beta > 0$ and that the mixing cannot be arbitrary large for fixed $\beta$ and $\Sigmatwo$. Since $\beta$ is strictly positive, it follows that $\Sigmatwo$ must be sufficiently large, such that $\Sigmatwo > \frac{\delta^2}{\beta} > 0$, or equivalently that $\delta$ must be sufficiently small, $\delta^2 < \beta \Sigmatwo$, in order to ensure positivity of the full mass matrix. In other words, the contribution of PQ breaking in the direction orthogonal to QCD and the contribution in the mixed direction are not independent of the contribution along the QCD direction. In particular, the limit $\Sigmatwo \to 0$ implies that the mixing vanishes too i.e. $\delta \to 0$. 

Before turning to the sum rule in the two-axion system, let us first examine the different regimes and how the positivity condition is realised in each of them.
\\ \\
\textit{\textbf{Vanishing mixing ($\delta = 0$)}}: This is the regime where $\bm{N}$ and $\bm{\hat{N}}^\perp$ are eigenvectors of the full $\bm{M}^2$ matrix, and, in particular, of $\bm{M}_{\cancel{\rm PQ}}^2$ matrix. The eigenvalues of the mass matrix reduce to
\begin{equation}
m_{+}^2 = \max(\beta,\Sigmatwo)\,, \qquad 
m_{-}^2 = \min(\beta,\Sigmatwo)\,,
\end{equation}
so that the mass ordering is determined by whether the QCD-aligned direction ($\beta$) or the orthogonal direction ($\Sigmatwo$) provides the dominant contribution.
\\ \\
\textit{\textbf{Vanishing contribution along $\rm{QCD}_\perp$ ($\Sigmatwo  \to 0$)}}: 
This is the limiting case $\Sigmatwo \to 0$, which, from the positivity condition, implies vanishing mixing ($\delta \to 0$). In this limit, the eigenvalues reduce to
\begin{equation}
m_+^2 \to \beta\,, \qquad m_-^2 \to 0\,,
\end{equation}
so that one state remains massive, while the orthogonal direction becomes parametrically massless due to the absence of explicit breaking along $\rm{QCD}_\perp$.
\\ \\
\textit{\textbf{Degenerate diagonal limit ($\Sigmatwo \to \beta$)}}: Imposing $\Sigmatwo=\beta$ leads to the condition
Imposing $\Sigmatwo = \beta$, the eigenvalues reduce to
\begin{equation}
m_\pm^2 = \beta \pm |\delta|\,,
\end{equation}
so that the degeneracy is lifted by the off-diagonal entry. In this limit, the mixing angle becomes maximal,
\begin{equation}
\theta' \to \frac{\pi}{4}\,\mathrm{sign}(\delta)\,,
\end{equation}
implying that the mass eigenstates are equal combinations of the two directions.
\\ \\
\textit{\textbf{$\beta$-dominated regime}}: 
In the limit $\beta\to\infty$ with $\Sigmatwo$ and $\delta$ fixed, the eigenvalues behave as
\begin{equation}
m_+^2 \simeq \beta \,,
\qquad
m_-^2 \simeq \Sigmatwo \,,
\end{equation}
up to corrections of order $\delta^2/\beta$. Moreover, if the QCD contribution dominates over all PQ-breaking scales, $\Lambda_{\rm QCD}^4 \gg \Lambda_i^{\prime 4}$, one has
\begin{equation}
\beta \simeq \Lambda_{\rm QCD}^4\,\frac{f_1^2+f_2^2}{f_1^2f_2^2}\,,
\qquad
\Sigmatwo \simeq 0\,,
\qquad
\delta \simeq 0 \,.
\end{equation}
Therefore the mass matrix is approximately already diagonal in the $\{\hat{\bm N},\hat{\bm{N}}^\perp\}$ basis, with eigenvalues
\begin{equation}
m_+^2 \simeq \Lambda_{\rm QCD}^4\,\frac{f_1^2+f_2^2}{f_1^2f_2^2}\,,
\qquad
m_-^2 \simeq 0 \,.
\end{equation}
In other words, one eigenstate is a heavy QCD-aligned axion, while the orthogonal combination remains parametrically light and only acquires a mass from the subleading PQ-breaking effects. 

Conversely, in the regime $\Lambda_{\rm QCD}^4 \ll \Lambda_i^{\prime 4}$, the QCD contribution to $\beta$ is subleading, and both eigenvalues are generically set by the PQ-breaking scales. In this case, the orthogonal combination is no longer parametrically light, and a light state can only arise if the PQ-breaking matrix itself exhibits an approximate flat direction.
\\ \\
\textit{\textbf{$\Sigmatwo$-dominated regime}}: 
In the limit $\Sigmatwo\to\infty$ with $\beta$ and $\delta$ fixed, the eigenvalues behave as
\begin{equation}
m_+^2 \simeq \Sigmatwo \,,
\qquad
m_-^2 \simeq \beta \,,
\end{equation}
up to corrections of order $\delta^2/\Sigmatwo$. In this regime, the heavy eigenstate is predominantly aligned with the $\hat{\bm{N}}^\perp$ direction, while the lighter one remains approximately aligned with the QCD direction. Therefore, large PQ-breaking effects along $\mathrm{QCD}_\perp$ can raise the orthogonal eigenvalue without affecting either the QCD-aligned contribution or the mixing.
\\ \\
\textit{\textbf{Maximal mixing regime}}:
Unlike in the previous cases, the mixed contribution cannot be increased arbitrarily. Positivity of the mass matrix requires
\begin{equation}
|\delta| < \sqrt{\beta\Sigmatwo}\,.
\end{equation}
The maximal mixed contribution is therefore approached as this bound is saturated,
\begin{equation}
|\delta| \to \sqrt{\beta\Sigmatwo}\,,
\end{equation}
for which the determinant tends to zero and the eigenvalues become
\begin{equation}
m_-^2 \to 0\,,
\qquad
m_+^2 \to \beta+\Sigmatwo \,.
\end{equation}
Thus, the mixed contribution can be made sizeable only as the mass matrix approaches an approximate flat direction, beyond which positivity is lost.
\subsubsection{QCD-PQ system}
\label{app:rank1story}
Let us now consider the case in which the PQ-breaking matrix has rank one, so that its determinant vanishes,
\begin{equation}
\det
\begin{pmatrix}
\dfrac{\Lambda_1^{\prime 4}}{f_1^2} & \dfrac{\Lambda_{12}^{\prime 4}}{f_1f_2}\\
\dfrac{\Lambda_{12}^{\prime 4}}{f_1f_2} & \dfrac{\Lambda_2^{\prime 4}}{f_2^2}
\end{pmatrix}
=0
\quad \Longleftrightarrow\quad
\Lambda_1^{\prime 4}\Lambda_2^{\prime 4}-(\Lambda_{12}^{\prime 4})^2=0 \,.
\end{equation}
This implies that only two of the three PQ-breaking scales are independent, and that the PQ-breaking contribution is aligned along a single direction in field space. Since the determinant is basis-independent, the same condition holds in the $\{\hat{\bm N},\hat{\bm{N}}^\perp\}$ basis. Defining
\begin{equation}
\beta_{\cancel{\rm PQ}} \equiv \beta - \Lambda_{\rm QCD}^4 \frac{f_1^2+f_2^2}{f_1^2 f_2^2}\,,
\end{equation}
the rank-one condition implies
\begin{equation}
\beta_{\cancel{\rm PQ}}\,\Sigmatwo - \delta^2 = 0
\qquad\Longleftrightarrow\qquad
\delta^2 = \beta_{\cancel{\rm PQ}}\,\Sigmatwo \,.
\end{equation}
Thus, in this limit, the three quantities $\beta$, $\Sigmatwo$ and $\delta$ are no longer independent, and the off-diagonal entry is entirely fixed by the diagonal ones.

As a consequence, the PQ-breaking matrix alone is positive semidefinite, with one vanishing eigenvalue. Its nonzero eigenvalue is given by
\begin{equation}
m_{\cancel{{\rm PQ}},+}^2 = \beta_{\cancel{\rm PQ}} + \Sigmatwo \,, 
\qquad
m_{\cancel{{\rm PQ}},-}^2 = 0 \,.
\end{equation}
Including the QCD contribution, the determinant of the full mass matrix becomes
\begin{equation}
\det \bm M^2 = \beta \Sigmatwo - \delta^2
= \Lambda_{\rm QCD}^4 \frac{f_1^2+f_2^2}{f_1^2 f_2^2}\,\Sigmatwo \,,
\end{equation}
and positivity of the full mass matrix  requires $\Sigmatwo > 0$.

If $\Lambda_{\rm QCD}^4 \gg \Lambda_i^{\prime 4}$, the eigenvalues of the full mass matrix simplify to
\begin{equation}
m_+^2 \simeq \Lambda_{\rm QCD}^4\,\frac{f_1^2+f_2^2}{f_1^2 f_2^2}\,,
\qquad
m_-^2 \simeq \Sigmatwo \,,
\end{equation}
up to corrections suppressed by $\beta_{\cancel{\rm PQ}}/\Lambda_{\rm QCD}^4 \left| \bm{N} \right|^2$. Thus, although the PQ-breaking matrix is rank one and would by itself leave one direction massless, the QCD contribution lifts this flat direction. The heavy eigenstate is aligned with the QCD direction, while the lighter eigenstate corresponds to the orthogonal direction and is controlled by $\Sigmatwo$.

Conversely, in the regime $\Lambda_{\rm QCD}^4 \ll \Lambda_i^{\prime 4}$, the full mass matrix remains close to rank one, and the QCD contribution only weakly lifts the flat direction. The eigenvalues then behave as
\begin{equation}
m_+^2 \simeq \beta_{\cancel{\rm PQ}} + \Sigmatwo \,,
\qquad
m_-^2 \simeq 
\frac{\Lambda_{\rm QCD}^4}{\beta_{\cancel{\rm PQ}}+\Sigmatwo}
\left|\bm{N}\right|^2\,\Sigmatwo \,.
\end{equation}
Hence, the lighter eigenstate is parametrically suppressed and originates from the QCD-induced lifting of the flat direction of the PQ-breaking matrix.

\subsubsection{Sum rule of the two-axion system}
The general sum rule \cref{eq:newbasissumrule} takes a particular simple form in the two-axion system. First, we decompose the photon coupling along the two directions, according to the notation in \cref{eq:photon_coupl_decomp}
\begin{align}
    &g_\parallel = \frac{1}{\sqrt{f_1^2+f_2^2}}\left(\frac{f_2}{f_1} C_{a_1 \gamma}+\frac{f_1}{f_2} C_{a_2 \gamma}\right)\,,\\
    &g_\perp = \frac{C_{a_1 \gamma}-C_{a_2 \gamma}}{\sqrt{f_1^2+f_2^2}} \,.
\end{align}
In the mass basis, we have
\begin{equation}
\tilde{g}_{a_{+} \gamma}=g_{\|} \cos \theta^{\prime}+g_{\perp} \sin \theta^{\prime}, \quad \tilde{g}_{a_{-} \gamma}=g_{\perp} \cos \theta^{\prime} -g_{\|} \sin \theta^{\prime}\,.
\end{equation}
The sum rule of the two-axion system becomes
\begin{widetext}
\begin{equation}
\sum_{i=1}^2 \frac{\tilde{g}_{a_i \gamma}^2}{m_i^2}=\frac{1}{f_1^2+f_2^2} \frac{\beta\left(C_{a_1 \gamma}-C_{a_2 \gamma}\right)^2-2 \delta\left(C_{a_1 \gamma}-C_{a_2 \gamma}\right)\left(\frac{f_2}{f_1} C_{a_1 \gamma}+\frac{f_1}{f_2} C_{a_2 \gamma}\right)+\Sigmatwo\left(\frac{f_2}{f_1} C_{a_1 \gamma}+\frac{f_1}{f_2} C_{a_2 \gamma}\right)^2}{\beta \Sigmatwo-\delta^2}
\end{equation}
\end{widetext}
which makes the QCD-aligned, orthogonal, and mixed contributions particularly transparent in connection with the $C_{a_i\gamma}$. The correspondence\footnote{It is worth stressing what we mean here by ``correspondence''. The basis used in \cref{eq:sum_rule_photon_2axion} is different from the one used here, and is obtained after decomposing $\bm{M}^2 = \LambdaQCDbar^4 \bm{N} \bm{N}^T + \bm{M}_A^2$ starting from the explicit axion potential and diagonalising the non-fully-aligned part $\bm{M}_A^2$, as explained in \cref{sec:Lambda_bar_definition}. In this appendix, we directly work in a basis defined by $\hat{\bm{N}}$ and its orthogonal subspace. To match the two descriptions, we simply identify here how the parameterisation of the main text can be recovered from the parameterisation of this appendix by appropriately choosing the parameters so as to reproduce its structure, without actually performing a change of basis. The exact correspondence is worked out in \cref{eq:correspondancebeta}, \cref{eq:correspondancesigmadelta} and \cref{eq:correspondancemariotomarko}.} with \cref{eq:sum_rule_photon_2axion} is immediate upon setting $\Lambda_{12}\to 0$, replacing $\beta$, $\Sigmatwo$ and $\delta$ using \cref{eq:beta2x2}, and performing the redefinition
\begin{equation}
\beta \;\to\; \beta - \beta_A \,, \qquad 
\beta_A \;\equiv\; \frac{\left(f_1^2+f_2^2\right)\left( \LambdaQCD^4-\LambdaQCDbar^4\right)}{f_1^2 f_2^2}\,.
\end{equation}
The regimes identified in the previous section can be studied for the particularly simple $2 \times 2 $ case:
\begin{itemize}
\item \emph{\underline{Non-universal anomaly coefficients}: } In this regime, if there are light states the sum rule is dominated by
\begin{equation}
\sum_i \frac{\tilde g_{a_i\gamma}^2}{m_i^2}
\approx
\frac{g_\perp^2}{\Sigmatwo}\,.
\end{equation}
Here there is light state (which trivially becomes approximately orthogonal) that dominates the sum rule and is to the left of the QCD line. The other state is QCD-like, but that generalised QCD line can move in parameter space via general-PQ effects. 
\item \emph{\underline{Universal Anomaly Regime}: } In this regime,
\begin{equation}
g_\perp=0\,,
\end{equation}
namely $C_{a_1\gamma}=C_{a_2\gamma}$, so that the photon coupling is fully aligned with the QCD direction $g_\parallel=C_{a\gamma} \sqrt{f_1^2+f_2^2}/f_1f_2$. The sum rule becomes
\begin{equation}
\sum_i \frac{\tilde g_{a_i\gamma}^2}{m_i^2}
=
\frac{\left(\frac{\alpha_{{\rm em}}}{2\pi}|\bm{N}|(E/N-C_{\chi})\right)^2}{\LambdaQCD^4|\bm{N}|^2+\beta_{\cancel{\mathrm{PQ}}}-\frac{\delta^2}{\Sigmatwo}}\,.
\end{equation}
Here either there is a QCD axion, or this is displaced by general-PQ effects, as discussed before. The other state is either light and orthogonal (photophobic) or heavy.
\end{itemize}

\section{Useful formulae for the two-axion system} \label{app:Two-axion_app}
\subsection{Exact diagonalisation}
\label{section:exactdiag}
In this section, we provide more details on the diagonalization of the two-axion system and consider the following matrix:
\begin{equation}
\bm{M}^2=\left(\begin{array}{cc}
\dfrac{\Lambda_1^4+\Lambda_{\rm QCD}^4 }{f_1^2} & \dfrac{\Lambda_{\rm QCD}^4+\Lambda_{12}^4}{f_1 f_2} \\
\dfrac{\Lambda_{\rm QCD}^4+\Lambda_{12}^4}{f_1 f_2} & \dfrac{\Lambda_2^4+\Lambda_{\rm QCD}^4 }{f_2^2}
\end{array}\right) \,.
\label{eq:generic12case}
\end{equation}
We work in the basis obtained after performing the procedure outlined in \cref{sec:Lambda_bar_definition}, while allowing for off-diagonal terms $\Lambda_{12}^4$ for completeness and fully general formulas. It is worth stressing that the procedure guarantees that one can always go to a basis where $\bm{M}_A^2$ is diagonal, and the reader interested in matching to the original formulation of the main text in \cref{eq:2x2text} can do so by setting $\Lambda_{12} \to 0$ and replacing $\Lambda_{\mathrm{QCD}} \to \bar{\Lambda}_{\mathrm{QCD}}$ in the rest of this section.
\\ \\
The eigenvectors that diagonalise $\bm{M}^2$ can be parametrised by an angle $\theta$
\begin{equation}
\bm{u}_{+}=(\cos \theta \,\,\sin \theta)^T, \quad \bm{u}_{-}=(-\sin \theta \,\, \cos \theta)^T\,,
\end{equation}
given by
\begin{align}
    &\cos\theta &&\equiv \sqrt{\frac{f_2^2\left(\Lambda_1^4+\LambdaQCD^4\right)-f_1^2\left(\Lambda_2^4+\LambdaQCD^4\right)+\Delta}{2\Delta}}  \nonumber \\
    & &&=\sqrt{\frac{f_1^2f_2^2 \left([\bm{M}^2]_{11} - [\bm{M}^2]_{22}\right)+\Delta}{2\Delta}} \label{eq:cosrotgen}\\
    &\sin\theta &&\equiv  \frac{f_1f_2\left(\LambdaQCD^4+ \Lambda_{12}^4\right)}{\Delta\sqrt{\frac{f_2^2\left(\Lambda_1^4+\LambdaQCD^4\right)-f_1^2\left(\Lambda_2^4+\LambdaQCD^4\right)+\Delta}{2\Delta}}} \nonumber\\
    & &&=\frac{f_1f_2\left(\LambdaQCD^4+ \Lambda_{12}^4\right)}{\Delta \cos \theta} \label{eq:sinandcosgen}
\end{align}
where we defined
\begin{align}
\Delta \equiv &
\Big[\left(f_2^2 \Lambda_1^4 - f_1^2 \Lambda_2^4 - (f_1^2-f_2^2)\Lambda_{\rm QCD}^4\right)^2 \nonumber\\
&+4 f_1^2 f_2^2 \left(\Lambda_{12}^4 + \Lambda_{\rm QCD}^4\right)^2\Big]^{1/2} \,. \label{eq:deltagen}
\end{align}
The eigenvalues are given by
\begin{align}
    m_+^2 = &\frac{\Lambda_1^4+\LambdaQCD^4}{f_1^2}\cos^2\theta+\frac{\Lambda_2^4+\LambdaQCD^4}{f_2^2}\sin^2\theta \nonumber \\ &+\frac{\LambdaQCD^4+\Lambda_{12}^4 }{f_1 f_2} \sin2\theta\,, \\
        m_-^2 = &\frac{\Lambda_1^4+\LambdaQCD^4}{f_1^2}\sin^2\theta+\frac{\Lambda_2^4+\LambdaQCD^4}{f_2^2}\cos^2\theta \nonumber\\&-\frac{\LambdaQCD^4+\Lambda_{12}^4}{f_1 f_2}  \sin2\theta  \,,
\end{align}
or directly in terms of matrix entries
\begin{equation}
       m_{\pm}^2 \equiv \frac{f_1^2\left(\Lambda_2^4+\LambdaQCD^4\right)+f_2^2\left(\Lambda_1^4+\LambdaQCD^4\right)\pm\Delta}{2f_1^2f_2^2} \,.
\end{equation}
In order to have $m_-^2 > 0$, the following condition must be fulfilled
\begin{equation}
     \left[f_1^2\left(\Lambda_2^4+\LambdaQCD^4\right)+f_2^2\left(\Lambda_1^4+\LambdaQCD^4\right)\right]^2 \overset{!}{>} \Delta^2 \,,
\end{equation}
which is only valid if
\begin{equation}
    \Lambda_{12} \overset{!}{<}\left[\sqrt{\left(\Lambda_1^4+\LambdaQCD^4\right)\left(\Lambda_2^4+\LambdaQCD^4\right)}-\LambdaQCD^4 \right]^{1/4}
    \, .
    \label{eq:positivem-}
\end{equation} 
We can now compute the contributions to the axion sum rule:
\begin{widetext}
    \begin{align}
\frac{\tilde{g}_{a_+ \gamma}^2}{m_+^2}&=\frac{(C_{a_1\gamma} f_2\cos\theta +C_{a_2\gamma} f_1\sin\theta )^2}{f_2^2 \Lambda_1^4 \cos^2{\theta}+f_1^2 \Lambda_2^4 \sin^2\theta +f_1f_2 \Lambda_{12}^4\sin 2\theta+(f_2 \cos\theta+ f_1\sin\theta)^2\LambdaQCD^4} \, , \\
     \frac{\tilde{g}_{a_- \gamma}^2}{m_-^2}&=\frac{(C_{a_2\gamma} f_1\cos\theta -C_{a_1\gamma} f_2\sin\theta)^2}{f_2^2 \Lambda_1^4 \sin^2{\theta}+f_1^2 \Lambda_2^4 \cos^2\theta -f_1f_2 \Lambda_{12}^4\sin 2\theta+(f_2 \sin\theta- f_1\cos\theta)^2\LambdaQCD^4} \, .
\end{align}
\end{widetext}
In the limit $\LambdaQCD \gg \Lambda_{1,2,12}$, the QCD contribution dominates the mass matrix and enforces alignment along the QCD direction. The heavy eigenstate $a_+$ becomes the QCD axion, while $a_-$ is the orthogonal combination that remains insensitive to QCD at leading order. We obtain:
\begin{align}
    \frac{\tilde{g}_{a_+ \gamma}^2}{m_+^2}& \approx \frac{\left(C_{a_2\gamma} f_1^2+C_{a_1\gamma}f_2^2\right)^2}{\left(f_1^2+f_2^2\right)^2\LambdaQCD^4}\,,\\
        \frac{\tilde{g}_{a_- \gamma}^2}{m_-^2}& \approx \frac{\left(C_{a_1\gamma} -C_{a_2\gamma}\right)^2}{\Lambda_1^4+\Lambda_2^4-2\Lambda_{12}^4}\,.
\end{align}
In this regime, the sum rule cleanly separates into a QCD-controlled contribution and a purely orthogonal PQ sector. The orthogonal mode is sensitive only to the \emph{relative misalignment} of anomaly coefficients and decouples if $C_{a_1\gamma}=C_{a_2\gamma}$.

In the limit $\Lambda_1 \gg \Lambda_{2,12,\rm QCD}$, only one eigenvalue becomes parametrically large. The heavy eigenstate aligns with the first axion direction, $a_+\simeq a_1$, while the orthogonal combination remains finite and inherits the residual QCD curvature. We find:
\begin{align}
    \frac{\tilde{g}_{a_+ \gamma}^2}{m_+^2}& \approx\frac{C_{a_1\gamma}^2}{\Lambda_1^4}\,,\\
        \frac{\tilde{g}_{a_- \gamma}^2}{m_-^2}& \approx\frac{C_{a_2\gamma}^2}{\Lambda_2^4+\LambdaQCD^4}\,.
\end{align}
The first contribution exhibits genuine decoupling, suppressed by the large PQ-breaking scale $\Lambda_1$. In contrast, the light eigenstate remains sensitive to both its intrinsic PQ-breaking scale and QCD effects, showing that the QCD curvature is effectively transferred to the surviving light direction.

In the limit $\Lambda_2 \gg \Lambda_{1,12,\rm QCD}$, the situation is reversed: only one eigenstate becomes heavy, aligning with the second axion direction, $a_+\simeq a_2$, while the orthogonal state remains finite and approximately aligned with $a_1$. We obtain:
\begin{align}
    \frac{\tilde{g}_{a_+ \gamma}^2}{m_+^2}& \approx 0\,,\\
        \frac{\tilde{g}_{a_- \gamma}^2}{m_-^2}& \approx\frac{C_{a_1\gamma}^2}{\Lambda_1^4+\LambdaQCD^4}\,.
\end{align}
More precisely, the first contribution scales as $C_{a_2\gamma}^2/\Lambda_2^4$ and vanishes in the strict decoupling limit. This regime therefore exhibits a \emph{screening effect}, where the heavy state becomes invisible to photons, and the sum rule is entirely saturated by the remaining light axion.

Finally, in the limit $\LambdaQCD \to 0$, the QCD contribution vanishes and the mass matrix is entirely controlled by PQ-breaking effects. In this case, the sum rule reduces to
\begin{equation}
    \sum_i \frac{\tilde{g}_{a_i\gamma}^2}{m_i^2}
    \simeq \frac{ C_{a_2 \gamma }^2\Lambda_{1}^4
    +C_{a_1 \gamma }^2\Lambda_{2}^4-2C_{a_1 \gamma }C_{a_2 \gamma }\Lambda_{12}^4}{\Lambda_1^4\Lambda_2^4-\Lambda_{12}^8}
\, .
\end{equation}
In this regime, the sum rule is entirely controlled by the PQ-breaking sector and the interplay between diagonal and off-diagonal entries. In the absence of mixing ($\Lambda_{12}\to 0$), the expression factorizes and the two axions contribute independently. In contrast, as $\Lambda_{12}^4 \to \Lambda_1^2 \Lambda_2^2$, the denominator vanishes and the sum rule is enhanced, signalling the emergence of an approximate flat direction and a parametrically light eigenstate (see also \cref{app:rank1story}).

Overall, these limits illustrate how the sum rule is dynamically redistributed depending on the hierarchy of PQ-breaking scales: QCD dominance enforces alignment along a single direction, while hierarchical PQ scales remove one degree of freedom and project the phenomenology onto the remaining light sector, with non-trivial suppression or enhancement controlled by mixing and anomaly structure.

The limit $\cos \theta \to \frac{f_2}{f_1}\sin \theta$, which gives an axion to the left of the QCD line provided that the QCD scale dominates, can be expressed using \cref{eq:sinandcosgen} as
\begin{equation}
    \cos \theta \to \frac{f_2\sqrt{\Lambda_{12}^4+\LambdaQCD^4}}{\sqrt{\Delta}}\,.
\end{equation}
Replacing with the expression for $\cos \theta$ given by \cref{eq:cosrotgen}, we get an equation for $\Delta$:
\begin{equation}
    f_1^2(\Lambda_2^4+\LambdaQCD^4)+f_2^2(\LambdaQCD^4-\Lambda_1^4+2\Lambda_{12}^4) \overset{!}{=} \Delta\,,
    \label{eq:deltageneric}
\end{equation}
provided that
\begin{equation}
\Lambda_{12}^4 >
\frac{1}{2}\left(
\Lambda_1^4-\Lambda_{\mathrm{QCD}}^4
-\frac{f_1^2}{f_2^2}\left(\Lambda_2^4+\Lambda_{\mathrm{QCD}}^4\right)
\right)\,.
\label{eq:forL12}
\end{equation}
In this case, we can solve \cref{eq:deltageneric} for e.g.~$f_1$, by replacing with the explicit form of $\Delta$ from \cref{eq:deltagen}. We obtain
\begin{equation}
   \cos \theta \to \frac{f_2}{f_1}\sin \theta \Leftrightarrow  f_1 \to \sqrt{\frac{f_2^2\Lambda_{12}^4-f_2^2\Lambda_1^4}{\Lambda_{12}^4-\Lambda_2^4}} \,,
\label{eq:solcosforf1}
\end{equation}
under the assumption that either of the following conditions is satisfied:
\begin{align}
    &1.\quad\Lambda_{12} < \Lambda_1 \,\,\text{and}\,\,\Lambda_{2} > \Lambda_{12}\,, \label{eq:cond1m-positive}\\
    &2.\quad\Lambda_{2} < \Lambda_{12} \,\,\text{and}\,\,\Lambda_{12} > \Lambda_{1}\label{eq:cond2m-positive} \,.
\end{align}
Condition $2.$ is however incompatible with the requirement that $m_-^2$ is always positive i.e.~\cref{eq:positivem-}, so only the regime imposed by condition $1.$ is valid for this limit. The allowed region for $\Lambda_{12}$ is therefore:
\begin{align}
&
\,
\frac{1}{2}\left(
\Lambda_1^4-\Lambda_{\mathrm{QCD}}^4
-\frac{f_1^2}{f_2^2}\left(\Lambda_2^4+\Lambda_{\mathrm{QCD}}^4\right)
\right)
 \\
&<
\Lambda_{12}^4
<
\min\!\left(\Lambda_1^4,\Lambda_2^4\right)\,,
\end{align}
which is non-empty only if
\begin{align}
&\frac{1}{2}\left(\Lambda_1^4-\Lambda_{\mathrm{QCD}}^4-\frac{f_1^2}{f_2^2}\left(\Lambda_2^4+\Lambda_{\mathrm{QCD}}^4\right)\right) \\ &<\min \left(\Lambda_1^4, \Lambda_2^4\right) \,.
\end{align}
If these conditions are satisfied and in the limit \cref{eq:solcosforf1}, we get
\begin{equation}
\begin{aligned}
    \frac{\tilde{g}_{a_- \gamma}^2}{m_-^2} &= \frac{(C_{a_2\gamma}-C_{a_1\gamma})^2}{\left(\Lambda_1^4 +\Lambda_2^4-2\Lambda_{12}^4\right)} \\ & = \left(\frac{\alphaem}{2\pi}\right)^2\frac{(E_1/N_1-E_2/N_2)^2}{\left(\Lambda_1^4 +\Lambda_2^4-2\Lambda_{12}^4\right)}\, ,
    \end{aligned}
\end{equation}
which is positive, again because of condition $1.$. In the limit of equal photon coupling i.e.~$C_{a_+\gamma}=C_{a_-\gamma}=C_{a\gamma}$ and $\LambdaQCD \gg \Lambda_{1,2,12}$, we obtain:
\begin{align}
    &\frac{\tilde{g}_{a_- \gamma}^2}{m_-^2} \approx\nonumber \\ &\frac{C_{a\gamma}^{2}\left(f_{2}^{2}\Lambda_{1}^{4}-f_{1}^{2}\Lambda_{2}^{4}+(f_1^2-f_2^2)\Lambda_{12}^4\right)^{2}}{\left(f_1^2+f_2^2\right)^2\left(\Lambda_{1}^4+\Lambda_2^4-2\Lambda_{12}^4\right)\Lambda_{\rm QCD}^{8}} .
\end{align}
The two-axion systems simplifies in the limit $\LambdaQCD \gg \Lambda_{1,2,12}$ in the same way as in \cref{eq:TwoAxionLimitplus}, except for $m^2_-$ which becomes
\begin{equation}
    m_-^2 \simeq \frac{\Lambda_{1}^4+\Lambda_{2}^4-2\Lambda_{12}^4}{f_1^2+f_2^2} \,.
\end{equation}

As discussed at the beginning of this appendix, matching to \cref{eq:2x2text} requires taking the limit $\Lambda_{12} \to 0$, and all the relevant expressions can be directly obtained from \cref{section:exactdiag} by setting $\Lambda_{12}=0$. Notice that the condition for $m_-^2 > 0$ in \cref{eq:positivem-} becomes trivially satisfied. The limit $\cos \theta \to \frac{f_2}{f_1}\sin \theta$ is obtained by a similar derivation, and the only condition to satisfy is \cref{eq:forL12} with $\Lambda_{12}=0$. We obtain
\begin{equation}
   \cos \theta \to \frac{f_2}{f_1}\sin \theta \Leftrightarrow  f_1 \to \frac{f_2 \Lambda_1^2}{\Lambda_2^2} \,.
   \label{eq:imrunningoutofnames}
\end{equation}
Note that this limit does not assume any hierarchy between $\LambdaQCD$ and $\Lambda_{1,2}$. In terms of the structure of the multi-axion theory, as discussed in \cref{sec:Twoaxion} and \cref{eq:limit_universal_contribution}, the mass matrix induced by the additional PQ-breaking potential is universal for both axion states, and the corresponding eigenvectors are given by a linear combination aligned with the gluon anomaly vector and an orthogonal combination. 

Additionally, it is straightforward to check that in the regime $\LambdaQCD \gg \Lambda_{1,2}$, we have
\begin{align}
    \Delta \xrightarrow[]{\LambdaQCD \gg \Lambda_{1,2}}  \,\,&(f_1^2+f_2^2)\LambdaQCD^4  \nonumber \\& +\frac{\left(f_1^2-f_2^2\right)\left(f_1^2\Lambda_2^4-f_2^2\Lambda_1^4\right)}{f_1^2+f_2^2} \\ &+ \mathcal{O}(1/\LambdaQCD^{4}) \nonumber \,,
\end{align}
and we recover \cref{eq:TwoAxionLimitplus,eq:TwoAxionLimitminus}. In the limit of $f_1 \to f_2 = f$ and $\Lambda_1 \to \Lambda_2 = \Lambda$, the rotation is maximal i.e.~$\cos \theta,\, \sin \theta \to 1/\sqrt{2}$ and, we have
\begin{align}
    & m_+^2 \to \frac{2\LambdaQCD^4+\Lambda^4}{f^2} \xrightarrow[]{\LambdaQCD \gg \Lambda_{1,2}}\frac{2\LambdaQCD^4}{f^2}\,,\\
    & m_-^2 \to \frac{\Lambda^4}{f^2}\xrightarrow[]{\LambdaQCD \gg \Lambda_{1,2}}0 \,.
\end{align}
As it can be seen, if we further assume that $\LambdaQCD \gg \Lambda_{1,2}$, we obtain an axion with a vanishing mass and a QCD axion whose mass is shifted from the QCD line by a factor $\sqrt{2}$, as can be appreciated in \cref{fig:Two_axions} (left).

In the case for two axions, it is interesting to study the value that $\Lambda_1$ must have to exhibit this maxion behaviour for non-universal photon couplings, in the limit where $\Lambda_2\rightarrow 0$ (analogously for $\Lambda_1\rightarrow 0$), that is, when only one $\UPQ$ symmetry remains anomalous under QCD. In this case, $\Lambda_1$ must satisfy the following relation:
\begin{widetext}
\begin{small}
\begin{equation}
\label{eq:Lambda1_2maxions}
   \Lambda_1^4= \frac{\Lambda_{\text{QCD}}^4}{4}\left(\frac{C_{a_1\gamma}^2}{C_{a_2\gamma}^2}\frac{f_1^2}{f_2^2}+2\frac{C_{a_1\gamma}-C_{a_2\gamma}}{C_{a_2\gamma}}\pm\sqrt{\frac{f_1^4}{f_2^4}-2\frac{f_1^2}{f_2^2}\frac{C_{a_1\gamma}^2-6C_{a_1\gamma}C_{a_2\gamma}+4C_{a_2\gamma}^2}{C_{a_2\gamma}^2}+\frac{C_{a_1\gamma}^2}{C_{a_2\gamma}^2}\frac{C_{a_1\gamma}^2+4C_{a_1\gamma}C_{a_2\gamma}-4C_{a_2\gamma}^2}{C_{a_2\gamma}^2}}\right)
\end{equation}
\end{small}
\end{widetext}

\subsection{Solving the two-axion system}
\label{app:explit22rotation}
For the two-axion system, it is instructive to highlight how the parametrisation of the main text can be explicitly derived and how it connects to the one outlined in \cref{app:generaldecomposition}. As detailed in \cref{sec:Lambda_bar_definition}, the decomposition of the mass matrix $\bm{M}^2$ is done by separating the contribution fully-aligned with the QCD direction i.e. $\bm N$. One writes
\begin{equation}
\bm M^2=\bar\Lambda_{\rm QCD}^4\,\bm N\bm N^T+\bm M_A^2\,,
\end{equation}
where we use the notation introduced in \cref{sec:Lambda_bar_definition}. In the orthonormal basis $\{\hat{\bm N},\bm N^\perp\}$, the two contributions take the form
\begin{equation}
\LambdaQCDbar^4\text{-term}=
\begin{pmatrix}
\bar\Lambda_{\rm QCD}^4\left| \bm{N} \right|^2 & 0\\
0 & 0
\end{pmatrix},
\,\,
\bm V^T\bm M_A^2\bm V=
\begin{pmatrix}
\beta_A & \delta\\
\delta & \Sigmatwo
\end{pmatrix},
\end{equation}
where
\begin{equation}
\beta_A=\hat{\bm N}^T\bm M_A^2\hat{\bm N},
\,\,
\delta=\hat{\bm N}^T\bm M_A^2\hat{\bm{N}}^\perp,
\,\,
\Sigmatwo=\hat{\bm{N}}^{\perp T}\bm M_A^2\hat{\bm{N}}^\perp \,,
\label{eq:betandco2x2}
\end{equation}
and $\bm{V} = ( \bm{\hat{N}} \,\,\bm{\hat{N}}^\perp )$. The matrix $\bm M_A^2$ is diagonalised by
\begin{equation}
\bm R_A=
\begin{pmatrix}
\cos\theta_A & -\sin\theta_A\\
\sin\theta_A & \cos\theta_A
\end{pmatrix},
\qquad
\theta_A=\frac12\arctan\frac{2\delta}{\beta_A-\Sigmatwo}\,,
\end{equation}
such that
\begin{equation}
\bm R_A^T
\begin{pmatrix}
\beta_A & \delta\\
\delta & \Sigmatwo
\end{pmatrix}
\bm R_A
= \text{diag}\left(\Lambda_{+}^4/f_{+}^2, \Lambda_{-}^4/f_{-}^2\right)\,,
\label{eq:newdefparanewbases}
\end{equation}
with
\begin{equation}
\frac{\Lambda_{\pm}^4}{f_{\pm}^2}
=
\frac{\beta_A+\Sigmatwo}{2}
\pm
\frac12\sqrt{(\beta_A-\Sigmatwo)^2+4\delta^2}\,.
\end{equation}
The $\LambdaQCDbar$-term is also rotated and becomes
$\bar\Lambda_{\rm QCD}^4\,\bm N^\prime\bm N^{\prime T}$, with $\bm{N}^\prime=\bm R_A^T \bm V^T \bm N$.

Since we diagonalise part of the full mass matrix in \cref{sec:Lambda_bar_definition} i.e. the non-fully-aligned part $\bm{M}_A^2$, it is also useful to provide explicit and general formulas starting from a generic $2 \times 2$ symmetric matrix as in \cref{eq:generic12case}, such that the procedure outline in \cref{sec:Lambda_bar_definition} can be directly implemented for any $2 \times 2$ system. The full mass matrix can be put into the form
\begin{widetext}
\begin{align}
\bm{R}^T\bm{M}^2\bm{R} = &\begin{pmatrix}
\dfrac{f_2^2 \Lambda_1^4 + f_1^2 \Lambda_2^4
+\Delta^\prime}
{2 f_1^2 f_2^2}
&
0
\\[1.2em]
0
&
\dfrac{f_2^2 \Lambda_1^4 + f_1^2 \Lambda_2^4
-\Delta^\prime}
{2 f_1^2 f_2^2}
\end{pmatrix}
\nonumber\\ &+\bar{\Lambda}_{\rm QCD}^4
\begin{pmatrix}
\dfrac{\left(f_2 \cos\phi + f_1 \sin\phi\right)^2}{f_1^2f_2^2}
&
\dfrac{\left(f_2\cos \phi + f_1 \sin \phi\right)\left(f_1 \cos \phi -f_2 \sin \phi \right)}
{ f_1^2f_2^2}
\\[1.2em]
\dfrac{\left(f_2\cos \phi + f_1 \sin \phi\right)\left(f_1 \cos \phi -f_2 \sin \phi \right)}
{ f_1^2f_2^2}
&
 \dfrac{\left(f_1 \cos\phi - f_2 \sin\phi\right)^2}{f_1^2f_2^2}
\end{pmatrix}\,.
\label{eq:matdecomposed}
\end{align}
\end{widetext}
where the rotation matrix is given by 
\begin{equation}
   \bm{R}=\begin{pmatrix}
        \cos \phi & -\sin \phi \\
        \sin \phi & \cos \phi
    \end{pmatrix}\,,\quad \phi \equiv \frac{1}{2}\arctan\!\left(
\frac{2 f_1 f_2 \Lambda_{12}^4}{f_2^2 \Lambda_1^4- f_1^2 \Lambda_2^4}
\right)\,,
\label{eq:defanglecanonic}
\end{equation}
and where we also define
\begin{equation}
    \Delta^\prime \equiv \sqrt{f_2^4 \Lambda_1^8 + f_1^4 \Lambda_2^8 + 2 f_1^2 f_2^2 \left(2\Lambda_{12}^8-\Lambda_1^4 \Lambda_2^4\right)}\,.
\end{equation}
The correspondence between the original parameters $\{f_i,\Lambda_i\}$ in \cref{eq:generic12case}  and the effective parameters in the rotated basis is given by:
\begin{equation}
\frac{1}{f_+}
\equiv
\frac{f_2\cos\phi+f_1\sin\phi}{f_1f_2},
\quad
\frac{1}{f_-}
\equiv
\frac{f_1\cos\phi-f_2\sin\phi}{f_1f_2},
\end{equation}
and
\begin{align}
\Lambda_+^4
&\equiv
\frac{
f_2^2\Lambda_1^4+f_1^2\Lambda_2^4+\Delta'
}
{2\left(f_2\cos\phi+f_1\sin\phi\right)^2}\,, \, \nonumber \\ \Lambda_-^4 
&\equiv
\frac{
f_2^2\Lambda_1^4+f_1^2\Lambda_2^4-\Delta'
}
{2\left(f_1\cos\phi-f_2\sin\phi\right)^2}.
\end{align}
With these definitions, we can write 
\begin{equation}
    \bm{R}^T\bm{M}^2\bm{R} = \text{diag}\left(\Lambda_+^4/f_+^2,\Lambda_-^4/f_-^2 \right) + \LambdaQCDbar^4 \bm{N}^\prime \bm{N}^{\prime T} \,,
\end{equation}
with $\bm{N}^\prime =\bm{R}^T \bm{N}\equiv (1/f_+,1/f_-)^T$.
\subsection{Explicit example}
We can workout the particular example of two-axion system in Ref.~\cite{Agrawal:2017cmd}.\footnote{Other models, such as those in Refs.~\cite{Lee:2025zpn,Choi:2020rgn}, can be mapped to these expressions.} Consider the following axion model
    \begin{equation}
    \begin{aligned}
    \mathcal{L}\supset \frac{1}{8\pi} \Big[&\alpha_h \left(c_{h1} \frac{a_1}{f_1}+ c_{h2}\frac{a_2}{f_2} \right)H\widetilde{H}+\alphas\left( \frac{a_1}{f_1} +\frac{a_2}{f_2} \right)G\widetilde{G} \\ +&\alphaem\left(c_{a_1\gamma} \frac{a_1}{f_1} + c_{a_2\gamma}\frac{a_2}{f_2} \right)F\widetilde{F}\Big]\,,
\end{aligned}
\end{equation}
where, following again the notation of \cref{eq:formulaformass}, we have 
\begin{equation}
   \{ \bm{v}_I\}=\{\bm{N}, \left(c_{h1}/f_1, c_{h2}/f_2\right)^T\}\, ,\,\,\{\Lambda_I\}=\{\LambdaQCD\,,\Lambda_h\}\,,
\end{equation}
giving the mass matrix
\begin{equation}
    M^2= \Lambda_h^4 \begin{pmatrix}
        \dfrac{c_{h1}^2}{f_1^2} & \dfrac{c_{h2} c_{h1}}{f_1f_2}\\
        \dfrac{c_{h2} c_{h1}}{f_1f_2}& \dfrac{c_{h2}^2}{f_2^2}
    \end{pmatrix} +\LambdaQCD^4 \begin{pmatrix}
        \dfrac{1}{f_1^2} & \dfrac{1}{f_1f_2}\\
        \dfrac{1}{f_1f_2}& \dfrac{1}{f_2^2}
    \end{pmatrix} \, .
\end{equation}
Following the procedure introduced in \cref{sec:Lambda_bar_definition}, we immediately identify $\bm{v}_h$ as non-fully-aligned\footnote{In the case $c_{h1}=c_{h2}$, $\bm{v}_h$ is fully aligned with QCD and the orthogonal state is massless.} to QCD and, after diagonalising the $\Lambda_h$ contribution to the mass matrix, we obtain
\begin{widetext}
    \begin{equation}
      M^2= \Lambda_h^4 \begin{pmatrix}
        0 & 0\\
        0& \dfrac{c_{h1}^2}{f_1^2}+\dfrac{c_{h2}^2}{f_2^2}
    \end{pmatrix} +\dfrac{\LambdaQCD^4}{f_1^2 c_{h2}^2+f_2^2 c_{h1}^2} \begin{pmatrix}
(c_{h1}-c_{h2})^2 &
\dfrac{(c_{h1}-c_{h2})\left(c_{h2} f_1^2 + c_{h1} f_2^2\right)}{f_1 f_2} \\
\dfrac{(c_{h1}-c_{h2})\left(c_{h2} f_1^2 + c_{h1} f_2^2\right)}{f_1 f_2} &
\dfrac{\left(c_{h2} f_1^2 + c_{h1} f_2^2\right)^2}{f_1^2 f_2^2}
    \end{pmatrix} \, .
    \label{eq:oneLdec}
\end{equation} 
The QCD contribution remains rank one, while the new QCD and EM vectors are
\begin{equation}
    \bm{N}'=\dfrac{1}{\sqrt{f_1^2 c_{h2}^2+f_2^2 c_{h1}^2}}\begin{pmatrix}
        c_{h1}-c_{h2}  \dfrac{c_{h2}f_1^2+c_{h1} f_2^2}{f_1 f_2 }
    \end{pmatrix}\,, \qquad \bm{E}'=\begin{pmatrix}
        c_{a_2 \gamma} c_{h1}-c_{a_1 \gamma}c_{h2}  \dfrac{c_{a_2 \gamma}c_{h2}f_1^2+c_{a_1 \gamma}c_{h1} f_2^2}{f_1 f_2 }
    \end{pmatrix} \, .
    \end{equation}
\end{widetext}
In the limit where $\Lambda_h\gg \LambdaQCD$ the axion $a_1$ decouples and we are left with a QCD-axion given by
\begin{equation}
    m_{a_- }^2 \simeq\LambdaQCD^4\dfrac{(c_{h1}-c_{h2})^2}{f_1^2 c_{h2}^2+f_2^2 c_{h1}^2}, \, \, g_{a_-\gamma}\simeq\dfrac{c_{a_2 \gamma} c_{h1}-c_{a_1 \gamma}c_{h2}}{\sqrt{f_1^2 c_{h2}^2+f_2^2 c_{h1}^2}}
\end{equation}
so the photon-mass ratio becomes
\begin{equation}
  \LambdaQCD^4 \dfrac{g_{a_-\gamma}^2}{m_{a_-}^2}\simeq\dfrac{(c_{a_2 \gamma} c_{h1}-c_{a_1 \gamma}c_{h2})^2}{(c_{h1}-c_{h2})^2} \, .
\end{equation}
In the case in which $\Lambda_h\ll \LambdaQCD$, the lightest axion is associated to the perpendicular vector of the new $\bm{N}'$ and hence the mass and coupling\footnote{Note that in this limit the photon coupling can be obtained from the original basis $\tilde{g}_{a_-\gamma}=\bm{E}'^T\cdot \hat{\bm{N}}'_\perp=\bm{E}\cdot \hat{\bm{N}}_\perp$. } are given by
\begin{align}
     m_{a_-}^2 =\Lambda_h^4\frac{(c_{h1}-c_{h2})^2}{f_1^2+f_2^2}, \, \, g_{a_-\gamma}=\dfrac{c_{a_2 \gamma} -c_{a_1 \gamma}}{\sqrt{f_1^2+ f_2^2}}   
\end{align}
where if we write the photon-mass ratio
\begin{equation}
    \LambdaQCD^4 \dfrac{g_{a_-\gamma}^2}{m_{a_-}^2}=\left(\dfrac{c_{a_2 \gamma} -c_{a_1 \gamma} }{c_{h1} -c_{h2} } \right)^2\frac{\LambdaQCD^4}{\Lambda_h^4} \, ,
\end{equation}
we note that this is the same result as in \cref{eq:left_of_theQCDline} with the replacement of $\Lambda_1^4+\Lambda_2^4 \to \Lambda_h^4\left(c_{h1}-c_{h2}\right)^2$, which is induced by the reparameterisation of the matrices.

Note that the coefficients $c_{h_i}=N'_i/N_i$ depend on the ratio of anomaly coefficients. Generating hierarchical coefficients via having large fermion represations in the hidden confining group has been put forward as a method to increase the photon couplings of the QCD axion \cite{Agrawal:2017cmd}. For this to occur, however, the EM anomaly vector must be non-aligned. The reason is that if the electromagnetic anomaly vector is universal, it is aligned with the QCD anomaly vector, so both vectors are rotated identically and no deviation from the QCD line is generated. This can be understood directly from the sum rule in \cref{eq:photon_sumrule} and its generalisation in \cref{eq:THE_sumRule}, where it is clear that shifts of this kind require a non-universal electromagnetic anomaly vector. In particular, the ratios $N_i'/N_i$ can be absorbed into a redefinition of the corresponding scales $\tilde{\Lambda}_i$. As a corollary, modifying the anomaly coefficients entering $V_{\cancel{\mathrm{PQ}}}$ cannot evade the previous condition on the photon coupling vector: misalignment of the electromagnetic anomalies remains the crucial ingredient for generating axion states to the left of the QCD line.

This example corresponds to $\Lambda_{12} \to \sqrt{\Lambda_1 \Lambda_2}$ in \cref{eq:matdecomposed}, with the inverse ordering for the eigenvalues. Because of this different ordering, we work here with 
\begin{equation}
   \bm{R}=\begin{pmatrix}
        \cos \phi & \sin \phi \\
        -\sin \phi & \cos \phi
    \end{pmatrix}\, , \quad \phi \equiv \frac{1}{2}\arctan\!\left(
\frac{2 f_1 f_2 \Lambda_{12}^4}{f_1^2 \Lambda_2^4- f_2^2 \Lambda_1^4}
\right)\,.
\end{equation}
Matching \cref{eq:oneLdec} with our parameterisation, we have
\begin{align}
c_{h1}
&=
\frac{2f_1f_2\cos\phi+(f_2^2-f_1^2)\sin\phi}{f_2(f_1^2+f_2^2)},
\\[0.4em]
c_{h2}
&=
\frac{(f_1^2-f_2^2)\cos\phi+2f_1f_2\sin\phi}{f_1(f_1^2+f_2^2)}\,,
\end{align}
which satisfy
\begin{align}
c_{h2}^2f_1^2+c_{h1}^2f_2^2 = 1 \,.
\end{align}
In the parametrisation of \cref{eq:betandco2x2}, we have
\begin{equation}
\begin{aligned}
&\beta-\Lambda_{\rm QCD}^4\,\frac{f_1^2+f_2^2}{f_1^2f_2^2}\equiv \beta_A \\&
=
\kappa_h \Lambda_h^4\,(\hat{\bm N}^T\cdot\bm v_h)^2
=
\kappa_h \Lambda_h^4\,\frac{(c_{h1}+c_{h2})^2}{f_1^2+f_2^2}\,, \\
\delta
&=
\kappa_h \Lambda_h^4\,(\hat{\bm N}^T\cdot\bm v_h)(\hat{\bm N}^{\perp T}\cdot\bm v_h)
=
\kappa_h \Lambda_h^4\,\frac{c_{h1}^2-c_{h2}^2}{f_1^2+f_2^2}\,,
\\[4pt]
\Sigmatwo
&=
\kappa_h \Lambda_h^4\,(\hat{\bm N}^{\perp T}\cdot\bm v_h)^2
=
\kappa_h \Lambda_h^4\,\frac{(c_{h1}-c_{h2})^2}{f_1^2+f_2^2}\,.
\end{aligned}
\end{equation}
Notice that positivity forces $\Lambda_1 \neq \Lambda_2 \Leftrightarrow c_{h1} \neq c_{h2}$, which is equivalent to enforcing $\Sigmatwo > 0$, as required for $\bm{M}^2$ to be positive definite (see \cref{eq:positivity2x2case}). In this setup with $\Lambda_{12} = \sqrt{\Lambda_1 \Lambda_2}$, there are no (real) solutions for $\cos \theta \to \frac{f_2}{f_1}\sin \theta$ in terms of $f_1$ or $f_2$, contrary to the fully generic setup. This can be seen for example by evaluating \cref{eq:solcosforf1} at $\Lambda_{12} = \sqrt{\Lambda_1 \Lambda_2}$, which  implies that e.g. $f_1$ becomes complex. Equivalently, when $\Lambda_{12} = \sqrt{\Lambda_1 \Lambda_2}$, it is not possible to satisfy neither condition 1. in \cref{eq:cond1m-positive} nor condition 2. in \cref{eq:cond2m-positive}. One way to see this is to think of $\sqrt{\Lambda_1\Lambda_2}$ as a geometric mean, and the mean cannot be smaller nor bigger than both numbers separately. Therefore, the condition $\cos \theta \to \frac{f_2}{f_1}\sin \theta$ can only be realised in the regime $\LambdaQCD \gg \Lambda_{1,2}$, where QCD dominates the mixing and enforces alignment along the QCD direction. Expanding in this limit, we obtain
\begin{equation}
    \frac{g_{a_-\gamma}^2}{m_-^2}\approx\frac{\left(C_{a_1}-C_{a_2}\right)^2}{\left(\Lambda_1^2-\Lambda_2^2\right)^2} \,,
\end{equation}
for $C_{a_1} \neq C_{a_2}$. In this case, the orthogonal state remains sensitive only to the relative misalignment of the anomaly coefficients, while its mass is set purely by PQ-breaking scales. In contrast, for aligned photon couplings ($C_{a_1} = C_{a_2} \equiv C_a$), the leading contribution cancels and the result becomes
\begin{equation}
    \frac{g_{a_-\gamma}^2}{m_-^2}\approx \frac{C_a^2\left(f_2^2\Lambda_1^2+f_1^2\Lambda_2^2\right)^2}{\left(f_1^2+f_2^2\right)^2\LambdaQCD^8}  \,.
\end{equation}
This suppression reflects an approximate alignment in coupling space i.e. the orthogonal mode becomes weakly coupled and its contribution is controlled by higher-order effects in the QCD-dominated expansion.

\section{Axion fermion couplings}\label{app:fermion_couplings}

In this appendix, we discuss the behaviour of other couplings relevant for axion phenomenology, such as the couplings to nucleons. The axion Lagrangian at low energies in \cref{eq:gluon_photn_Lagrangian} can be extended with the fermion couplings
\begin{equation}
    \mathcal{L} \supset \frac{\partial_\mu \bm{a}^T}{2}\cdot\left( \bm{c}_{au}\bar u \gamma_\mu \gamma_5 u+\bm{c}_{ad}\bar d \gamma_\mu \gamma_5 d+\bm{c}_{ae}\bar e \gamma_\mu \gamma_5 e \right) \, ,
\end{equation}
where in this notation \begin{equation}
\label{eq:fermion_couplings}
    \bm{c}_{af}=\begin{pmatrix}
    \dfrac{c_{a_1f}}{N_1 f_1} &\dots &  \dfrac{c_{a_N f}}{N_N f_{N}}
\end{pmatrix}^T \, .
\end{equation}
In order to compute the couplings to nucleons, it is convenient to perform a chiral field redefinition of the light quark fields. Following Ref.~\cite{DiLuzio:2020wdo}, we consider the generic transformation
\begin{equation}
    q \;\to\; \exp\!\left(i\gamma_5\,\mathcal{Q}_a\,\frac{\bm{a}^T\!\cdot \bm{N}}{2}\right)\,q \, ,
\end{equation}
where $\mathcal{Q}_a$ is a matrix satisfying $\Tr \mathcal{Q}_a = 1$, so as to remove the gluonic anomalous coupling. A convenient choice is
\begin{equation}
    \mathcal{Q}_a=\frac{M_q^{-1}}{\Tr M_q^{-1}}\,,
\end{equation}
which eliminates axion--pion mass mixing, with $M_q=\mathrm{diag}(m_u,\,m_d)$ the light-quark mass matrix. Under this rotation, the fermion-coupling vectors in \cref{eq:fermion_couplings} are shifted according to
\begin{equation}
    \bm{c}_{af}\;\to\; \bm{c}_{af}-\delta_{fq}\,\mathcal{Q}_a\,\bm{N}\, .
\end{equation}

At low energies, the derivative couplings to nucleons can be written as~\cite{DiLuzio:2020wdo}
\begin{equation}
\mathcal{L}\supset\frac{1}{2}\partial_\mu \bm{a}^T\cdot\bm{c}_{aN}\,\bar{N}\, \gamma^\mu\gamma_5\,N\,,
\end{equation}
with the proton and neutron couplings given by
\begin{widetext}
\begin{align}
\bm{c}_{ap} &= -\left(\frac{m_d}{m_u+m_d}\,\Delta u+\frac{m_u}{m_u+m_d}\,\Delta d\right)\bm{N}
        + \Delta u \, \bm{c}_{au}+ \Delta d\, \bm{c}_{ad} \,,
\\
\bm{c}_{an} &= -\left(\frac{m_u}{m_u+m_d}\,\Delta u+\frac{m_d}{m_u+m_d}\,\Delta d\right)\bm{N}
       +\Delta d\,  \bm{c}_{au} + \Delta u \,\bm{c}_{ad}\,.
\end{align}  
\end{widetext}

Another relevant effect is the RGE evolution of the electron coupling, which receives contributions induced by the photon anomaly vector. Following Refs.~\cite{Chang:1993gm,Srednicki:1985xd,DiLuzio:2020wdo}, in a multi-axion setup one finds
\begin{align}
    &\bm{\delta c}_{ae}
    =\nonumber\\
    &\frac{3\alphaem^2}{4\pi^2}\left[
    \log\!\left(\frac{f_a}{\mu_{\rm IR}}\right)\bm{E}
    -\frac{2}{3}\,\frac{4m_d+m_u}{m_u+m_d}\,
    \log\!\left(\frac{\Lambda_\chi}{\mu_{\rm IR}}\right)\bm{N}
    \right] \, ,
\end{align}
with $\Lambda_\chi \simeq 1\,$GeV.

Altogether, the axion nucleon-couplings contain a component aligned with $\bm{N}$ induced by the chiral rotation, and the electron coupling a running contribution $\bm{\delta c}_{ae}$. Both of these pieces vanish for axion eigenstates orthogonal to the QCD direction. Once again, if the ratios $c_{a_i f}/N_i$ are universal, the corresponding fermion-coupling vector remains aligned with the QCD direction. However, it is arguably more natural for the coefficients $c_{a_if}$ to be universal instead. In that case, if the $N_i$ anomalies differ, the fermion couplings are generically not aligned with the QCD direction.

Finally, it is worth stressing that the running of the electron coupling can itself induce misalignment through the photon anomaly vector. In particular, whenever $\bm{E}\not\propto \bm{N}$, the RGE contribution $\bm{\delta c}_{ae}$ is not aligned with the QCD direction, providing an additional source of non-universality.

\section{Pseudo-Nambu Goldstone bosons with \texorpdfstring{$N=0$}{N=0}}
\label{app:N_0}
Here we consider the existence of pseudo-Nambu Goldstone bosons that originate
from $\Uone$ global symmetries not anomalous under QCD, i.e.~with
QCD anomaly coefficient $N=0$. If they do not mix with axions (which
originate from $\Uone$ anomalous under QCD), then they do not belong
to the multi-axion system, and can trivially appear in the entire
$(m_{a},g_{a\gamma})$ parameter space via their independent mass
scales and QED coefficients $E$. The only non-trivial scenario is
when these states mix with axions. To understand their effect, let
us consider a test system with potential
\begin{equation}
V=-\Lambda_{{\rm QCD}}^{4}\cos\!\left(\frac{a_{1}}{f_{1}}-\bar{\theta}\right)-\Lambda_{h}^{4}\cos\!\left(c_{1}\frac{a_{1}}{f_{1}}+c_{2}\frac{a_{2}}{f_{2}}-\theta_{h}\right).
\end{equation}
The QCD cosine defines the vector $\boldsymbol{N}^{T}=(1/f_{1},0)$,
while the non-QCD cosine defines $v_{h}^{T}=(c_{1}/f_{1},c_{2}/f_{2})$.
To solve strong CP, it is sufficient to impose $c_{2}\neq0$. Expanding
around the minimum, the quadratic mass matrix is 
\begin{equation}
\bm{M}^{2}=\Lambda_{{\rm QCD}}^{4}\begin{pmatrix}\frac{1}{f_{1}^{2}} & 0\\[2pt]
0 & 0
\end{pmatrix}+\Lambda_{h}^{4}\begin{pmatrix}\frac{c_{1}^{2}}{f_{1}^{2}} & \frac{c_{1}c_{2}}{f_{1}f_{2}}\\[2pt]
\frac{c_{1}c_{2}}{f_{1}f_{2}} & \frac{c_{2}^{2}}{f_{2}^{2}}
\end{pmatrix}.
\end{equation}
Thus the system is genuinely mixed only if $c_{1}\neq0$, otherwise
one state becomes a decoupled ALP. The regime $\Lambda_{h}\ll\Lambda_{\mathrm{QCD}}$ trivially
resembles the light axion scenario of the main text: the heavier mass
eigenstate is the QCD axion, approximately aligned to $\boldsymbol{N}$,
while the lighter state gets a suppressed coupling to $G\tilde{G}$
via the small mixing $\sim\Lambda_{h}/\Lambda_{\mathrm{QCD}}$. The
main difference is that the photon coupling of the light state will
be dominated by $E_{2}$, rather than by the difference $E_{1}/N_{1}-E_{2}/N_{2}$
in the main text. Qualitatively, this scenario mimics the light axion
scenarios to the left of the QCD line in the main text.

Naively, the regime $\Lambda_{h}\gg\Lambda_{\mathrm{QCD}}$ seems
non-trivial, because here the off-diagonal blocks dominate and $v_{h}$
is approximately the heavier eigenstate. We define the heavy and light-state
decay constants by 
\begin{equation}
\frac{1}{f_{H}^{2}}\equiv\frac{c_{1}^{2}}{f_{1}^{2}}+\frac{c_{2}^{2}}{f_{2}^{2}},\qquad\frac{1}{f_{L}^{2}}\equiv\frac{f_{H}^{2}}{f_{1}^{2}f_{2}^{2}}.
\end{equation}
The approximate mass eigenvalues and eigenstates are
\begin{equation}
m_{H}^{2}\simeq\frac{\Lambda_{h}^{4}}{f_{H}^{2}},\qquad m_{L}^{2}\simeq\Lambda_{{\rm QCD}}^{4}\frac{c_{2}^{2}}{f_{L}^{2}}.
\end{equation}
\begin{equation}
\boldsymbol{u}_{H}\simeq f_{H}\begin{pmatrix}c_{1}/f_{1}\\[2pt]
c_{2}/f_{2}
\end{pmatrix},\qquad\boldsymbol{u}_{L}\simeq\frac{1}{f_{L}}\begin{pmatrix}-\,c_{2}f_{1}\\[2pt]
c_{1}f_{2}
\end{pmatrix}.
\end{equation}

Crucially, the $G\tilde{G}$ coupling of a mass eigenstate is obtained
by projecting the original QCD anomaly vector onto the corresponding
eigenvector, 
\begin{equation}
\mathcal{L}\supset\frac{\alpha_{s}}{8\pi}\left(\boldsymbol{N}\cdot\boldsymbol{u}_{i}\right)a_{i}\,G\tilde{G}=\frac{\alpha_{s}}{8\pi}\left(\frac{c_{1}f_{H}}{f_{1}^{2}}a_{H}-\frac{c_{2}}{f_{L}}a_{L}\right)\,G\tilde{G}.
\end{equation}
The equation above contains the full story: the heavy state inherits a $G\tilde{G}$ coupling through its $a_{1}$ component, that goes to zero when $c_{1}=0$. The lighter state naturally also gets a $G\tilde{G}$ coupling, unless $c_{2}=0$, when it becomes massless. Both couplings are controlled by effective decay constants that define their own mass-coupling line. We can finally write the $G\tilde{G}$ coupling-mass relations for both states
\begin{equation}
\Lambda_{{\rm QCD}}^{4}\frac{(\boldsymbol{N}\cdot\boldsymbol{u}_{H})^{2}}{m_{H}^{2}}\simeq c_{1}^{2}\left(\frac{f_{H}}{f_{1}}\right)^{4}\left(\frac{\Lambda_{{\rm QCD}}}{\Lambda_{h}}\right)^{4}\ll1.
\end{equation}
\begin{equation}
\frac{(\boldsymbol{N}\cdot\boldsymbol{u}_{L})^{2}}{m_{L}^{2}}\simeq\frac{c_{2}^{2}/f_{L}^{2}}{\Lambda_{{\rm QCD}}^{4}c_{2}^{2}/f_{L}^{2}}=\frac{1}{\Lambda_{{\rm QCD}}^{4}}.
\end{equation}

Here it is clear that the heavy state is generally to the right of
the QCD line, while the light state indeed recovers the usual QCD
axion relation. Therefore, in general, this scenario is qualitatively
similar to the heavy axion scenario described in the main text. Since
both axions inherit $G\tilde{G}$ coupling in general, the presence
of the $E_{2}$ and $E_{2}/N_{2}$ typically leads to just $\mathcal{O}(1)$
variations in the photon coupling (up to fine-tuning).

We have explored in more complicated systems (with three states, combining axions and ALPs and with non-trivial mixing structures among them) that the qualitative phenomenological scenarios remain unchanged, and hence we conclude that the existence of ALPs with $N=0$ can trivially populate
the entire $(m_{a},g_{a\gamma})$ plane (if they do not mix), or mimic the phenomenology of the multi-axion systems from the main text. As such, one may generalise our framework from the main text to include the presence of these states, and then our general framework will cover scenarios that consider this kind of states, like the clockwork mechanism \cite{Farina:2016tgd,Agrawal:2017cmd}.

However, from the model-building point, these states seem less motivated. Even though they can inherit $G\tilde{G}$ couplings through mixing, they do not appear in the QCD cosine so they cannot participate in the relaxation of $\bar{\theta}$. Apart from specific mechanisms to modify the phenomenology of the QCD axion, or from a potential string theory origin, they do not seem particularly well-motivated in the context of the strong CP problem. Compared to extra-dimensional axions, KK towers or product gauge groups, the $N=0$ ALPs that mix with the QCD state seem more model-engineered and less predictive than the QCD-only anomaly structure.

\newpage

\newpage

\newpage
\newpage
%-----------------------------------------------
%-----------------------------------------------------------------------------
%%
\bibliographystyle{BiblioStyle}
\bibliography{refs}

\end{document}